\documentclass[a4paper,11pt]{article}
\usepackage{jcappub}
\usepackage{lineno}
\usepackage{color}
\usepackage{xspace}
\usepackage{manyfoot}
\usepackage{url}
\usepackage{bm}
\usepackage{makecell}
\usepackage{soul}
\usepackage{multirow}
\usepackage{subcaption}
\usepackage{array}
\usepackage{colortbl,xcolor,xfp}
\usepackage[normalem]{ulem}

\definecolor{ForestGreen}{rgb}{0.3, 0.7, 0.3}
\newcommand{\cca}[1]{
  \cellcolor{ForestGreen!#1}
  \ifnum\pdfstrcmp{\fpeval{#1 > 101}}{1}=0 \color{white}\fi#1
}

\newcommand{\quijote}{\textsc{Quijote}\xspace}
\newcommand{\pmwd}{\texttt{pmwd}\xspace}
\newcommand{\quijotepng}{\textsc{QuijotePNG}\xspace}
\newcommand{\pmwdsuite}{\textsc{PNG}-\texttt{pmwd}\xspace}
\newcommand{\deepsets}{\texttt{DeepSets}\xspace}

\newcommand{\lptpng}{\textsc{2LPT-PNG}\xspace}
\newcommand{\cnn}{\texttt{CNN}\xspace}

\newcommand{\perslay}{\texttt{PersLay}\xspace}

\newcommand{\xgb}{\texttt{XGB}\xspace}

\newcommand{\fnl}{f_{\rm NL}}
\newcommand{\fnlloc}{f_{\rm NL}^{\rm loc}}
\newcommand{\fnleq}{f_{\rm NL}^{\rm equil}}

\title{Primordial non-Gaussianity -- Fast simulations and persistent summary statistics}
\author[a]{Juan Calles,}
\author[b,c]{Gabriella Contardo,}
\author[d]{Jorge Nore\~na,}
\author[e]{Jacky H. T. Yip,}
\author[e]{and Gary Shiu}
\affiliation[a]{Instituto de F\'isica y Astronom\'ia, Universidad de Valpara\'iso,\\Avda. Gran Bretaña 1111, Valpara\'iso, Chile}
\affiliation[b]{Center for Astrophysics and Cosmology, University of Nova Gorica,\\Ajdovščina I-5270, Slovenia}
\affiliation[c]{Theoretical and Scientific Data Science, Scuola Internazionale Superiore di Studi Avanzati,\\Trieste 34136, Italy}
\affiliation[d]{Instituto de F\'isica, Pontificia Universidad Cat\'olica de Valpara\'iso, \\Casilla 4950, Valpara\'iso, Chile}
\affiliation[e]{Department of Physics, University of Wisconsin-Madison,\\Madison, WI 53706, USA}

\abstract{
We investigate the sensitivity of topological and traditional summary statistics to primordial non-Gaussianity (PNG) using two suites of simulations. First, we introduce a new simulation suite for PNG, \pmwdsuite, comprising more than $20{,}000$ halo catalogs that vary individually local and equilateral shapes, together with variations in $\Omega_m$ and $\sigma_8$. Second, we carry out a systematic comparison of topological descriptors, as well as powerspectrum and bispectrum measurements, evaluating their constraining power on both local and equilateral $\fnl$ and how this sensitivity varies with halo mass. This dataset enables likelihood-free neural regression of $\fnl$ across multiple halo mass bins for a wide range of summary statistics. Third, we assess the transferability of these learned mappings by testing whether models trained on fast \pmwd simulations can robustly infer on simulations from the \quijotepng suite. We find that a combination of simple descriptive statistics of the topological features (PD-statistics) leads to the best performance to constrain equilateral PNG. We observe that the constraining power of these summaries comes from large-mass halos, with small-mass halos adding noise and degrading performance. Similarly, we find that the transferability of the learned mappings, for both topological and powerspectrum plus bispectrum, degrades if small scales or small-mass halos are included.

}

\begin{document}
\maketitle
\flushbottom

\section{Introduction}

The origin and statistical nature of the primordial curvature perturbation is a central open question in cosmology. Although adiabatic perturbations are known to be approximately Gaussian, many inflationary scenarios predict small but potentially observable deviations from Gaussianity, termed primordial non-Gaussianity (PNG). This is encoded in the primordial bispectrum through both its amplitude, $\fnl$, and its shape. Different inflationary mechanisms have bispectrum signals that peak in different triangle configurations; for example, the squeezed limit for local-type PNG, characteristic of multi-field models, and the equilateral configuration for models with non-canonical kinetic terms. The cosmic microwave background currently provides the most stringent constraints on PNG~\cite{Jung:2025nss,2020A&A...641A...9P}, but its information content is approaching a cosmic-variance limit (see e.g.~\cite{CORE:2016ymi,Kalaja:2020mkq}). This motivates the use of large-scale structure as a complementary probe. Upcoming surveys such as Euclid~\cite{2025A&A...697A...1E}, LSST~\cite{2009arXiv0912.0201L}, and SPHEREx~\cite{2014arXiv1412.4872D} will map the matter distribution with unprecedented precision. The projected constraint from Euclid are at the level of $\sigma(\fnlloc)=2.2$ and $\sigma(\fnleq)=108$, offering a unique opportunity to distinguish between single and multi-fields inflationary models~\cite{Euclid:2025hlc}.

Traditional analyses of large-scale structure (LSS) have relied heavily on the powerspectrum, which fully characterizes Gaussian random fields. However, as we push toward percent-level precision in cosmological inference, it becomes clear that two-point statistics compress the data, discarding much of the structural richness at small scales. Higher-order statistics, such as the bispectrum (whose shapes for single field inflation have been fully classified in \cite{Chen:2006nt}), encode crucial information about PNG, but extracting and interpreting these features is challenging~\cite{Philcox:2021ukg,Jung:2022rtn,Coulton:2022qbc,Bakx:2025pop}. Non-linear gravitational evolution and galaxy biasing also become non-linear introducing degeneracies that contaminate the primordial signal, requiring advanced statistical techniques to disentangle early-universe physics from late-time effects.

This observational leap motivates the development of novel statistical and computational frameworks capable of robustly extracting higher-order information from LSS at smaller scales in an interpretable manner. Among the many approaches explored in recent work are hierarchical statistics~\cite{White:2016yhs,Massara:2020pli,Massara:2022zrf,Marinucci:2024bdq,Massara:2024cvu,Jung:2024esv,Cowell:2024wyl,Cowell:2025mov,Vislosky:2025udw}, skew-spectra~\cite{Hou:2022rcd,Schmittfull:2020hoi,Chen:2024bdg,Hou:2024blc}, the Wavelet Scattering Transform~\cite{Eickenberg:2022qvy,Valogiannis:2022xwu,Valogiannis:2023mxf,Valogiannis:2024rvt,Peron:2024xaw}, k-nearest neighbor methods~\cite{Banerjee:2020umh,Coulton:2023ouk,Ouellette:2025nll}, and Minkowski functionals~\cite{Lippich:2020vpy,Liu:2023qrj,Jiang:2023nzz,Liu:2025haj}. One can also bypass the necessity of building summary statistics, working directly on the ``raw data'' in a field-level inference setting. Such approaches can, for instance, use density fields with CNNs~\cite{SimBIG:2023ywd,Bairagi:2025ytq,Cagliari:2025guv,  Bayer:2025ija, Bairagi:2025ytq}, or leverage neural architectures designed for graphs \cite{Shao:2022mzk,deSanti:2023zzn, Chatterjee:2024eur}, or point-clouds ~\cite{Anagnostidis:2022rbs, huang2025cosmobench}, where neural architectures tailored for set-inputs such as \deepsets~\cite{2017arXiv170306114Z}, \texttt{PointNet}~\cite{2016arXiv161200593Q}, or \texttt{PointMLP}~\cite{2022arXiv220207123M} ensure permutation invariance. 
Some of these novel statistics have been studied for PNG \cite{Coulton:2023ouk,Jung:2024esv,Peron:2024xaw,Vislosky:2025udw,Cagliari:2025guv}.

Topological Data Analysis (TDA), and in particular persistent homology (PH), provides a complementary framework for extracting meaningful features from complex data. 
Persistent homology encodes multi-scale,
non-local correlations in a dataset that are not captured by traditional higher-point statistics. For this reason, it was proposed as a powerful means to probe primordial non-Gaussianity in the CMB \cite{Cole:2017kve} and the LSS \cite{Biagetti:2020skr}.
It was further shown in \cite{Yip:2023vud,Calles:2024cxl} that vectorized topological statistics derived from PH are well-suited for machine learning inference.
In the context of LSS, persistent homology quantifies the shape and connectivity of matter across multiple scales, capturing topological invariants such as connected components, loops, and cavities; corresponding to clusters, filaments, and voids. However, topological information is not directly mapped into a Hilbert space but rather into tuples of birth and death scales for these features, requiring a vectorization postprocessing to enable standard and machine learning-based analyses. Neural designs like \deepsets and \perslay~\cite{carriere2020perslay} offers an automated framework in processing raw topological inputs, while handcrafted summaries (e.g., Betti curves, persistence images) offers an interpretable representations suitable for downstream analysis which can also be processed with more traditional ML methods. Recent work~\cite{2022arXiv221209703A} have systematically compared different approaches for classification of human shapes, images of clothes, and images of textures highlighting trade-offs between expressiveness, stability, and computational cost. Nevertheless, the choice of an appropriate vectorization strategy requires careful consideration of the task and data characteristics, as no universal guideline exists.

The choice of tracers used to build a summary statistics is also crucial \cite{Desjacques:2016bnm}. These tracers are biased relative to the underlying matter distribution, as they preferentially form in high-density regions thus clustering more strongly. While clustering bias can be corrected at large scales using $n$-point correlation functions, which respond differently to biasing, its behavior at nonlinear scales is shaped by the complex physics of galaxy formation. 

Topological summaries present a subtler challenge~\cite{bermejo2024topologicalbiashaloestrace}. The morphology of the cosmic web is not uniquely defined, but instead depends intrinsically on the mass and nature of the tracer population used to reconstruct it. Even after accounting for two-point clustering, different halo populations vary in how they connect and delineate the web, a ``topological bias.'' For instance, low-mass halos tend to scatter along extended filaments and lie near the resolution threshold of observations or halo finders, while higher mass halos track large-scale structures. Additionally, halos of identical mass can reside in vastly different environments, a halo in a dense cluster may display richer substructure and connectivity than one near a void. 

In this paper, we study the sensitivity of various summary statistics, including topological-based summaries, clustering statistics, and their combinations, derived from halo catalogs, to the PNG amplitude ($\fnl$). We evaluate these statistics across a range of halo mass bins in a likelihood-free setting, using neural networks to map data vectors directly to $\fnl$ values as a test for their information content.

To support this work, we release the \pmwdsuite simulation suite, introducing $22{,}410$ new halo catalogs sampled via Latin hypercube design, spanning both local and equilateral PNG types. A key question of observational relevance is whether models trained on simulations can generalize to real data analysis~\cite{Heitmann:2004gz,Heitmann:2007hr,Schneider:2015yka,Angulo:2021kes}. As a first step, to carry out this task, we test the robustness of our models when applied to different simulations not seen during training. At its core, \pmwdsuite was designed to evaluate such robustness with respect to the \quijotepng suites, which vary in their N-body solvers and numerical approximations.

The paper is structured as follow: Section~\ref{sec:ph} showcase a lightning review of the persistent homology framework applied to LSS. Section~\ref{sec:pngpmwd_suite} introduces the new \pmwdsuite suite of halo catalogs, and Section~\ref{sec:datasets} details the datasets used in this work along with the post-processing pipeline, including the construction of persistence diagrams, topological summaries, and the powerspectrum and bispectrum vectorizations. In Section~\ref{sec:training_hp}, we define the likelihood-free inference framework. Section~\ref{sec:results} and Section~\ref{sec:transfer} presents and discusses our main results, comparing the performance of all summary statistics across mass bins and evaluating their ability to constrain $\fnl$ and generalize to other datasets. Finally, Section~\ref{sec:conclusions} broadens the implications of our findings and outlines future research directions. To complement the main text, the appendices offer comprehensive technical details and implementation notes.

\section{The topological framework}
\label{sec:ph}

Topological Data Analysis (TDA) offers a framework for capturing the global structure of complex datasets by analyzing their connectivity and organization across multiple scales. TDA extracts a rich set of topological features (such as connected components, loops, and cavities) that provide deeper insight into the underlying processes generating the data. These techniques are rooted in algebraic topology, particularly concepts introduced by Henri Poincaré and later formalized through persistent homology.

In the following, we provide a concise overview of the core concepts to understand persistent homology and their relevance to cosmological applications. For rigorous definitions and formalism, we refer the reader to standard textbooks and comprehensive reviews~\cite{books/daglib/0025666,carlsson2021topological,wasserman2016topologicaldataanalysis}.

\subsection{Persistent homology overview}
\label{sec:ph_overview}
A central tool in TDA is persistent homology, which tracks the emergence and disappearance of topological features as a scale parameter varies, and establishes a principled framework to recover the underlying topology. In the context of large scale structure of the universe, we typically observe a finite set of dark matter tracers (e.g. galaxies) distributed on or near this latent structure\footnote{These tracers can be viewed as realizations sampled from the underlying continuous dark matter density field.}. To study its connectivity, one builds a simplicial complex, defined as a collection of vertices (0-simplices), edges (1-simplices), triangles (2-simplices), and tetrahedron (3-simplices), which encode proximity relationships among points. Rather than selecting a single fixed scale to define connections, persistent homology invokes a filtration: imagine placing spheres around each point and gradually expanding their radii. As the spheres grow, they begin to intersect, generating edges, triangles, and higher-dimensional simplices. Each radius value (i.e., a non-negative scale parameter) produces a different simplicial complex, with new simplices added as this parameter increases, and the entire filtration is a nested sequence of these sub-complexes. This progression reveals topological invariant features that appear and vanish throughout the filtration, connected components merge (0-cycles), loops form and fill in (1-cycles), and cavities become enclosed (2-cycles).

The output of this process is often visualized as a persistence diagram, which records the birth and death scales of each topological feature in the filtration. Each feature corresponds to a point in a two-dimensional scatter plot, where the horizontal axis marks the scale at which the feature appears (birth) and the vertical axis marks the scale at which it disappears (death)\footnote{An alternative convention sets birth on the $x$-axis and persistence (i.e., death minus birth)-scales on the $y$-axis, emphasizing the longevity of each feature.}. Features that lie far from the diagonal in the persistence diagram are long-lived and typically represent robust topological signatures of the underlying field, while those near the diagonal, short-lived, are generally interpreted as noisy or small-scale fluctuations. However, there is no particular reason for treating points near the diagonal as less relevant than others for learning tasks. These diagrams encapsulate morphological traits of the cosmic web that are sensitive to the cosmological model and early-universe physics~\cite{Cole:2020gzt,Biagetti:2020skr}. They have been successfully applied across a diverse range of cosmological datasets, cosmic shear maps~\cite{DES:2025akz,Heydenreich:2022dci}, magneto-hydrodynamical simulations~\cite{Tsizh:2023hdh}, CMB~\cite{Feldbrugge_2019}, and topological classification of the cosmic web~\cite{2010MNRAS.408.2163A,Sousbie2011,10.1093/mnras/stw2862}. They have also been used in Fisher forecasts~\cite{Biagetti:2022qjl,Yip:2024hlz} and for cosmological parameter inference~\cite{Calles:2024cxl,Yip:2023vud}.

\subsection{Filtration}
\label{sec:filtration}
In Euclidean space, a common filtration choice is the $\alpha$-filtration (or $\alpha$-shape)~\cite{1056714}, constructed from the Delaunay triangulation of a point cloud. It is implemented in several libraries, including \texttt{CGAL}~\cite{cgal:eb-19a}, \texttt{GUDHI}~\cite{gudhi:urm}, \texttt{scikit-tda}~\cite{scikittda2019}, and has also been applied to the analysis of large-scale cosmic structures~\cite{2010MNRAS.408.2163A,vandeweygaert2013alphabettimegaparsecuniverse}. The Delaunay triangulation partitions space into simplices (vertices, edges, triangles, and tetrahedra) such that no point lies inside the circumsphere of any simplex. At each scale, $\alpha$, the $\alpha$-complex contains all simplices whose circumscribed balls have Euclidean radius less than or equal to $\alpha$. Increasing $\alpha$ produces the nested sequence of simplicial complexes that captures the topological connectivity across scales.

Despite its geometric appeal, the standard $\alpha$-filtration is sensitive to noise and outliers. This can lead to a misrepresentation of the sizes of topological features, particularly in sparsely populated regions. To improve robustness against noise, \cite{10.5555/3122009.3242016} introduced the Distance-to-Measure (DTM) function as a density-aware alternative to raw Euclidean distances. Given a point cloud \(X \subset \mathbb{R}^d\) and a location \(x \in \mathbb{R}^d\), the DTM is defined as:
\begin{equation}
\label{eq:DTMdistance}
    \mathrm{DTM}(x) = \sqrt{ \frac{1}{k} \sum_{x_i \in \mathcal{N}_k(x)} \|x - x_i\|^2 },
\end{equation}
where $\mathcal{N}_k(x)$ denotes the set of the $k$ nearest neighbors of $x$ in $X$. Larger values of $k$ produce smoother DTM fields that suppress local noise but reduce sensitivity to fine-grained structures.

The $\alpha$-DTM-$\ell$ filtration, implemented to halo catalogs in~\cite{Biagetti:2020skr,Biagetti:2022qjl,Yip:2024hlz}, extends the standard $\alpha$-complex by incorporating local density information through the DTM function while preserving their simplex structure. For a simplex $\sigma$ located at $x$, the filtration scale $\nu$ defines a modified radius function:
\begin{equation}
\label{eq:dtmradii}
    r(\nu) = \sqrt{\nu^2 - \mathrm{DTM}^2(x)}.
\end{equation}
A simplex is included in the filtration if $\nu \geq \mathrm{DTM}(x)$, with the $\alpha$-DTM adjustment affecting only the entry of vertices and edges\footnote{For example, an edge connecting vertices $x_1$ and $x_2$ is added to the complex once their radius functions intersect, assuming the vertices are already included. The corresponding filtration scale can be computed as:
\begin{equation}
    \nu_{\sigma_{x_1 x_2}} = \sqrt{ \frac{ \left[(DTM(x_1) + DTM(x_2))^2 + d_{x_1 x_2}^2\right] \left[(DTM(x_1) - DTM(x_2))^2 + d_{x_1 x_2}^2\right] }{2\, d_{x_1 x_2}} },
\end{equation}
where $d_{x_1 x_2}$ is their Euclidean distance. This expression clarifies that the filtration parameter $\nu$ generalizes $\alpha$, adjusting for local variations in sampling density rather than simply representing a circumsphere radius.}; higher-dimensional simplices follow the standard $\alpha$-complex rule and are added once all their faces are present.

\begin{figure}[ht]
    \centering
    \includegraphics[width=1\textwidth]{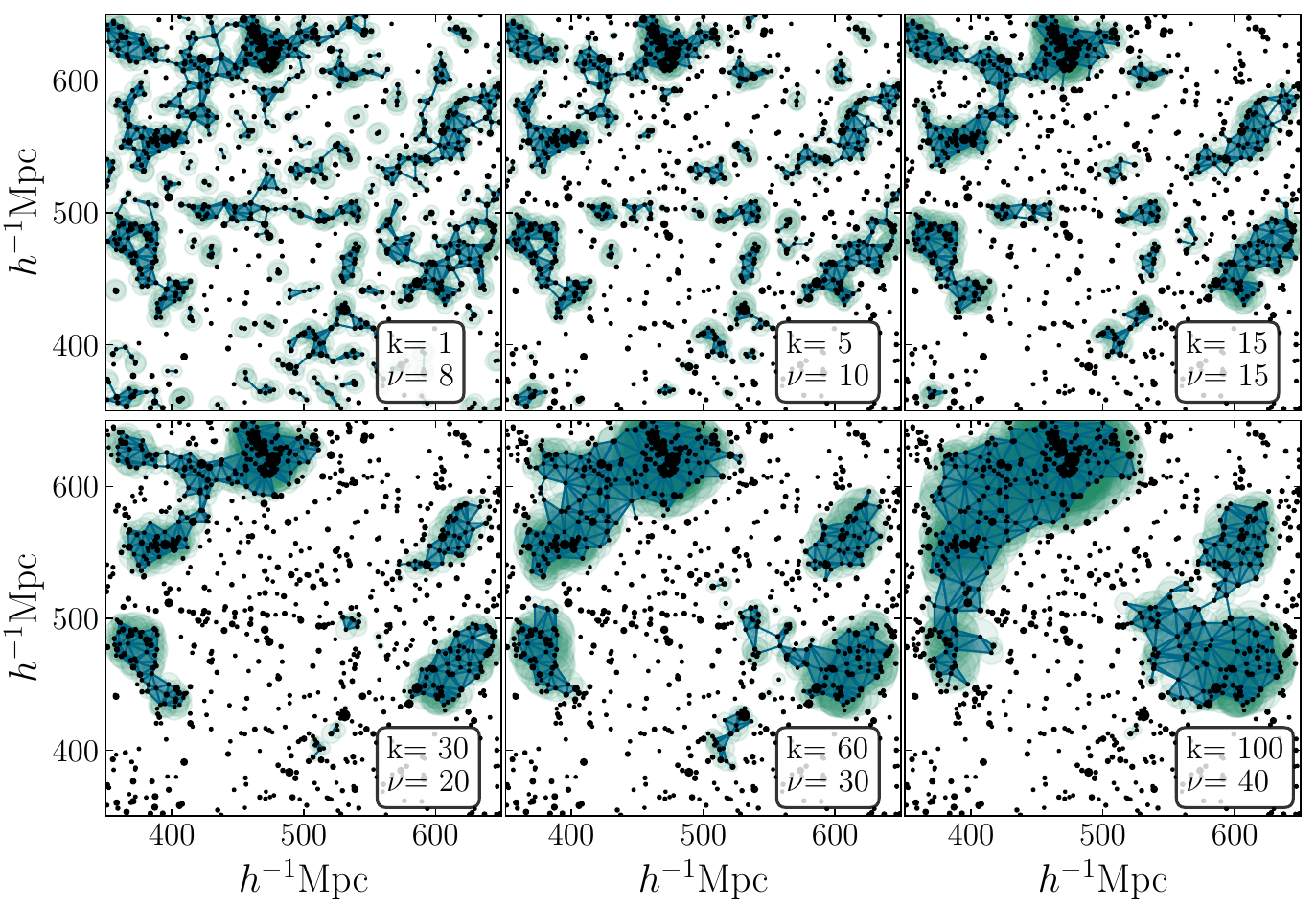}
    \caption{Impact of varying the number of nearest neighbors in the $\alpha$-DTM-$\ell$ filtration, shown for the fiducial cosmology and a fixed seed. Black dots indicate halo positions in redshift space, with the size of the dots proportional to the mass of the halo; green circles denote radii computed via Eq.~\eqref{eq:dtmradii}. Dark blue lines represent 1-simplices (edges), while light blue shaded regions correspond to 2-simplices (triangle faces) present at this filtration scale.}
    \label{fig:filtration_varyk}
\end{figure}

To illustrate how the $\alpha$-DTM-$\ell$ filtration probes the topology of the halo distribution across scales, Figure~\ref{fig:filtration_varyk} shows different maps for $k$ ranging from $k = 1$ to $k = 100$. At low $k$, the filtration explores fine-grained filaments and small-scale features. As $k$ increases, it reveals progressively larger coherent patterns, attenuating smaller features. At the highest $k$, the field is dominated by broad, continuous regions, with only the most persistent, large-scale topological features remaining visible. Because the typical DTM amplitude grows with $k$, we adjust $\nu$ so that each panel displays features at comparable filtration levels.

\subsection{Topological Summaries}
\label{sec:topological_summaries}

We apply a suite of vectorization techniques that transform persistence diagrams, combinatorial and non-Euclidean objects, into fixed-size, interpretable representations suitable for statistical analysis and machine learning applications. In the following, we summarize several methods and compare their ability to predict $\fnl$. Additional details and references are provided in Appendix~\ref{appx:topological_summaries}.

\begin{itemize}
    \item Persistence Images (PI): Smooth each birth-death point using a weighted kernel and aggregated over a grid to produce a 2D grayscale image.
    \item Histogram of Counts (PD-Histogram): Bin birth and death values separately for each homology dimension ($0$-, $1$-, and $2$-cycles), producing six histograms per diagram. Captures marginal distributions across scales.
    \item Betti Curves (PD-Betti): Record the number of topological features alive at each filtration scale for $p$-cycles.
    \item Persistent Betti Curves(PD-PerBetti): Filtered Betti curves that retain only the most persistent features.
    \item Persistence Statistics (PD-statistic): Compute descriptive statistics (mean, median, range, percentiles, entropy) from birth, death, midpoint, and persistence values. No parametrization required.
    \item Persistence Landscapes (PD-Landscape): Encode diagrams as a sequence of piecewise-linear functions representing feature prominence across scales.
    \item Persistence Silhouettes (PD-Silhouette): Weighted average of landscape functions, emphasizing persistent features in a lower-dimensional summary.
\end{itemize}

\section{PNG-pmwd Suite}
\label{sec:pngpmwd_suite}

We present the PNG‑pmwd suite\footnote{Full documentation and data access are available at \url{https://png-pmwd-suite.readthedocs.io/en/latest/}}, a novel dataset comprising a large collection of dark matter halo catalogs designed to advance studies of primordial non‑Gaussianity. With thousands of realizations spanning varied cosmological and non-Gaussian initial conditions, the suite constitutes a substantial resource for data-driven analyses and provides the foundation for the investigations developed in this work.

The \pmwdsuite suite comprises a complementary set of $22{,}410$ halo catalogs specifically designed for machine learning analyses of the impact of primordial non-Gaussian initial conditions. The suite is motivated by three key goals: (i) to probe small, observationally relevant values of $\fnl$ that align with the sensitivity of forthcoming cosmological surveys; (ii) to vary both standard cosmological parameters $(\Omega_m, \sigma_8)$ and primordial non-Gaussianity parameters of the local and equilateral types $(\fnlloc, \fnleq)$, enabling the study of parameter degeneracies; and (iii) to increase the number of simulations available for training modern deep learning pipelines.

We highlight some of the features of \pmwdsuite suite:
\begin{itemize}
    \item Six Latin hypercube designs evaluated at redshift $z = 0.5030475$: four with fixed cosmological parameters $\Omega_m$ and $\sigma_8$ and incorporating non-Gaussian initial conditions $(\fnlloc,\fnleq)$ individually with two different ranges of variation, two that also vary cosmological parameters.
    \item Two additional datasets, one corresponding to a fiducial flat $\Lambda$CDM cosmology, following the parameters in Table~\ref{tab:quijote-lh} with $\fnlloc=0$, and a ``1P'' design that varies one parameter at a time, in small increments, with the others parameters being kept at their fiducial values.
    \item The suite tracked the evolution of approximately $3$ trillion particles producing $22,410$ dark matter halo catalogs.
    \item A total of $22{,}421$ CPU hours dedicated to simulation generation.
\end{itemize}

The halo catalogs in the \pmwdsuite suite were generated from initial conditions set at redshift $z = 9$, using $512^3$ dark matter particles in a periodic cubic box of $(1\,\mathrm{Gpc}/h)^3$. Primordial non-Gaussianity was incorporated via the same template-based prescription implemented in the \lptpng code, using second-order Lagrangian displacements (2LPT) and a linear powerspectrum derived from the Eisenstein \& Hu transfer function~\cite{Eisenstein:1997ik}. The particle ensemble was evolved through 20 linear time steps to redshift $z = 0.5030475$ using the \textit{particle-mesh with derivatives}\footnote{\url{https://github.com/eelregit/pmwd}} (\pmwd) solver~\cite{Li:2022qlf}, a fast N-body code implemented in JAX.

Since the original \pmwd codebase does not natively support non-Gaussian initial conditions, we extended its functionality to include both local and equilateral-type non-Gaussianity templates. The modified version is publicly available\footnote{The updated codebase with local and equilateral-type support can be found at \url{https://github.com/jcallesh/pmwd}} and builds upon an earlier implementation of local-type PNG\footnote{\url{https://github.com/tsfloss/pmwd}}.

Halos are identified using the Friends-of-Friends algorithm, implemented in the \texttt{nbodykit} library~\cite{Hand:2017pqn}, with a linking length of $b = 0.2$. Only halos containing at least $20$ particles are retained in the final catalogs. Particle snapshots themselves were not stored.

Table~\ref{tab:png-pmwd-datasets} provides an overview of the cosmological parameters varied across each dataset in the \pmwdsuite suite, detailing the parameter ranges and number of realizations.

\begin{table}[ht]
    \centering
    \begin{tabular}{lccccr}
    \multicolumn{6}{c}{\textbf{ Latin Hypercubes in the \pmwdsuite Suite}} \\
    \hline\hline
    \textbf{Dataset} & $\Omega_m$ & $\sigma_8$ & $f_{\mathrm{NL}}^{\mathrm{local}}$ & $f_{\mathrm{NL}}^{\mathrm{equil}}$ & \#Realizations \\ 
    \hline\hline
    \texttt{fiducial}       & $0.3175$ & $0.834$ & $0$             & $0$             & $2000$ \\
    \texttt{1P}             & $[0.2825,0.3525]$ & $[0.804,0.864]$ & $[-50,50]$ & $[-250,250]$  & $410$  \\
    \hline
    \texttt{LH\_LC300}         & $0.3175$ & $0.834$ & $[-300,300]$  & $0$               & $2000$ \\
    \texttt{LH\_LC50}       & $0.3175$ & $0.834$ & $[-50,50]$    & $0$               & $2000$ \\
    \texttt{LH\_LC\_om\_s8} & $[0.2825,0.3525]$ & $[0.804,0.864]$ & $[-50,50]$ & $0$ & $6000$ \\
    \hline
    \texttt{LH\_EQ600}         & $0.3175$ & $0.834$ & $0$             & $[-600,600]$  & $2000$ \\
    \texttt{LH\_EQ250}      & $0.3175$ & $0.834$ & $0$             & $[-250,250]$  & $2000$ \\
    \texttt{LH\_EQ\_om\_s8} & $[0.2825,0.3525]$ & $[0.804,0.864]$ & $0$          & $[-250,250]$  & $6000$ \\
    \hline
    \end{tabular}
    \caption{Summary of Latin Hypercube designs in the \pmwdsuite suite.
    }
    \label{tab:png-pmwd-datasets}
\end{table}

Finally, the ``\texttt{1P}'' dataset varies one parameter at a time using linearly spaced steps:
\begin{equation}
    \Delta\Omega_m = 0.007,\quad \Delta\sigma_8 = 0.006,\quad \Delta f_{\mathrm{NL}}^{\mathrm{local}} = 10,\quad \Delta f_{\mathrm{NL}}^{\mathrm{equil}} = 50.
\end{equation}
Each parameter is varied over $10$ steps ($5$ positive, $5$ negative), resulting in $40$ unique configurations. Each configuration includes $10$ independent realizations, for a total of $400$ simulations plus $10$ more fiducial realizations with the same seeding. Seeds for these runs are offset by $10000$ to ensure statistical independence from the main datasets.

\section{Experimental Protocol}
\label{sec:datasets}
We use two complementary suites of N-body simulations to study the sensitivity to primordial non-Gaussianity of the topological pipeline, the \quijotepng simulations~\cite{Jung:2023kjh,Coulton:2022qbc,Jung:2024esv} and the new \pmwdsuite suite. Both datasets consist of a large number of dark matter halo catalogs generated under well-defined cosmological models and initial conditions. In this section, we describe these datasets in detail and summarize the postprocessing summary statistics derived from them.

\subsection{Datasets}

\paragraph{\pmwdsuite dataset:} For the basis of our data analysis, we rely on the \pmwdsuite Suite of halo catalogs. Specifically, we use the following Latin hypercube designs-\{\texttt{LH\_LC300}, \texttt{LH\_LC50}, \texttt{LH\_EQ600}, \texttt{LH\_EQ250}\}- which sample only the amplitude of either local or equilateral PNG while keeping all cosmological parameters fixed; and \{\texttt{LH\_LC\_om\_s8}, \texttt{LH\_EQ\_om\_s8}\}, which additionally sample standard cosmological parameters.

\paragraph{\quijotepng dataset:} The \quijotepng simulations\footnote{\url{https://quijote-simulations.readthedocs.io/en/latest/png.html}} is an extension of the well-known \quijote suite~\cite{Villaescusa-Navarro:2019bje}. These are full N-body dark matter simulations designed to probe the large-scale structure of the Universe. Each simulation evolves $512^3$ dark matter particles in a cubic volume of $(1\,h^{-1}\mathrm{Gpc})^3$ using the TreePM code \texttt{Gadget-III}~\cite{Springel:2005mi}. Particles are initialized at redshift $z=127$ with second-order Lagrangian Perturbation Theory (2LPT), incorporating non-Gaussian initial conditions through the \lptpng code\footnote{\url{https://github.com/dsjamieson/2LPTPNG}}, based on the separable formalism described in~\cite{Scoccimarro:2011pz}. The linear transfer function is computed using CAMB~\cite{Lewis:1999bs}. And halo catalogs are generated via the Friends-of-Friends (FoF) algorithm~\cite{Davis:1985rj}, with a linking length of $b=0.2$.

The \quijote suite includes several tailored subsets of simulations for different analysis goals. In this work, we focus on the Latin hypercube subset that varies the amplitude of local-type primordial non-Gaussianity, $\fnlloc$. This subset consists of $1{,}000$ simulations that uniformly sample $\fnlloc$ around zero while keeping all other cosmological parameters fixed. The fiducial cosmology and sampling range are summarized in Table~\ref{tab:quijote-lh}.

\begin{table}[ht]
    \centering
    \begin{tabular}{l c c c c c c r}
        \hline\hline
        Dataset & $\Omega_{m}$ & $\Omega_{b}$ & $h$ & $n_{s}$ & $\sigma_8$ & $f_{\mathrm{NL}}^{\mathrm{local}}$ & \#Realizations \\
        \hline\hline
        \texttt{Q-LH}\_$\fnlloc$ & 0.3175 & 0.049 & 0.6711 & 0.9624 & 0.834 & $[-300,300]$ & 1000\\
        \hline
    \end{tabular}
    \caption{Fiducial cosmological parameters and sampling range for the Latin hypercube subset of the \quijotepng simulations used in this work.}
    \label{tab:quijote-lh}

\end{table}

\subsection{Post-processed measurements}

All measurements are performed in redshift space, with redshift-space distortions modeled using the distant-observer approximation, assuming the line of sight is aligned with the $z$-axis. Under this approximation, the redshift-space position $\boldsymbol{x}_{\text{rs}}$ of each halo is given by:
\begin{equation}
    \boldsymbol{x}_{\text{rs}} = \boldsymbol{x} + \frac{(\boldsymbol{v} \cdot \hat{\boldsymbol{z}})}{a(z) H(z)} \hat{\boldsymbol{z}},
\end{equation}
where $\boldsymbol{v}$ is the peculiar velocity, $a(z) = (1 + z)^{-1}$ is the scale factor, and $H(z)$ is the Hubble expansion rate at redshift $z$. The Hubble parameter is evaluated assuming a flat $\Lambda$CDM cosmology, $H(z) = 100 \sqrt{\Omega_m a^{-3} + \Omega_\Lambda}$ km/s/(Mpc/h), with $\Omega_\Lambda = 1 - \Omega_m$ due to spatial flatness.

To investigate how halo population influences the cosmological information extracted from the topological pipeline, we partition each halo catalog into seven mass bins, all based in the fiducial cosmology. The first bin, \texttt{HLow}, includes all halos above a fixed lower mass threshold of $3.28 \times 10^{13}\,M_\odot/h$, corresponding to halos with at least $50$ dark matter particles. This cut ensures reliable halo identification and avoids biases from poorly resolved low-mass halos.

The next three bins, \texttt{HLow-A}, \texttt{HLow-B}, \texttt{HMid}, are constructed to contain approximately equal numbers of halos, ensuring balanced sampling across the full mass range. In particular, \texttt{HMid} marks a secondary mass resolution threshold, containing halos composed of more than $107$ particles. To further probe this regime, we subdivide \texttt{HMid} into three narrower bins, \texttt{HMid-A}, \texttt{HMid-B}, \texttt{HHigh}, each with comparable halo number densities. The \texttt{HHigh} bin also serves as a third mass threshold, containing halos with more than $201$ particles.

\begin{table}[ht]
    \centering
    \begin{tabular}{l |c | c | l}
    \textbf{Label} & \quijote $(\bar{n}_h\;[h^3\text{Mpc}^{-3}])$ & \pmwdsuite $(\bar{n}_h\;[h^3\text{Mpc}^{-3}])$ &\textbf{Mass Range [$10^{13}\,M_\odot/h$]} \\
    \hline\hline
    \texttt{HLow} & $9.88\times10^{-5}$ & $8.93\times10^{-5}$ &$[3.28,\ \infty)$ \\
    \texttt{HLow-A} & $3.45\times10^{-5}$ & $3.03\times10^{-5}$ &$[3.28,\ 4.46)$ \\
    \texttt{HLow-B} & $3.25\times10^{-5}$ & $2.93\times10^{-5}$ &$[4.46,\ 7.09)$ \\
    \texttt{HMid} & $3.17\times10^{-5}$ & $2.98\times10^{-5}$ &$[7.09,\ \infty)$ \\
    \texttt{HMid-A} & $1.06\times10^{-5}$ & $9.80\times10^{-6}$ &$[7.09,\ 9.06)$ \\
    \texttt{HMid-B} & $1.05\times10^{-5}$ & $9.87\times10^{-6}$ &$[9.06,\ 13.26)$ \\
    \texttt{HHigh} & $1.06\times10^{-5}$ & $1.01\times10^{-5}$ &$[13.26,\ \infty)$ \\
    \end{tabular}
    \caption{Mean halo number densities and mass ranges for the seven mass-bin definitions used in this work, computed from $2{,}000$ realizations of the fiducial cosmology of \pmwdsuite compared to \quijote.}
    \label{tab:mass_bins}
\end{table}

The binning thresholds shown in Table~\ref{tab:mass_bins} were derived from $2{,}000$ fiducial realizations from the \pmwdsuite suite and are consistently applied to all subsequent datasets in our analysis.

\begin{figure}[ht]
    \centering
    \includegraphics[width=1\textwidth]{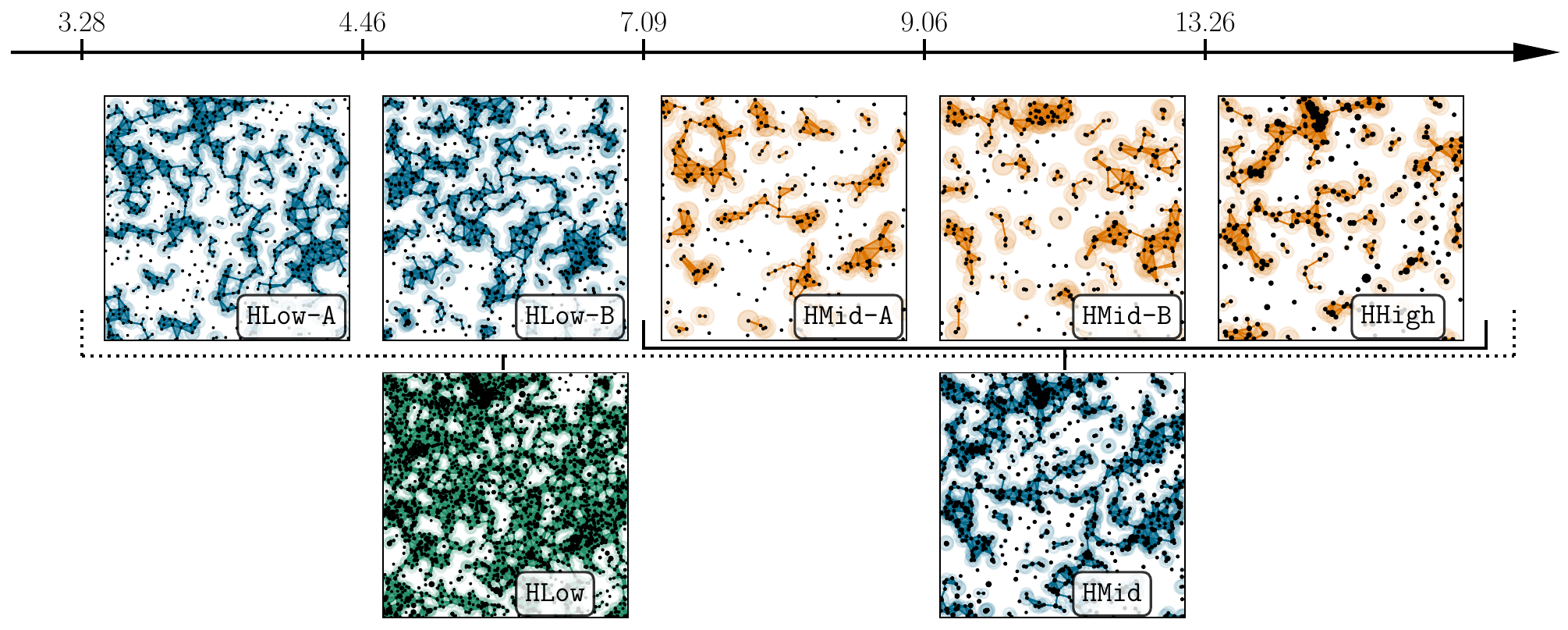}
    \caption{We show the simplicial complex in a slice from sub-boxes of size $350$ to $650\,h^{-1}\mathrm{Mpc}$ in the \texttt{LH\_LC300} dataset of the \pmwdsuite suite with $\fnlloc = 294.45$. Each panel represent a different mass cut. The arrow indicates the mass binning scheme used in Table~\ref{tab:mass_bins} in units of $[10^{13}\,\rm{M}_\odot/h]$. Color coding correspond to groups with similar mean halo density.}
    \label{fig:filtration_massbins}
\end{figure}

Figure~\ref{fig:filtration_massbins} shows the spatial distribution of halos across the seven mass bins, where the color indicates similar halo number densities. For visualization, we overlay the simplicial complex at a fixed filtration scale with $k=1$, choosing $\nu$ separately for each density group to account for their sparsities. Although the panels are not intended to reveal topological differences by eye, they illustrate how mass selection alters the sampling of the density field and, consequently, the structures that persistent homology probes.

\subsubsection{Persistence pipeline}
 \label{sec:persistencepipeline}

\begin{figure}[ht]
    \centering
    \includegraphics[width=1\textwidth]{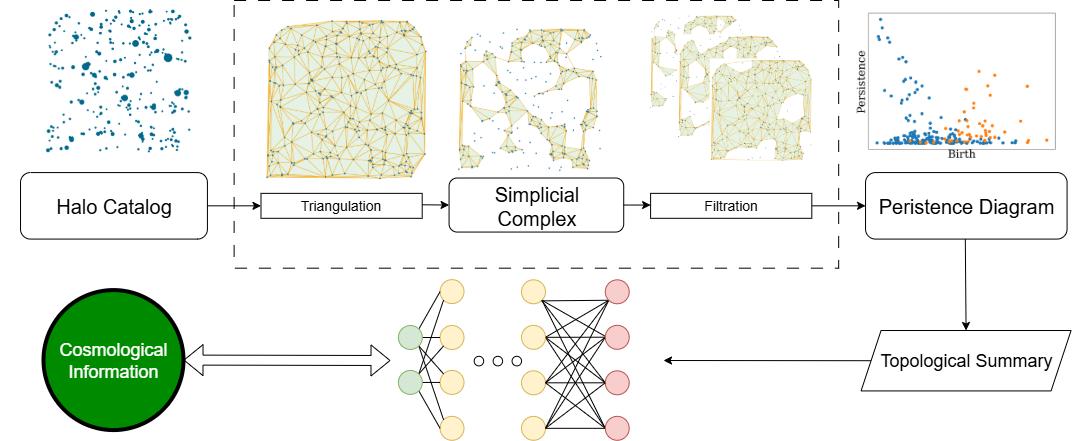}
    \caption{Cosmological models guide LSS N-body simulations, from which we identify dark matter halo catalogs. Persistent homology extracts topological features as persistence diagrams, which are vectorized or directly feed into machine learning models to infer cosmological parameters. }
    \label{fig:pipeline_plot}
\end{figure}

The conceptual flow illustrated in Figure~\ref{fig:pipeline_plot} captures the full arc of our pipeline. Starting from a cosmological model, we generate simulated data and identify dark matter halos. Their spatial distribution is triangulated to construct simplicial complexes, which serve as the backbone for the filtration. Persistent homology then distills geometric and topological information into persistence diagrams that encode multiscale features of the underlying dark matter field. These diagrams are vectorized, either through standard methods or neural-network-based compression, giving compact representations of the topological structure. Finally, these vectors are mapped to cosmological parameter predictions via a neural network, closing the loop from initial conditions to quantitative inference.

Persistence diagrams are computed using the $\alpha$-DTM-$\ell$ filtration described in Section~\ref{sec:filtration}. For each halo catalog, diagrams are generated across three homology dimensions (0-, 1-, and 2-cycles), while varying the nearest-neighbor smoothing parameter $k$ over $\{1, 5, 15, 30, 60, 100\}$. This multi-scale strategy captures complementary structural information from the halo distribution. Thus, each halo catalog produces $18$ diagrams (one for every combination of homology dimension (3) and $k$ (6)) for each mass bin specified in Table~\ref{tab:mass_bins}, resulting in a total of 126 diagrams per catalog.

Diagrams are standardized in two steps. First, global percentile filters (up to the 99.5th) are applied to birth and death coordinates per homology dimension across all realizations, mitigating outlier effects. Second, min-max scalers are learned per cycle and $k$ value using the training set, mapping coordinates to the unit square $(b, d) \in [0,1]^2$. These scalers are consistently applied across train, validation, and test sets to ensure compatibility and prevent data leakage.

The resulting filtered diagrams serve as input to our machine learning models, including \deepsets, \perslay, and summary-statistic approaches. For histogram-based descriptors (e.g., histogram of counts, Betti curves, etc.), we fix the binning to 35 equally spaced intervals over $[0,1]$. For persistence images, we use Epanechnikov kernel density estimation on $64 \times 64$ pixel grids, with bandwidth set to $5 \times$ the persistence-per-pixel scale (i.e., persistence range divided by $64$), implemented via \texttt{KernelDensity} from \texttt{scikit-learn}.

\subsubsection{Powerspectrum and bispectrum}

To measure the powerspectrum and bispectrum, we use the publicly available \texttt{PBI4} tool\footnote{\url{https://github.com/matteobiagetti/pbi4}}. Halo catalogs are interpolated onto a Fourier grid of $144^3$ points using a fourth-order mass assignment scheme with interlacing to reduce aliasing, as described in~\cite{Sefusatti:2015aex}. The powerspectrum is computed in $k$-bins\footnote{In this section only, $k$ denotes wave modes; elsewhere it refers to the nearest-neighbor parameter in the $\alpha$-DTM-$\ell$ filtration.} of width $\Delta k = 2k_f$, where the fundamental mode is $k_f = 0.006\,h\,\mathrm{Mpc}^{-1}$. Binning begins at $2k_f$ and extends to $k_{\max} \leq 0.3\,h\,\mathrm{Mpc}^{-1}$, producing 24 bins.

We include the monopole ($\ell=0$), quadrupole ($\ell=2$), and hexadecapole ($\ell=4$) moments of the redshift-space powerspectrum. For the monopole, we apply a Poisson shot noise correction, $P^{\{\ell=0\}}_{hh}(k) = P^{\{\ell=0\}}_{\rm tot}(k) - 1/\bar{n}$, where $\bar{n}$ is the mean halo number density.

The bispectrum is computed using the same $k$-binning scheme, considering all closed triangle configurations with $k_i \leq 0.3\,h\,\mathrm{Mpc}^{-1}$, resulting in 1522 unique triangle bins. For the bispectrum monopole, we subtract the leading-order shot noise contribution, $B^{\{\ell=0\}}_{hhh}(k_1,k_2,k_3) = B^{\{\ell=0\}}_{\rm tot}(k_1,k_2,k_3) - 1/\bar{n}[P^{\{\ell=0\}}_{hh}(k_1) + P^{\{\ell=0\}}_{hh}(k_2) + P^{\{\ell=0\}}_{hh}(k_3)] - 1/\bar{n}^2$.

The final data vector, denoted PSBS, combines the three powerspectrum multipoles with the bispectrum monopole, $\text{PSBS} = [P^{\{\ell=0\}}_{hh}, P^{\{\ell=2\}}_{hh}, P^{\{\ell=4\}}_{hh}, B^{\{\ell=0\}}_{hhh}]$, totaling 1594 features. For simplicity, we refer to this combined vector as PSBS throughout.

\section{Architectures and training}
\label{sec:training_hp}

Five types of architectures were employed, each matched to the structure of its input representation. Persistence images are processed with standard Convolutional Neural Networks (CNNs), while one-dimensional vectors, such as summary statistics and histogram-based encoding, are handled by fully connected multilayer perceptrons (MLPs). For raw persistence diagrams, we use the \deepsets and \perslay architectures, which handle permutation-invariant set inputs.  As a non-neural baseline, we also include gradient-boosted decision trees, well-suited for tabular data~\cite{grinsztajn2022tree}. 

All architectures were trained to predict a single value, $\fnl$. We adopt a fixed dataset split: 200 samples for testing, 200 for validation, and the remainder for training. 

Hyperparameter selection and model configuration were performed independently for each dataset and input representation. Full details of the architectures and the corresponding hyperparameter search are provided in Appendix~\ref{appx:architectures} and~\ref{appx:hyperparameter}, respectively.

\subsection{Likelihood-Free Parameter Inference}

All models were trained to jointly predict the mean and standard deviation of the target parameter $\fnl$. We adopt a likelihood-free framework~\cite{Jeffrey:2020itg}, which directly optimizes both outputs through a custom loss function.

Given a batch of $N$ samples, the loss is defined as:
\begin{equation}
    \mathcal{L} = \log\left( \sum_{i=1}^{N} (y_{i}-\mu_{i})^2 \right)  + \log\left( \sum_{i=1}^{N} \left[ (y_{i} - \mu_{i})^2 - \sigma_{i}^2 \right]^2  \right),
\end{equation}
where $\mu_i$ and $\sigma_i$ are the predicted mean and standard deviation for sample $i$, and $y_i$ is the ground truth.

The first term penalizes the squared error between predictions and targets, while the second term encourages consistency between predicted uncertainty $\sigma_i$ and the empirical error, promoting well-calibrated uncertainty estimates. This loss function has been widely adopted in the cosmological community across a variety of applications, as demonstrated in~\cite{Villaescusa-Navarro:2021pkb,Villaescusa-Navarro:2021cni,Villanueva-Domingo:2021dun,Wang:2022zpv,Villanueva-Domingo:2022rvn,Perez:2022nlv,Villaescusa-Navarro:2022twv,deSanti:2023zzn,Chawak:2023bil,deSanti:2023rsw,Jung:2024esv,Gondhalekar:2024iqm,Lee:2025tql,Bayer:2025ija,Chatterjee:2024ctp,Jo:2025ndh}.

\subsection{Metrics}

To validate the fidelity of our pipeline, we employ three complementary metrics, the root-mean-square error (RMSE) and coefficient of determination ($R^2$) to evaluate the accuracy of the predicted mean values, and the chi-squared statistic ($\chi^2$) for quantifying uncertainty calibration.

The \textit{RMSE} measures the average deviation between predicted means and ground truth values:
\begin{equation}
    {\rm RMSE} = \sqrt{\frac{1}{N} \displaystyle\sum_i (\theta_i - \hat{\mu}_i)^2},
\end{equation}
where $\theta_i$ and $\hat{\mu}_i$ denote the true and predicted values for the $i$-th sample, and $N$ is the batch size. 

The \textit{$R^2$ score}, or coefficient of determination, indicates how well the model captures the variance in the target data:
\begin{equation}
    R^2(\theta, \hat{\mu}) = 1 - \frac{\displaystyle\sum_i (\theta_i - \hat{\mu}_i)^2}{\displaystyle\sum_i (\theta_i - \bar{\theta})^2}\,.
\end{equation}
$R^2$ values close to $1$ signals perfect alignment between predictions and observations, whereas $0$ implies the model performs no better than simply predicting the overall mean in the sample. Negative values reflect scenarios where the model fits worse than a constant predictor. For convenience we report the $R^2\times100$ scores in the rest of the paper.

The \textit{$\chi^2$ statistic} evaluates the consistency between predicted uncertainties and empirical errors:
\begin{equation}
    \chi^2(\theta, \hat{\mu}, \hat{\sigma}) = \frac{1}{N} \sum_i{\left( \frac{(\theta_i - \hat{\mu}_i)^2}{\hat{\sigma}_i^2}\right)}.
\end{equation}
A value of $\chi^2 \approx 1$ signals well-calibrated uncertainty estimates. Values below $1$ overestimated variances, while values above $1$ indicate an underestimation.

\section{Results}
\label{sec:results}

We compared handcrafted topological summaries, neural compression methods, and standard clustering statistics, namely powerspectrum and bispectrum (PSBS) for inferring primordial non-Gaussianity (local and equilateral). Our analysis also spans multiple halo mass bins and simulation suites, allowing us to characterize how halo mass selection influences inference and provides insights into how biasing affects different statistics. This is a total of nearly $500$ model configurations evaluated.

The main results are summarized in Tables~\ref{tab:fnl_loc_metrics} and~\ref{tab:fnl_equil_metrics}. We include the intermediate mass bins results for each of the configurations in Appendix~\ref{appx:full_results}. We find the following:
\begin{itemize}
\item We found that among all methods tested, PD-Statistics (XGB) consistently delivered the best performance for both $\fnlloc$ and $\fnleq$ amplitudes. Combining PD-Statistics with PSBS gives a small improvement on the constraints, which suggests that they contain a small amount of complementary information.
\item We also studied how constraints on PNG depend on the halo masses included in the analysis. We found that smaller halos, which are not well-resolved by the halo finder, degrade the performance of the models by introducing noise. We find that most information about PNG is obtained from the largest halos.
\item Topological summary statistics that favor high-persistence features lead to poorer performance, which aligns with previous results showing that early-birth low-persistence features in 0-cycles seem to contain more critical information.
 \item In the dataset where $\fnl$, $\Omega_m$, and $\sigma_8$ are jointly varied, e.g. \texttt{LH\_LC\_om\_s8} and \texttt{LH\_EQ\_om\_s8}, marginalizing over $\Omega_m$ and $\sigma_8$ significantly weakens constraints on $\fnl$ across all vectorization methods. Known degeneracies appear as all three parameters increase the abundance of massive halos, their effects become difficult to disentangle~\cite{2011MNRAS.416.2527M}. By contrast, fixing $\Omega_m$ and $\sigma_8$ (low-amplitude datasets \texttt{LH\_LC50} and \texttt{LH\_EQ250}) substantially improves performance, in line with the increase in sensitivity to $\fnl$ of the halo mass function under fixed cosmology~\cite{LoVerde:2007ri,Jung:2023kjh}.
\end{itemize}

We now provide more details on the how different halo mass cuts affect the constraints, and on the comparison across different models and statistics.

\begin{table}[ht]
\small
    \centering
    \begin{tabular}{|c|l|c c|c c|c c|c c|}
    \hline
    Model & Mass bin 
    & \multicolumn{2}{c|}{\texttt{Q-LH}\_$\fnlloc$} 
    & \multicolumn{2}{c|}{\texttt{LH\_LC300}} 
    & \multicolumn{2}{c|}{\texttt{LH\_LC50}} 
    & \multicolumn{2}{c|}{\texttt{LH\_LC\_om\_s8}} \\
    & & RMSE & {R2} 
      & RMSE & {R2} 
      & RMSE & {R2} 
      & RMSE & {R2} \\[0.5ex] 
    \hline\hline
    \multirow{3}{*}{\makecell[t]{PSBS \\
    (\textbf{MLP})}}
    &\texttt{HLow}       & $49.0$ &$\cca{91.9}$  & $57.8$ &$\cca{90.2}$ & $27.7$ &$\cca{18.6}$ & $28.3$ &$\cca{9.0}$ \\ 
    &\texttt{HMid}       & $33.8$ &$\cca{96.1}$  & $35.1$ &$\cca{96.4}$ & $24.3$ &$\cca{37.0}$ & $27.5$ &$\cca{14.3}$ \\ 
    &\texttt{HHigh}       & $29.3$ &$\cca{97.1}$  & $28.4$ &$\cca{97.6}$ & $21.4$ &$\cca{51.6}$ & $27.9$ &$\cca{11.7}$ \\ 
    \hline
    \multirow{3}{*}{\makecell[t]{PD \\
    (\textbf{\deepsets})}}
    &\texttt{HLow}       & $70.5$ &$\cca{83.2}$  & $74.1$ &$\cca{83.7}$ & $30.7$ &$\cca{0.1}$ & $28.1$ &$\cca{1.1}$ \\ 
    &\texttt{HMid}       & $58.1$ &$\cca{88.6}$  & $48.7$ &$\cca{93.0}$ & $29.5$ &$\cca{7.8}$ & $28.0$ &$\cca{1.4}$ \\ 
    &\texttt{HHigh}       & $49.1$ &$\cca{91.8}$  & $35.9$ &$\cca{96.2}$ & $28.6$ &$\cca{13.2}$ & $27.8$ &$\cca{2.9}$ \\ 
    \hline
    \multirow{3}{*}{\makecell[t]{PI \\
    (\textbf{\cnn})}}
    &\texttt{HLow}       & $49.1$ &$\cca{91.9}$  & $52.1$ &$\cca{92.0}$ & $29.5$ &$\cca{7.8}$ & $27.8$ &$\cca{2.7}$ \\ 
    &\texttt{HMid}       & $36.8$ &$\cca{95.4}$  & $39.0$ &$\cca{95.5}$ & $27.1$ &$\cca{22.0}$ & $27.5$ &$\cca{5.0}$ \\ 
    &\texttt{HHigh}       & $34.1$ &$\cca{96.1}$  & $34.7$ &$\cca{96.4}$ & $25.9$ &$\cca{28.5}$ & $27.4$ &$\cca{5.9}$ \\ 
    \hline
    \multirow{3}{*}{\makecell[t]{PD-Statistics \\
    (\textbf{XGB})}}
    &\texttt{HLow}       & $35.5$ &$\cca{95.7}$  & $39.9$ &$\cca{95.3}$ & $23.7$ &$\cca{40.4}$ & $28.6$ &$\cca{7.4}$ \\ 
    &\texttt{HMid}       & $25.8$ &$\cca{97.8}$  & $27.5$ &$\cca{97.8}$ & $20.6$ &$\cca{55.0}$ & $28.1$ &$\cca{10.1}$ \\ 
    &\texttt{HHigh}       & $28.1$ &$\cca{97.3}$  & $26.7$ &$\cca{97.9}$ & $20.9$ &$\cca{53.8}$ & $27.8$ &$\cca{12.2}$ \\ 
    \hline
    \multirow{3}{*}{\makecell[t]{PD-Statistics/PSBS \\
    (\textbf{XGB})}}
    &\texttt{HLow}       & $33.7$ &$\cca{96.2}$  & $39.6$ &$\cca{95.4}$ & $23.3$ &$\cca{42.2}$ & $27.7$ &$\cca{12.9}$ \\ 
    &\texttt{HMid}       & $26.2$ &$\cca{97.7}$  & $27.2$ &$\cca{97.8}$ & $20.3$ &$\cca{56.2}$ & $27.1$ &$\cca{16.3}$ \\ 
    &\texttt{HHigh}       & $28.2$ &$\cca{97.3}$  & $26.0$ &$\cca{98.0}$ & $20.3$ &$\cca{56.3}$ & $27.5$ &$\cca{14.4}$ \\ 
    \hline
    \multirow{3}{*}{\makecell[t]{PD-Histogram \\
    (\textbf{MLP})}}
    &\texttt{HLow}       & $36.9$ &$\cca{95.4}$  & $42.4$ &$\cca{94.7}$ & $25.3$ &$\cca{32.0}$ & $28.5$ &$\cca{8.0}$ \\ 
    &\texttt{HMid}       & $27.9$ &$\cca{97.4}$  & $30.6$ &$\cca{97.2}$ & $23.0$ &$\cca{44.0}$ & $28.1$ &$\cca{10.2}$ \\ 
    &\texttt{HHigh}       & $29.8$ &$\cca{97.0}$  & $29.2$ &$\cca{97.5}$ & $22.7$ &$\cca{45.3}$ & $27.7$ &$\cca{13.0}$ \\ 
    \hline
    \multirow{3}{*}{\makecell[t]{PD-Betti \\
    (\textbf{MLP})}}
    &\texttt{HLow}       & $39.2$ &$\cca{94.7}$  & $41.5$ &$\cca{94.9}$ & $30.8$ &$-0.8$ & $29.7$ &$-0.2$ \\ 
    &\texttt{HMid}       & $28.7$ &$\cca{97.2}$  & $31.0$ &$\cca{97.2}$ & $30.3$ &$\cca{2.6}$ & $29.7$ &$-0.2$ \\ 
    &\texttt{HHigh}       & $31.4$ &$\cca{96.7}$  & $30.2$ &$\cca{97.3}$ & $22.9$ &$\cca{44.4}$ & $29.6$ &$\cca{0.3}$ \\ 
    \hline
    \multirow{3}{*}{\makecell[t]{PD-PerBetti \\
    (\textbf{MLP})}}
    &\texttt{HLow}       & $37.4$ &$\cca{95.3}$  & $41.3$ &$\cca{95.0}$ & $25.0$ &$\cca{33.8}$ & $28.0$ &$\cca{10.8}$ \\ 
    &\texttt{HMid}       & $28.2$ &$\cca{97.3}$  & $30.8$ &$\cca{97.2}$ & $22.6$ &$\cca{45.6}$ & $28.0$ &$\cca{11.1}$ \\ 
    &\texttt{HHigh}       & $31.3$ &$\cca{96.7}$  & $30.1$ &$\cca{97.3}$ & $22.7$ &$\cca{45.4}$ & $27.8$ &$\cca{12.2}$ \\ 
    \hline
    \multirow{3}{*}{\makecell[t]{PD-Silhouette \\
    (\textbf{XGB})}}
    &\texttt{HLow}       & $48.0$ &$\cca{92.2}$  & $50.5$ &$\cca{92.5}$ & $26.4$ &$\cca{25.9}$ & $28.8$ &$\cca{5.9}$ \\ 
    &\texttt{HMid}       & $34.7$ &$\cca{95.9}$  & $36.0$ &$\cca{96.2}$ & $24.5$ &$\cca{36.1}$ & $28.6$ &$\cca{7.3}$ \\ 
    &\texttt{HHigh}       & $33.8$ &$\cca{96.1}$  & $33.8$ &$\cca{96.6}$ & $23.8$ &$\cca{39.6}$ & $28.3$ &$\cca{9.2}$ \\ 
    \hline
    \multirow{3}{*}{\makecell[t]{PD-Landscape \\
    (\textbf{XGB})}}
    &\texttt{HLow}       & $65.3$ &$\cca{85.6}$  & $67.9$ &$\cca{86.4}$ & $29.7$ &$\cca{6.3}$ & $29.6$ &$\cca{0.2}$ \\ 
    &\texttt{HMid}       & $48.9$ &$\cca{91.9}$  & $49.3$ &$\cca{92.8}$ & $27.1$ &$\cca{21.9}$ & $29.2$ &$\cca{3.5}$ \\ 
    &\texttt{HHigh}       & $41.0$ &$\cca{94.3}$  & $38.1$ &$\cca{95.7}$ & $24.9$ &$\cca{33.9}$ & $28.8$ &$\cca{6.0}$ \\ 
    \hline
    \end{tabular}
    \caption{Performance of the different summary statistics used in this work for $\fnlloc$ inference. The mass bins here correspond to (in units of $10^{13}\ \text{M}_\odot/h$): \texttt{HLow} halos with masses in the range $[3.28,\ \infty)$, \texttt{HMid} corresponds to $[7.09,\ \infty)$, and \texttt{HHigh} to $[13.26,\ \infty)$. Bold labels indicate the lower RMSE values across mass bins and datasets.We display the RMSE and $R^2$ averaged over $10$ independent runs ($4$ in \deepsets models). \xgb models use cross-validation across $9$ splits ($4$ in the \texttt{Q-LH}\_$\fnlloc$ dataset) with fixed training size. Neural networks models are filtered by their $\chi^2$ performance, retaining only the realizations with $\chi^2_{\text{val}} \leq 1.5$ scores to ensure calibrated uncertainties estimates.  
    }
    \label{tab:fnl_loc_metrics}
\end{table}

\begin{table}[ht]
\small
    \centering
    \begin{tabular}{|c|l|c c|c c|c c|}
    \hline
    Model & Mass bin 
    & \multicolumn{2}{c|}{\texttt{LH\_EQ600}} 
    & \multicolumn{2}{c|}{\texttt{LH\_EQ250}} 
    & \multicolumn{2}{c|}{\texttt{LH\_EQ\_om\_s8}} \\
    & & RMSE & {R2} 
      & RMSE & {R2} 
      & RMSE & {R2} \\[0.5ex] 
    \hline\hline
    \multirow{3}{*}{\makecell[t]{PSBS \\
    (\textbf{MLP})}}
    &\texttt{HLow}       & $162.7$ &$\cca{80.5}$  & $120.2$ &$\cca{38.6}$ & $144.1$ &$\cca{5.7}$ \\ 
    &\texttt{HMid}       & $89.6$ &$\cca{94.1}$  & $80.3$ &$\cca{72.6}$ & $144.1$ &$\cca{5.7}$ \\ 
    &\texttt{HHigh}       & $76.6$ &$\cca{95.7}$  & $68.4$ &$\cca{80.1}$ & $144.6$ &$\cca{5.1}$ \\ 
    \hline
    \multirow{3}{*}{\makecell[t]{PD \\
    (\textbf{\deepsets})}}
    &\texttt{HLow}       & $369.6$ &$-0.8$  & $154.2$ &$-1.1$ & $141.1$ &$\cca{0.0}$ \\ 
    &\texttt{HMid}       & $179.4$ &$\cca{69.6}$  & $153.2$ &$\cca{0.2}$ & $141.0$ &$\cca{0.2}$ \\ 
    &\texttt{HHigh}       & $114.8$ &$\cca{90.1}$  & $126.4$ &$\cca{28.2}$ & $140.2$ &$\cca{1.3}$ \\ 
    \hline
    \multirow{3}{*}{\makecell[t]{PI \\
    (\textbf{\cnn})}}
    &\texttt{HLow}       & $156.9$ &$\cca{81.8}$  & $136.7$ &$\cca{20.6}$ & $140.6$ &$\cca{0.7}$ \\ 
    &\texttt{HMid}       & $117.5$ &$\cca{89.8}$  & $104.3$ &$\cca{53.8}$ & $139.6$ &$\cca{2.1}$ \\ 
    &\texttt{HHigh}       & $101.3$ &$\cca{92.4}$  & $94.1$ &$\cca{62.3}$ & $138.3$ &$\cca{3.9}$ \\ 
    \hline
    \multirow{3}{*}{\makecell[t]{PD-Statistics \\
    (\textbf{XGB})}}
    &\texttt{HLow}       & $74.9$ &$\cca{95.9}$  & $65.9$ &$\cca{81.6}$ & $134.5$ &$\cca{17.8}$ \\ 
    &\texttt{HMid}       & $64.3$ &$\cca{97.0}$  & $58.7$ &$\cca{85.3}$ & $141.5$ &$\cca{9.0}$ \\ 
    &\texttt{HHigh}       & $69.4$ &$\cca{96.4}$  & $65.2$ &$\cca{81.9}$ & $143.8$ &$\cca{6.1}$ \\ 
    \hline
    \multirow{3}{*}{\makecell[t]{PD-Statistics/PSBS \\
    (\textbf{XGB})}}
    &\texttt{HLow}       & $74.0$ &$\cca{96.0}$  & $67.8$ &$\cca{80.4}$ & $136.9$ &$\cca{14.9}$ \\ 
    &\texttt{HMid}       & $64.5$ &$\cca{96.9}$  & $58.1$ &$\cca{85.7}$ & $141.7$ &$\cca{8.7}$ \\ 
    &\texttt{HHigh}       & $70.8$ &$\cca{96.3}$  & $66.0$ &$\cca{81.5}$ & $143.7$ &$\cca{6.2}$ \\ 
    \hline
    \multirow{3}{*}{\makecell[t]{PD-Histogram \\
    (\textbf{MLP})}}
    &\texttt{HLow}       & $101.0$ &$\cca{92.5}$  & $121.7$ &$\cca{34.3}$ & $147.3$ &$\cca{1.4}$ \\ 
    &\texttt{HMid}       & $94.5$ &$\cca{93.4}$  & $80.5$ &$\cca{72.4}$ & $143.3$ &$\cca{6.7}$ \\ 
    &\texttt{HHigh}       & $88.1$ &$\cca{94.3}$  & $78.9$ &$\cca{73.5}$ & $143.6$ &$\cca{6.4}$ \\ 
    \hline
    \multirow{3}{*}{\makecell[t]{PD-Betti \\
    (\textbf{MLP})}}
    &\texttt{HLow}       & $102.2$ &$\cca{92.3}$  & $154.0$ &$-0.9$ & $148.7$ &$-0.4$ \\ 
    &\texttt{HMid}       & $95.3$ &$\cca{93.3}$  & $80.4$ &$\cca{72.5}$ & $145.4$ &$\cca{4.0}$ \\ 
    &\texttt{HHigh}       & $89.3$ &$\cca{94.1}$  & $80.8$ &$\cca{72.3}$ & $144.7$ &$\cca{4.9}$ \\ 
    \hline
    \multirow{3}{*}{\makecell[t]{PD-PerBetti \\
    (\textbf{MLP})}}
    &\texttt{HLow}       & $100.9$ &$\cca{92.5}$  & $113.9$ &$\cca{42.2}$ & $146.0$ &$\cca{3.0}$ \\ 
    &\texttt{HMid}       & $95.2$ &$\cca{93.3}$  & $80.2$ &$\cca{72.7}$ & $143.8$ &$\cca{6.1}$ \\ 
    &\texttt{HHigh}       & $91.3$ &$\cca{93.8}$  & $80.5$ &$\cca{72.5}$ & $144.5$ &$\cca{5.1}$ \\ 
    \hline
    \multirow{3}{*}{\makecell[t]{PD-Silhouette \\
    (\textbf{XGB})}}
    &\texttt{HLow}       & $135.0$ &$\cca{86.6}$  & $108.2$ &$\cca{50.3}$ & $143.5$ &$\cca{6.4}$ \\ 
    &\texttt{HMid}       & $108.5$ &$\cca{91.3}$  & $90.4$ &$\cca{65.3}$ & $144.1$ &$\cca{5.7}$ \\ 
    &\texttt{HHigh}       & $101.3$ &$\cca{92.4}$  & $86.0$ &$\cca{68.5}$ & $146.1$ &$\cca{3.0}$ \\ 
    \hline
    \multirow{3}{*}{\makecell[t]{PD-Landscape \\
    (\textbf{XGB})}}
    &\texttt{HLow}       & $195.4$ &$\cca{71.8}$  & $131.7$ &$\cca{26.3}$ & $147.5$ &$\cca{1.1}$ \\ 
    &\texttt{HMid}       & $128.6$ &$\cca{87.8}$  & $103.4$ &$\cca{54.5}$ & $145.0$ &$\cca{4.5}$ \\ 
    &\texttt{HHigh}       & $108.1$ &$\cca{91.4}$  & $97.1$ &$\cca{59.9}$ & $146.5$ &$\cca{2.5}$ \\ 
    \hline
    \end{tabular}
    \caption{Performance of the different summary statistics used in this work for $\fnleq$ inference. The mass bins shown here correspond to (in units of $10^{13}\ M_\odot/h$): \texttt{HLow} includes halos in the range $[3.28,\ \infty)$, \texttt{HMid} corresponds to $[7.09,\ \infty)$, and \texttt{HHigh} to $[13.26,\ \infty)$. Bold labels indicate the lower RMSE values across mass bins and datasets. $R^2$ and RMSE values as in Table~\ref{tab:fnl_loc_metrics}.
    }
    \label{tab:fnl_equil_metrics}
\end{table}

\subsection{Mass dependence on $\fnl$ constraints}

Halo mass binning plays an important role in shaping the sensitivity to primordial non-Gaussianity. Bins that include the high-mass tail, such as \texttt{HLow}, \texttt{HMid}, and \texttt{HHigh}, consistently achieved lower RMSE scores for both $\fnlloc$ and $\fnleq$. In the case of the intermediate mass ranges, see Tables~\ref{tab:fnl_loc_metrics_full_1} and~\ref{tab:fnl_equil_metrics_full_1}, the higher mass bin (e.g., \texttt{HMid-B}) improved performance compared to low-mass bins (\texttt{HLow-A}, \texttt{HLow-B}, \texttt{HMid-A}) by up to $30\%$ in RMSE, even in equilateral cases.

The overall trend shows that the constraining power is dominated by the high-mass end of the halo population, even for the topological summaries evaluated in this work. This behavior is expected for local- and equilateral-type PNG, which modulates the abundance of massive halos~\cite{LoVerde:2007ri}. This introduces a scale-dependent effect that enhances or suppresses halo formation proportional to $\fnl$ during early cosmic evolution~\cite{Matarrese:2000iz}. For the equilateral case, the best performance comes from \texttt{HMid}, which includes mid- and high-mass halos, though excluding the high-mass halos degrades the constraints (see Table~\ref{tab:fnl_equil_metrics_full_1}).

As shown in \cite{bermejo2024topologicalbiashaloestrace}, halo samples of different masses trace distinct topological regimes, low-mass halos delineate the fine-grained filamentary structure of the cosmic web, while high-mass halos probe a sparser, more clustered environment dominated by large-scale features. The stronger clustering of massive halos leads to larger voids and connected components (overdensities), which in turn shape the topological signal imprinted by $\fnl$. 

For the PSBS, it is well established that more massive (and thus more biased) halos exhibit a stronger scale-dependent bias effect, leading to tighter constraints on $\fnl$~\cite{Desjacques:2016bnm}. In contrast, the constraint from Betti curves, PD-statistics, PD-histogram, and Silhouette topological summaries showed to have stable performance between \texttt{HMid} and \texttt{HHigh}, but degrades in \texttt{HLow}. A simple interpretation is that \texttt{HMid} and \texttt{HHigh} sample broadly similar regions of the density field and therefore preserve comparable large-scale topological features that drive the $\fnl$ response. \texttt{HLow}, however, contains an excess of low-mass halos and effectively samples a different layer of the underlying density field. This introduces a much larger perturbation to the small-scale topology, injecting noise to the $\fnl$ signal in the persistence diagrams. Under this view, topological summaries may be sensitive to a different set of systematics than the PSBS, less sensitive to large-scale clustering bias, but more reponsive to changes in the fine-grained halo field.

\subsection{Model performance across architectures and summaries}

Of particular interest is the strong performance of PD-Statistics on the equilateral template, which is typically harder to detect due to its weaker scale dependence. Despite this challenge, PD-Statistics consistently achieved the lowest RMSEs and maintained high $R^2$ scores even in low-amplitude scenarios, outperforming PSBS and other topological summaries across all mass bins. This simple vectorization, built from a handful of elementary statistics, has also achieved the best performance in other classification tasks~\cite{2022arXiv221209703A,2019arXiv190407768C}. For local-type datasets, several vectorization methods performed well, including PSBS, Betti curves, persistent Betti, and histograms of counts.

To test whether this advantage was due to model architecture or the summary statistic itself, we compared Betti curves across three architectures using the \texttt{LH\_LC300} and \texttt{Q-LH}\_$\fnlloc$ datasets in \texttt{HMid}. The results were consistent, MLP achieved RMSE $\sim30$; XGB scored RMSE $\sim32$; and \perslay scored RMSE $\sim31$. A similar experiment with PSBS, comparing MLP with XGB, showed comparable results between architectures, thus the choice of the summary statistic had a greater impact on performance than the downstream model.

Hybrid models that combined PD-Statistics with PSBS offered modest but consistent improvements in the small $\fnl$ amplitude datasets. This seems to suggest that they contain a small ammount of complementary information.

\subsubsection*{Comparison with other topological summaries}

Regarding more complex topological summaries, we observed the following:

\begin{itemize} 

\item Persistence images paired with CNNs failed to outperform simpler models. To test whether this was an overparameterization issue, we trained on two image resolutions $128\times128$ and $64\times64$ on both \texttt{LH\_LC300} and \texttt{Q-LH}\_$\fnlloc$ datasets and the smaller image had a few percent improvement in performance. However, even the smaller images contained $\mathcal{O}(10^4)$ input features, while one-dimensional vectorizations (i.e. PD-histograms, Betti curves, silhouette, and landscape) had roughly $\mathcal{O}(10^3)$.

\item Set-based architectures such as \deepsets consistently underperformed relative to handcrafted summary statistics. Their training was highly sensitive to initialization and frequently failed to converge, while the computational overhead made hyperparameter tuning both challenging and inefficient. Input dimensionality was also a limiting factor, across the $18$ persistence diagrams, the combined feature size was on the order of $\mathcal{O}(10^5)$ points, yet each equivariant layer contained only $\mathcal{O}(10^3)$ trainable parameters. Scaling to such high-dimensional inputs typically requires larger neural networks and more extensive training datasets, which were beyond the scope of our current setup.

Combining pre-trained \deepsets and CNN networks performed comparable to one-dimensional summaries. Testing this set-up on \texttt{LH\_LC300} dataset in \texttt{HMid} achieved an improvement of $\sim 10\%$  over the individual \deepsets and CNN RMSE. Although promising, the added complexity and resource demands led us to exclude this hybrid configuration from the main analysis.

\item Although \perslay offered a more theoretically grounded alternative, it faced similar practical limitations, training in the sparsest bin, \texttt{HHigh}, required several hours per run, making full hyperparameter searches unrealistic. Nevertheless, when configured to emulate Betti curves using a constant $\phi$ function, linear learnable weights, and sum aggregation, \perslay achieved competitive results even without tuning.

\item A similar trend was observed for persistence landscapes, which underperformed relative to their more compact and smoother counterparts, persistence silhouettes. While silhouettes achieved reasonable performance, they, along with landscapes and persistence images, depend on tunable hyperparameters that control their expressiveness. In the case of landscapes, this includes the number of ``landscapes,'' whereas for persistence images, the bandwidth of the smoothing kernel plays a similar role. However, such tuning was not performed in this work.

\item In addition, several vectorization methods also incorporate weighting schemes to modulate the influence of features based on their persistence. For example, persistence images and silhouettes are weighted by a power of persistence, $(d_i - b_i)^p$, for each feature in the diagram. Larger values of $p > 1$ suppress short-lived features, while smaller values $p < 0.5$ amplify them. We tested $p = 1$ and $0.5$ for persistence images, and $p = 2$, $1$, and $0.5$ for silhouettes. The setting $p = 0.5$ performed slightly better for both methods. In particular, this aligns with the findings of our previous work~\cite{Calles:2024cxl}, where XGB-derived feature importance maps highlights that early-birth, low-persistence features in $0$-cycles carry significant cosmological information.
\\
This might hint on why summaries like persistence landscapes underperformed, as they tend to emphasize only the most persistent features in the diagram. A similar experiment was performed using histogram of counts vectorization, where we sorted each persistence diagram with the top $80\%$ most persistent features and constructed the histogram using only this sorted subset. Once again, this approach performed worse than using the full histogram, suggesting that less persistent features carry meaningful information in this dataset.

\end{itemize}

\section{Transferability}
\label{sec:transfer}

Machine learning methods require large datasets for training. In the case of cosmological observables, this means large simulation suites that are computationally very expensive. Thus, a key question is whether we can use cheaper ``fast'' simulations for training models that can then be compared with data, maybe using expensive simulations for validation. Although answering this question in general is beyond the scope of this work, we provide some examples in this direction.

For this, we train on the cheaper \pmwdsuite simulations \texttt{LH\_LC300}, and use them to predict the values from the \quijotepng suite; we conduct this experiment using the combined powerspsectrum (PSBS), and the Betti-curves topological statistics, using an MLP as models. Our results show that this transfer from fast to full simulations can work as long as the statistics do not include small scales or small halos, as we now discuss. 

\begin{figure}[ht]
    \centering
    \textbf{PSBS trained in \texttt{HMid}}\\[0.5em]
    \begin{subfigure}[b]{0.48\textwidth}
        \centering
        \includegraphics[width=\textwidth]{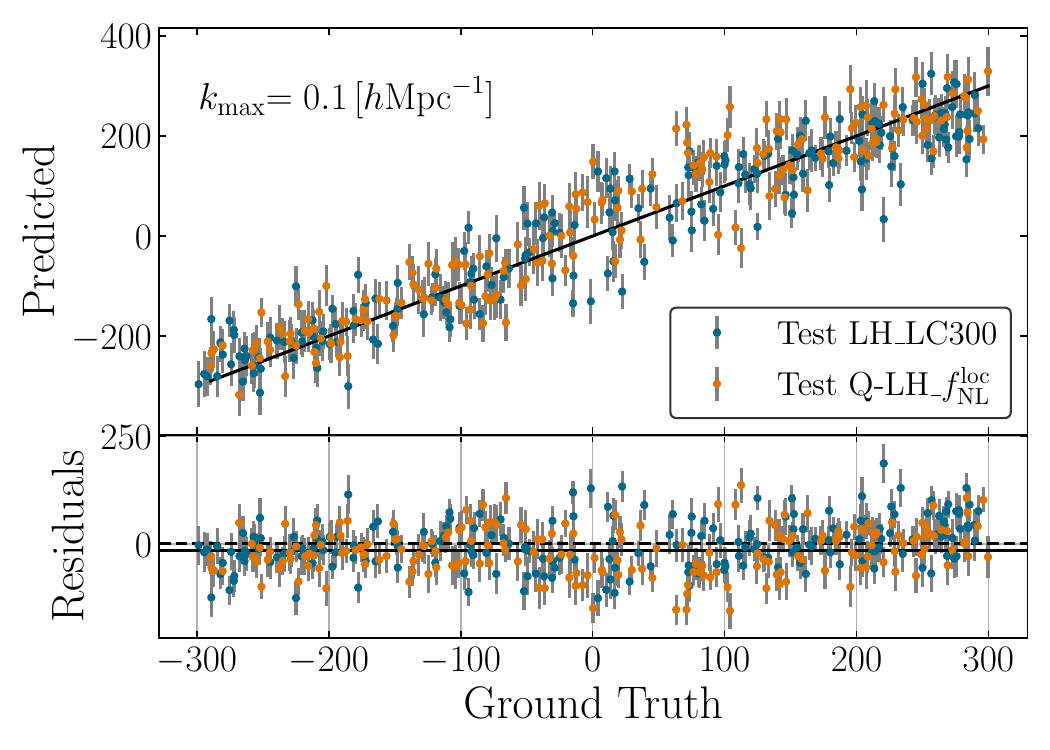}
    \end{subfigure}
    \hfill
    \begin{subfigure}[b]{0.48\textwidth}
        \centering
        \includegraphics[width=\textwidth]{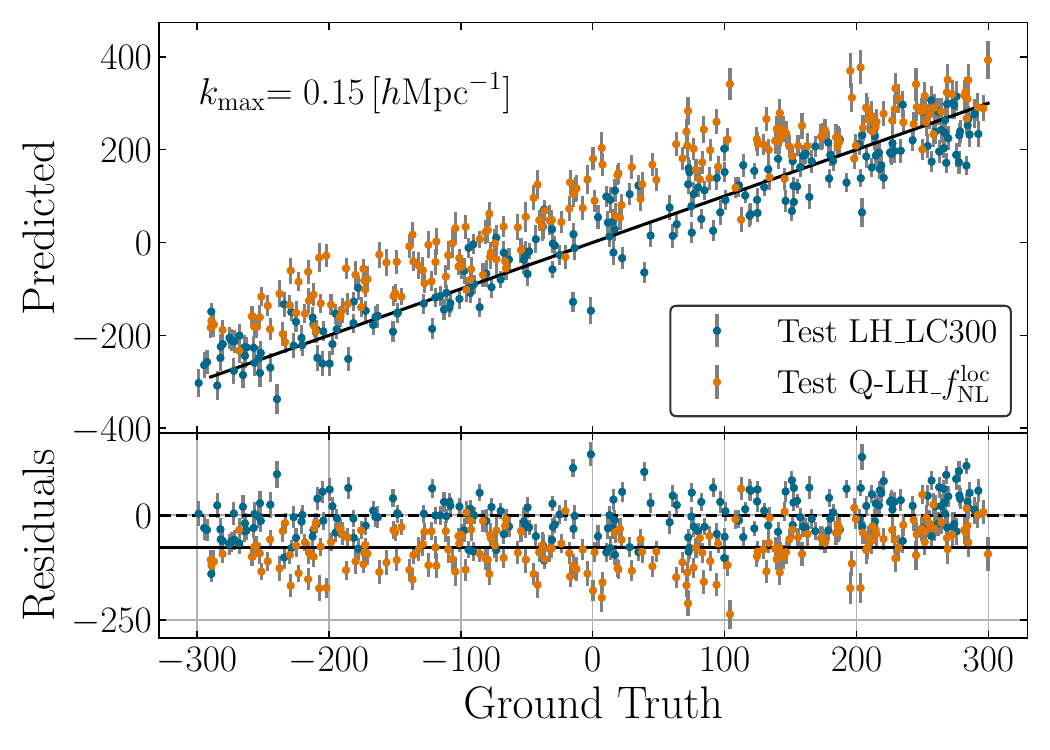}
    \end{subfigure}
    \begin{subfigure}[b]{0.48\textwidth}
        \centering
        \includegraphics[width=\textwidth]{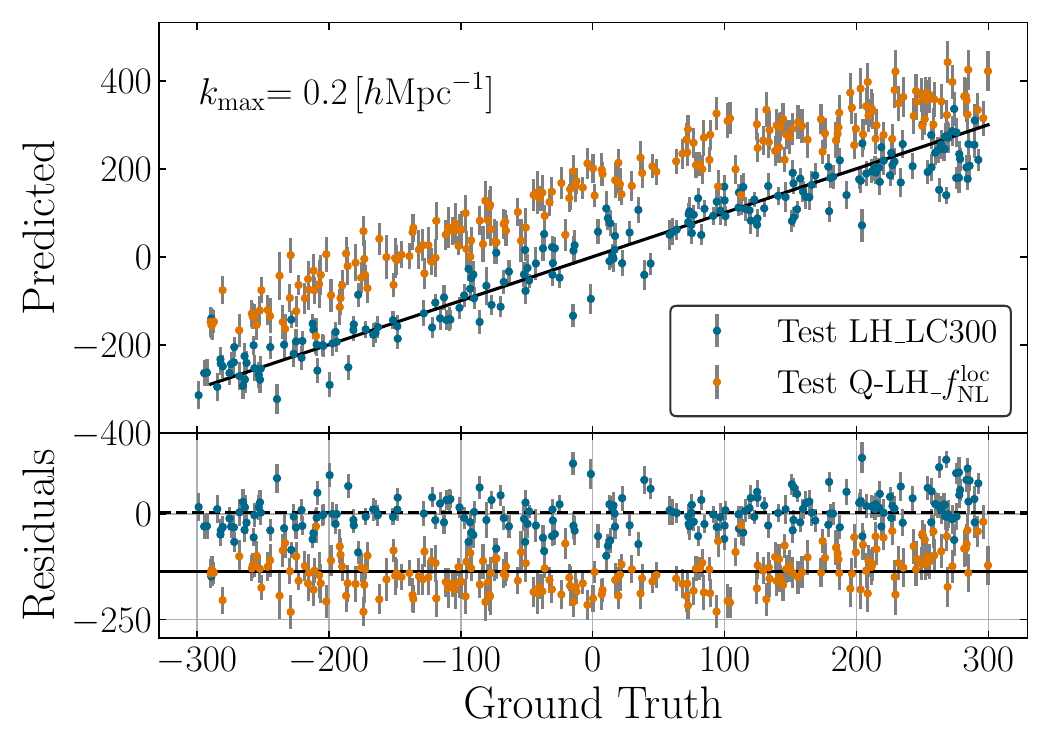}
    \end{subfigure}
    \hfill
    \begin{subfigure}[b]{0.48\textwidth}
        \centering
        \includegraphics[width=\textwidth]{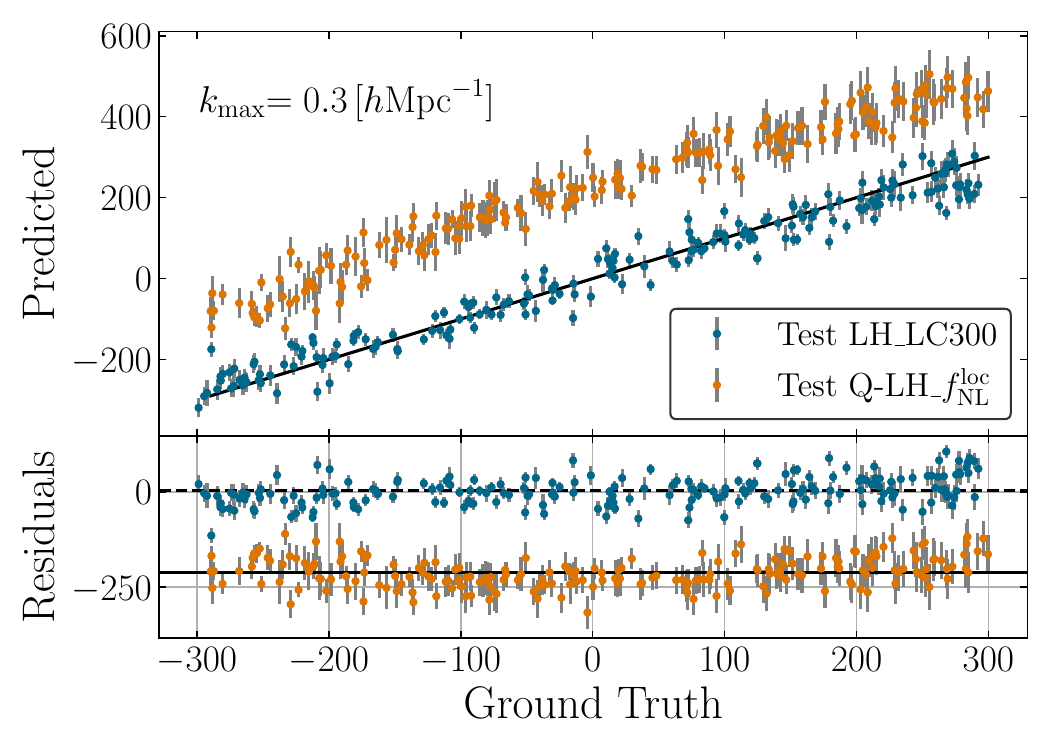}
    \end{subfigure}
    \caption{A MLP trained on PSBS vector from the \pmwdsuite \texttt{LH\_LC300} training split and evaluated on both \pmwdsuite (in blue) and \quijotepng (in orange) test sets to infer $\fnlloc$ in the \texttt{HMid} mass bin. Each panel shows the effect of including higher-wavelength bispectrum configurations, where all triangles containing a side with $k > k_{\rm max}$ are ignored. As $k_{\rm max}$ increases, predictions on the native test set become tighter due to the inclusion of more squeezed triangles, but robustness across simulations decreases.}
    \label{fig:PSBS_MLP_klambda}
\end{figure}

\begin{figure}[ht]
    \centering
    \textbf{PSBS trained in \texttt{HHigh}}\\[0.5em]
    \begin{subfigure}[b]{0.48\textwidth}
        \centering
        \includegraphics[width=\textwidth]{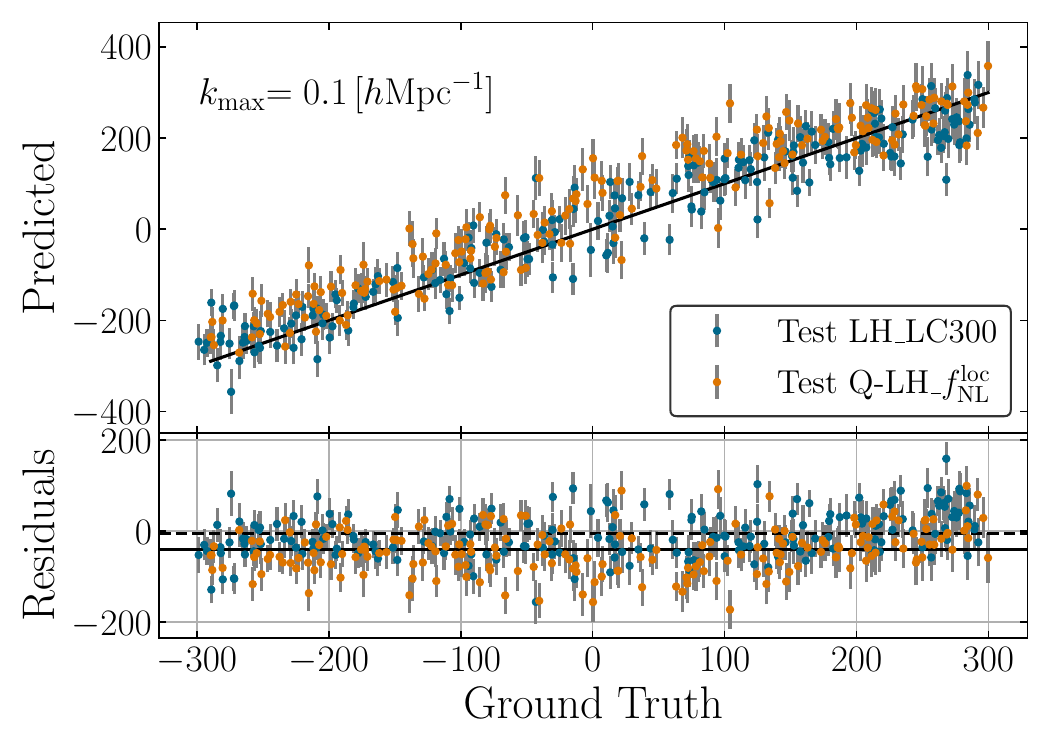}
    \end{subfigure}
    \hfill
    \begin{subfigure}[b]{0.48\textwidth}
        \centering
        \includegraphics[width=\textwidth]{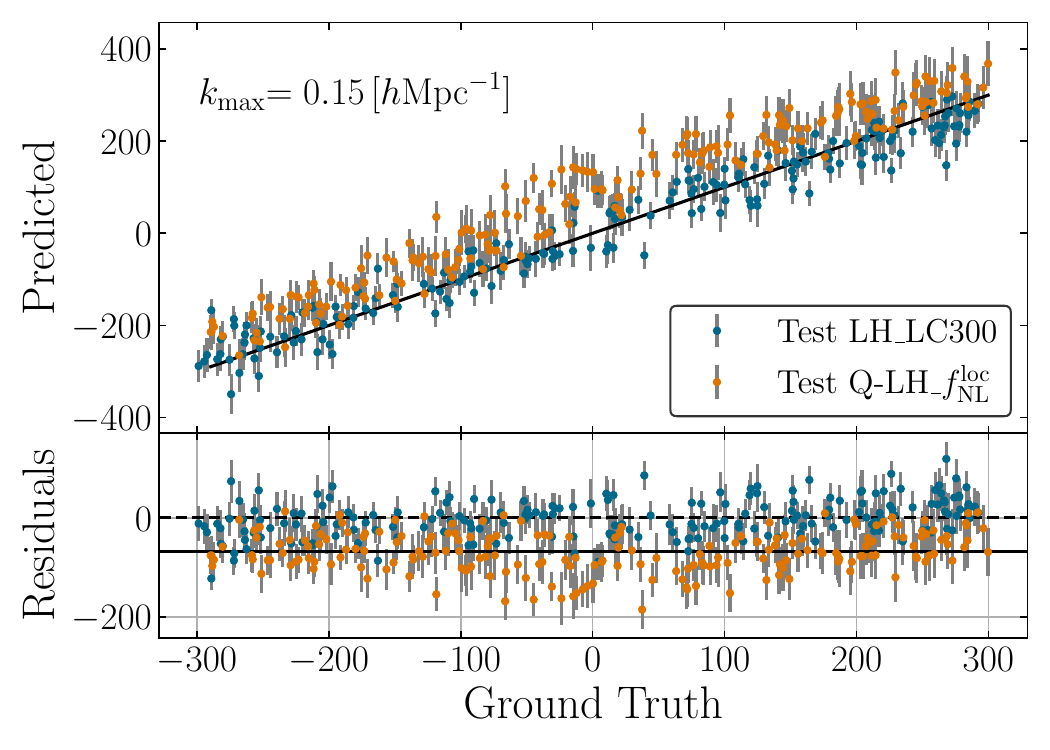}
    \end{subfigure}
    \begin{subfigure}[b]{0.48\textwidth}
        \centering
        \includegraphics[width=\textwidth]{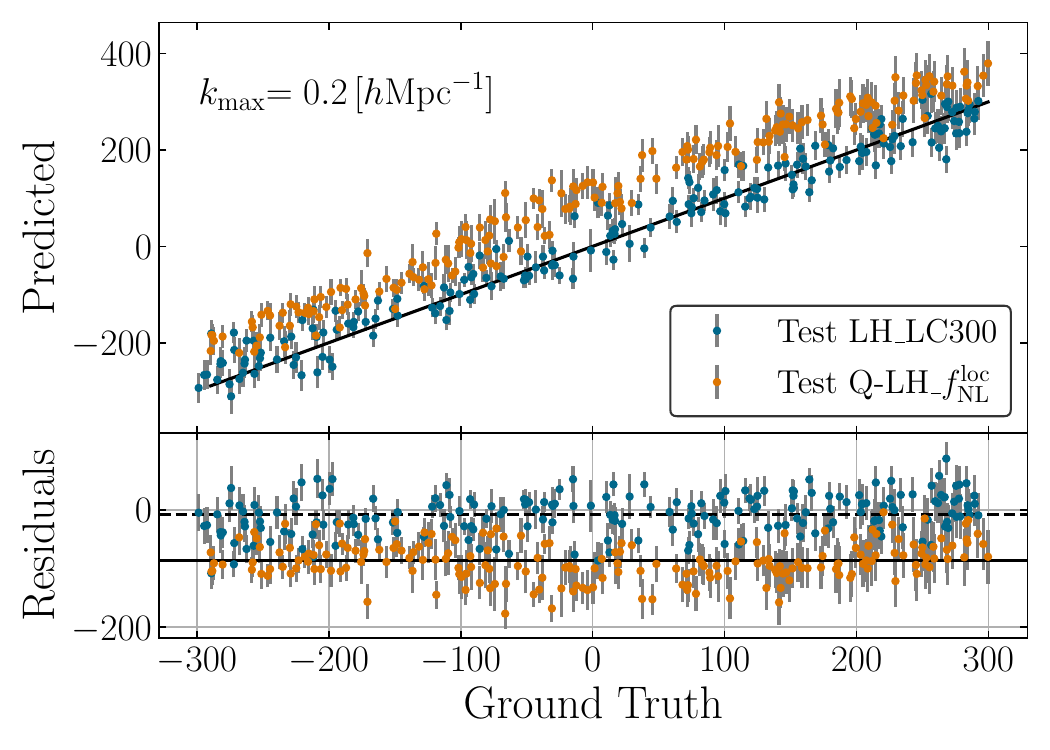}
    \end{subfigure}
    \hfill
    \begin{subfigure}[b]{0.48\textwidth}
        \centering
        \includegraphics[width=\textwidth]{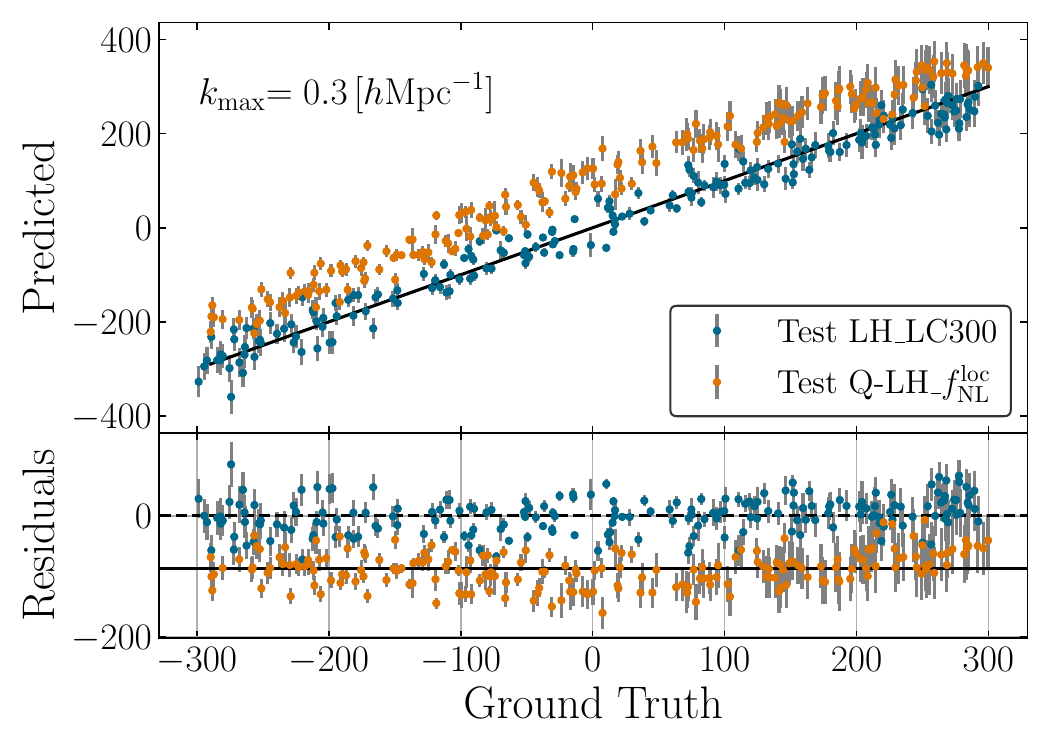}
    \end{subfigure}
    \caption{Same as Figure~\ref{fig:PSBS_MLP_klambda} but evaluated in the \texttt{HHigh} mass bin.}
    \label{fig:PSBS_MLP_klambda_mbin6}
\end{figure}

In the case of PSBS, we show in Figures~\ref{fig:PSBS_MLP_klambda},~\ref{fig:PSBS_MLP_klambda_mbin6} the ground truth \textit{versus} prediction obtained by these statistics in different settings (different wavelength used for PSBS and different mass bins) on a test set coming from \pmwdsuite (in blue) and on a test set coming from \quijotepng (in orange). We observe an upward offset in the predictions, especially when small-mass halos or small-scale features were included. Although models trained on the powerspectrum alone were robust in all mass bins, the inclusion of the bispectrum introduced an upward bias, induced by smaller scales. In this case, the offset arises from triangle configurations involving modes above $0.1\,h\,\rm{Mpc}^{-1}$ (see, e.g., bottom panels in Figure \ref{fig:PSBS_MLP_klambda}). This offset persisted in both the \texttt{HMid} (Figure \ref{fig:PSBS_MLP_klambda}) and \texttt{HHigh} mass bins (Figure \ref{fig:PSBS_MLP_klambda_mbin6}), though in the latter the constant offset was reduced by roughly a factor of two. We note that, conversely, we observed a similar effect when training on \quijotepng and evaluating on \pmwdsuite, resulting in a downward offset. A similar bias has been reported when transferring models across different simulation suites, where predictions tend to be biased toward higher values~\cite{Villaescusa-Navarro:2021cni,Villaescusa-Navarro:2021pkb,Villaescusa-Navarro:2022twv,Shao:2022mzk}.

In the case of Betti curves (and binned topological summary in general), a straightforward application of the same setup followed in Section~\ref{sec:results} in terms of normalization (i.e. where the statistics are normalized with respect to the full training dataset) leads to a large bias when predicting on the \quijote counterpart dataset. To build the Betti curves, we followed the choice from an earlier work and use a 35-bin uniform discretization, a setting that was found to optimize the Fisher information for histogram-based summaries in a dataset of comparable size~\cite{Yip:2024hlz}. Each bin counts the number of topological features in a range of scales. Changing the dataset potentially moves the number of features at each scale; thus, while the overall shape of the curve is consistent, it might be shifted. Using the binning definition from another dataset therefore changes the counts in each bin. 

To account for this, we propose to test a different standardization protocol. Instead of applying a global standardization with respect to the full (training) dataset, each summary is standardized individually. This approach should provide more robust statistics with respect to error introduced by the modeling of small mass halos, and is closer to a strategy one would follow in an actual observation scenario to reduce those effects from unreliable parts of the simulations (see Appendix~\ref{app:transfer} for additional details on the standardization protocols).

\begin{figure}[ht]
    \centering
    \textbf{Betti curves}\\[0.5em]
    \begin{subfigure}[b]{0.48\textwidth}
        \centering
        \includegraphics[width=\textwidth]{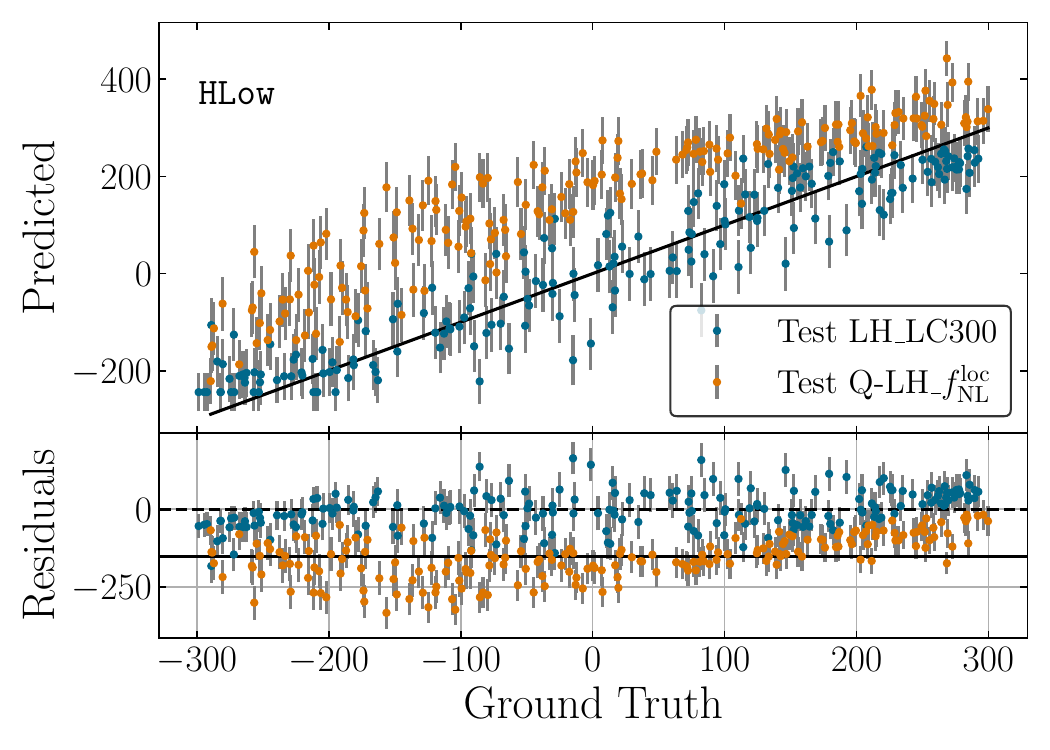}
    \end{subfigure}
    \hfill
    \begin{subfigure}[b]{0.48\textwidth}
        \centering
        \includegraphics[width=\textwidth]{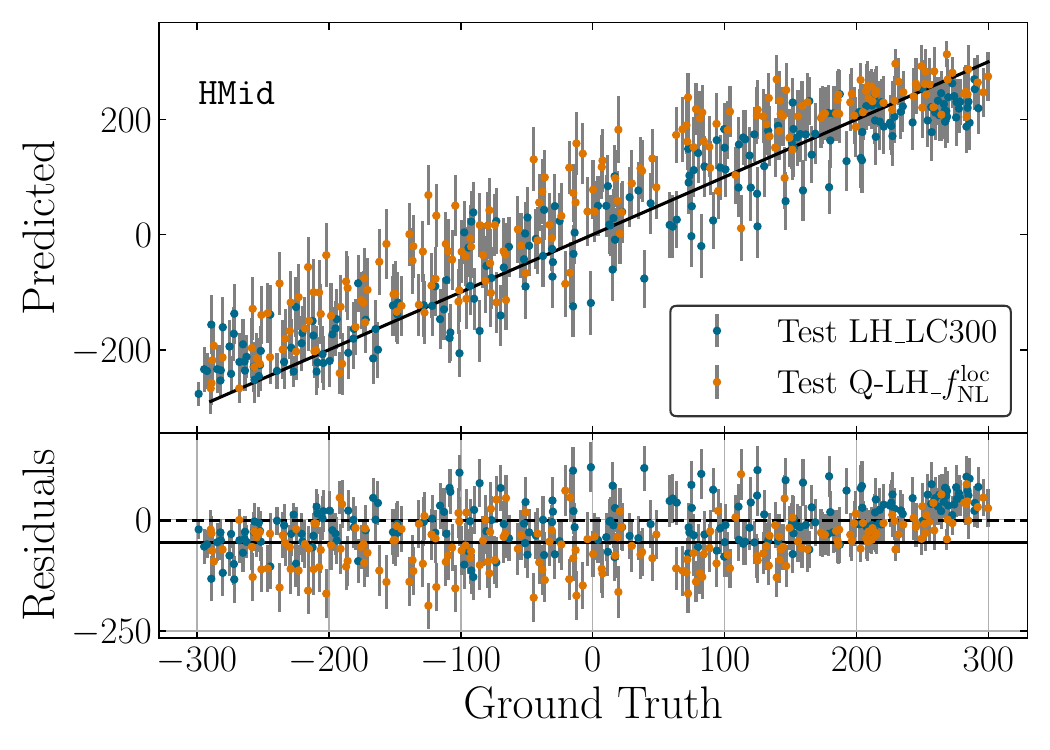}
    \end{subfigure}
    \begin{subfigure}[b]{0.48\textwidth}
        \centering
        \includegraphics[width=\textwidth]{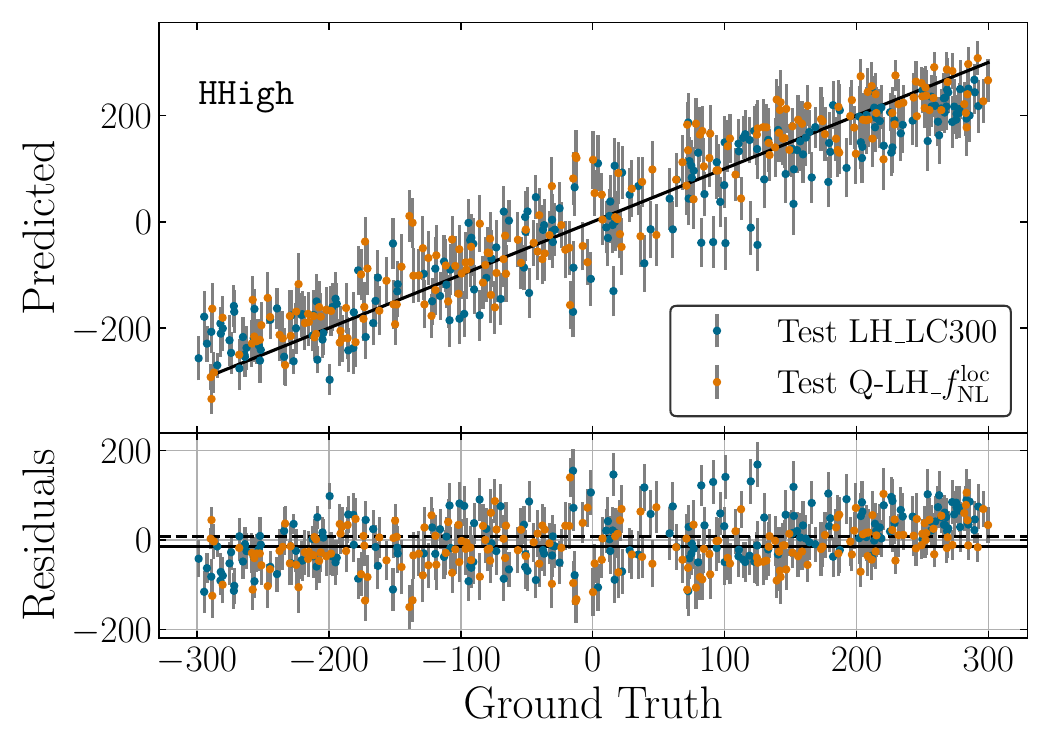}
    \end{subfigure}
    \caption{We train a MLP using Betti curves from the \pmwdsuite \texttt{LH\_LC300} training split and evaluate its performance on both \pmwdsuite (in blue) and \quijotepng (in orange) test sets across different mass bins to infer $\fnlloc$. Each panel shows the predicted mean values, while the lower subpanels display residuals with respect to the ground truth, including predicted error bars. Robustness cross-simulations Suite is achieved when using halos with at least $200$ particles (\texttt{HHigh}), as smaller halos tend to introduce significant systematic errors between simulations suites.}
    \label{fig:betti_MLP_massbins}
\end{figure}

The results obtained for Betti curves with this individual standardization protocol are shown in Figure~\ref{fig:betti_MLP_massbins}. Model transfer within the \texttt{HLow} (top-left panel) mass bin failed to generalize and exhibits a similar offset effect as the one observed with PSBS using smaller scales, with more variance in prediction. Transfers from \texttt{HMid} (top-right panel) showed better alignment with ground truth, although a small constant bias remained. In contrast, the model trained on \texttt{HHigh} (bottom panel) is robust and generalizes well across both test sets. However, this normalization protocol leads to reduced constraining power with less precise predictions (see Appendix~\ref{app:transfer} for a comparison with other normalizations protocols).

\begin{figure}[ht]
    \centering
    \textbf{PD-statistic with \xgb}\\[0.5em]
    \begin{subfigure}[b]{0.48\textwidth}
        \centering
        \includegraphics[width=\textwidth]{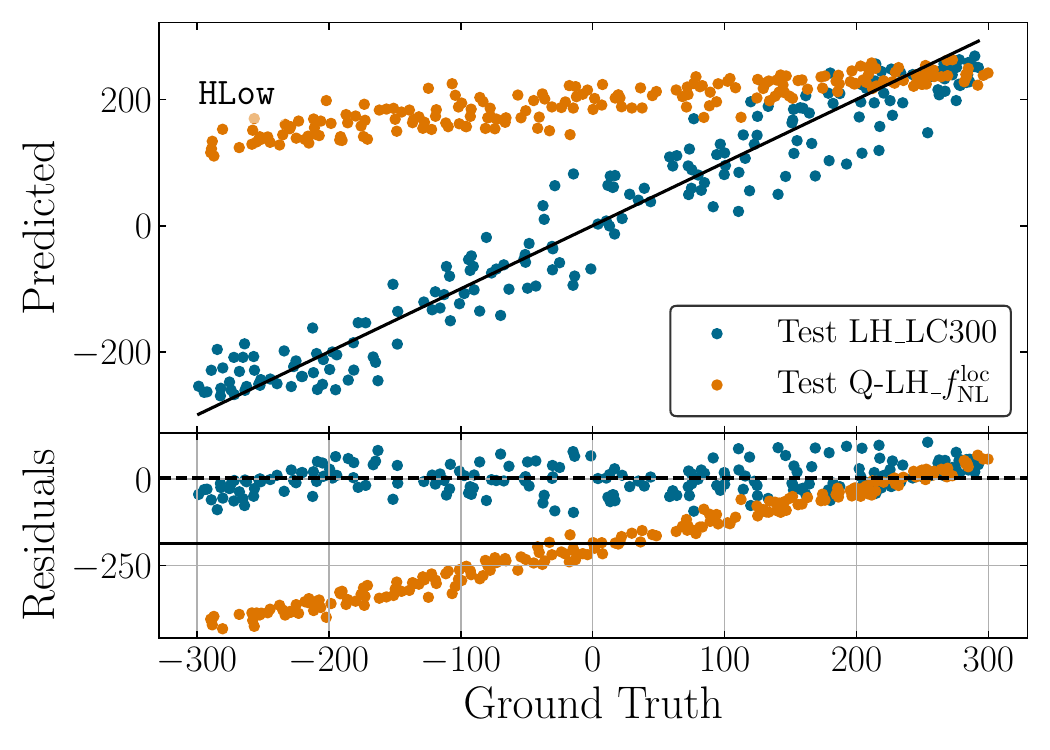}
    \end{subfigure}
    \hfill
    \begin{subfigure}[b]{0.48\textwidth}
        \centering
        \includegraphics[width=\textwidth]{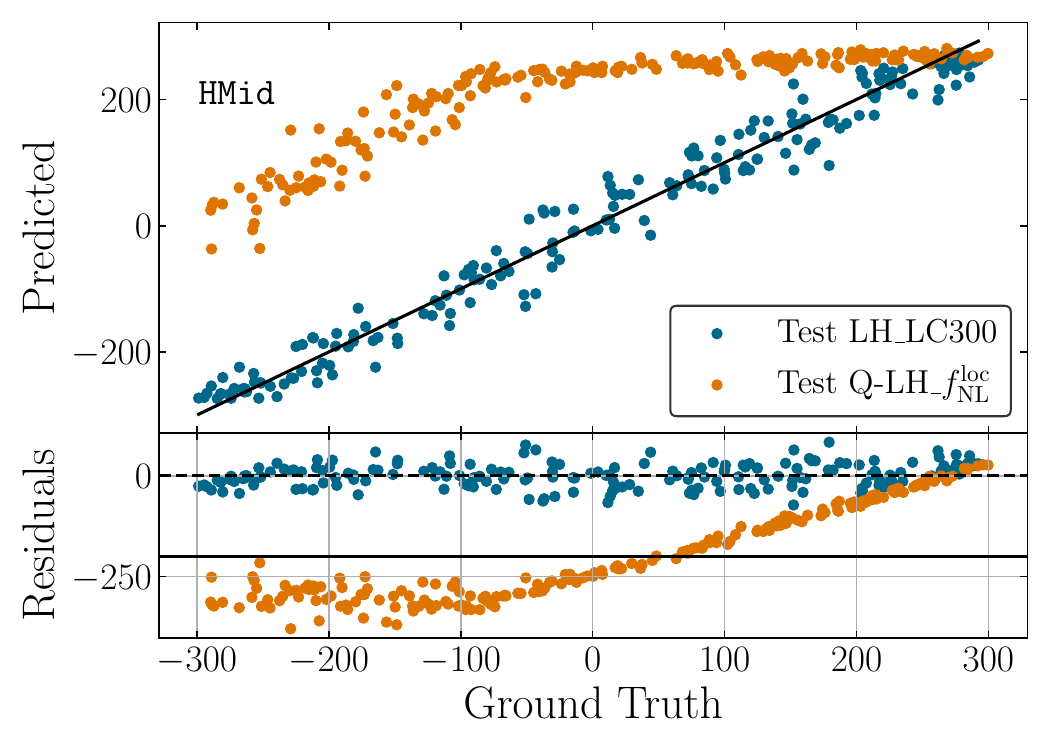}
    \end{subfigure}
    \begin{subfigure}[b]{0.48\textwidth}
        \centering
        \includegraphics[width=\textwidth]{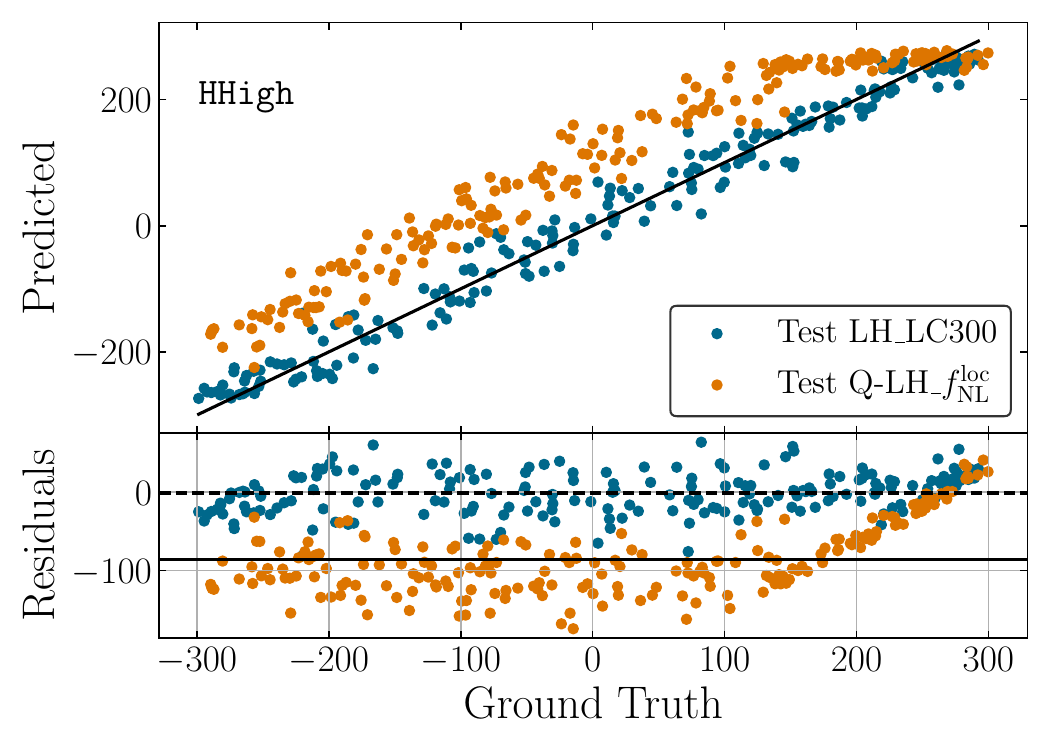}
    \end{subfigure}
    \caption{\xgb inference using PD-statistics from \pmwdsuite (\texttt{LH\_LC300}) and tested on both \pmwdsuite (in blue) and \quijotepng (in orange) across mass bins. Panels show mean predictions and residuals. Here, per-run rescaling has only a small effect on transferability, with discrepancies mainly driven by differences in mean halo density between the suites.}
    \label{fig:pdstat_xgb_massbins}
\end{figure}

We performed a similar experiment for the PD-statistic, with results shown in Figure~\ref{fig:pdstat_xgb_massbins}. Despite giving the tightest constraints, PD-statistic also lacked robustness for this transferability test, in all normalization protocols. We note that this summary provides simple statistics on the persistence features such as means and counts, where, for instance, the most important feature appears to be the total number of $0$-cycle in $k=1$. Those are directly tied to the number of haloes; thus, a local standardization solution of the statistics itself is not possible, and a local standardization of the Persistence Diagrams themselves will not effectively ``align'' the statistics, and the actual density will have an impact on the prediction. We indeed observed that restricting to higher mass haloes (i.e. \texttt{HHigh}), which provide more comparable densities across suites (see Table~\ref{tab:mass_bins},  provides a strong reduction in transfer bias. Results of the PD-statistics transferability experiments with different standardization protocols are provided in \ref{app:transfer}. Another possible strategy to investigate for transfer with this statistic could be to downsample the simulations to have similar mean halo number density.

We note ~\cite{Bayer:2025ija} also showed that smoothing the fields enabled more stable inference across simulation suites, pointing to small-scale and resolution differences from the solvers as primary sources of the constant offset. In our case, fast solvers are typically designed to reproduce two-point statistics and halo abundance at the percent level~\cite{Angulo:2021kes}, but tend to diffuse particles within halos, leading halo finders to fragment massive structures and underestimate smaller ones~\cite{Dai:2019fma}. This behavior might be the source of the bias observed in our setup.

We do not evaluate model transferability for $\fnleq$ since the \texttt{Q-LH}\_$\fnleq$ samples in \quijotepng vary multiple cosmological parameters simultaneously, including $\Omega_m$ and $\sigma_8$, alongside $\fnleq$. As discussed above, this setup induces strong parameter degeneracies that spoil the isolated impact of $\fnleq$, making direct comparisons across suites unreliable.

\section{Conclusions}
\label{sec:conclusions}

Our primary goal was to evaluate the information content of different summary statistics through a simple regression task within a likelihood-free framework, and to test their robustness when inferring on a simulation independent from the one used for training. 

To this end, we introduced the \pmwdsuite suite, a large-scale simulation suite specifically designed to support data-driven analyses of primordial non-Gaussianity. The suite provides $22{,}400$ dark matter halo catalogs, generated from Latin hypercube designs that vary primordial non-Gaussiniaties amplitudes $(\fnlloc, \fnleq)$ and standard cosmological parameters $(\Omega_m, \sigma_8)$ with different ranges. The suite, which is particularly useful in primordial non-Gaussianity analyses, is freely available. See the documentation at \url{https://png-pmwd-suite.readthedocs.io/en/latest/}.

Building on this new suite, we conducted a comprehensive benchmarking of common topological descriptors and clustering-based statistics across seven halo mass bins and five model architectures. PD-Statistics emerged as the most effective descriptor to infer both $\fnlloc$ and $\fnleq$ amplitudes.  Architectural complexity did not guarantee better performance: \deepsets, CNNs, and \perslay were computationally intensive and often failed to outperform simpler vectorization schemes. Between one-dimensional topological summaries (Betti curves, histogram of counts, persitence landscape, and persistence silhouettes), no particular method showed a clear advantage, as performance can be further improved through a dedicated hyperparameter optimization. 

While PD-Statistics retained predictive power even in the datasets with smaller ranges of $\fnl$, disentangling cosmological effects from $(\Omega_m, \sigma_8)$ remained challenging for all summary statistics studied without conditioning or assuming strong priors on other cosmological parameters. Such priors can come, for example, from previous observations, which already put stringent limits on these parameters. For this, we would need to combine the likelihood from these prior experiments with the likelihood-free frameworks studied here. It would be interesting to explore whether this can be achieved through novel simulation-based inference techniques.

Interestingly, our findings suggest that short-lived topological features carry more meaningful information than long-lived/persistent ones for $\fnl$ of both local and equilateral types. Weighting by $(death - birth)^p$ with $p \leq 0.5$ yielded better results than higher powers, suggesting that giving short-lived structures relatively more influence improves predictive power.

When we investigated whether we could use faster approximate algorithms such as \pmwd to train before applying to more accurate simulations like \quijotepng, we observed that the models tested (powerspectrum plus bispectrum, Betti curves, and PD statistics) are robust when restricted to halos containing at least $200$ particles or when focusing on scales around $\sim0.1\,h\,\mathrm{Mpc}^{-1}$, but exhibit biased predictions when either small-mass halos or small scales are included. This limitation may come from solver systematics as fast solvers tend to diffuse particles within halos, causing halo finders to fragment massive structures and artificially enhance clustering. One possible mitigation strategy is to filter out small-scale modes in Fourier space, or to discard features in persistence diagrams that die below the simulation's resolution scale, as these are not reliably resolved and may reflect numerical artifacts rather than physical structure. However, this comes at the cost of precision, since those small-scales carry relevant information. 

Since testing across fast and full simulations can be seen as a proxy for the difference between simulations and actual data, our results (which also concern the commonly used powerspectrum and bispectrum statistics) suggest that the crux lies in correctly modeling small scale/small-mass halos in order to gain constraining power. It would be interesting to further explore the transferability of statistics in general, across simulations and to actual data. Given that small scales are the hardest to describe, one might look for ways to improve the constraining power of the probes we study here in a way that is robust with respect to, or does not rely on, the modeling of the small scales.

Our results motivate several further investigations. First, it is important to obtain a better understanding  of the properties of the effect of primordial non-Gaussianity on topological summaries; this could, for instance, highlight if (and which part of) the summaries have a linear response to PNG on small ranges. Characterizing those responses (and their potential non-linearity) could motivate the investigation of different approaches for parameter inference, such as Fourier feature encoding \cite{tancik2020fourier}, that could improve the information extraction from the summaries. Second, investigating the combination of other types of probes (e.g. CMB) with the statistics investigated here is crucial, both to better understand the nature of the information they encode and to potentially improve the constraints on the parameters.

\section*{Acknowledgments}
Powered@NLHPC: This research was partially supported by the supercomputing infrastructure of the NLHPC (CCSS210001). J.C. is supported by FONDECYT de Postdoctorado, N° 3240444.  J.N. was supported for this work by ANID FONDECYT Regular 1251908. J.H.T.Y. and G.S. are supported in part by the U.S. Department of Energy, Office of Science, Office of High Energy Physics under Award Numbers DE-SC-0023719 and DE-SC-0017647. G.C. is supported by the European Union’s Horizon Europe research and innovation program under the Marie Sklodowska-Curie COFUND Postdoctoral Programme grant agreement No.101081355- SMASH and by the Republic of Slovenia and the European Union from the European Regional Development Fund. 
\\Disclaimer: Co-funded by the European Union. Views and opinions expressed are however those of the author(s) only and do not necessarily reflect those of the European Union or European Research Exacutive Agency. Neither the European Union nor the granting authority can be held responsible for them.

\section*{Code Availability}
We have developed and shared a modified version of the \pmwd code that incorporates equilateral-type non-Gaussianity. The updated codebase, which supports Gaussian and both local and equilateral non-Gaussian initial conditions, is available at \url{https://github.com/jcallesh/pmwd}.

Full documentation and data access to the \pmwdsuite suite can be found at \url{https://png-pmwd-suite.readthedocs.io/en/latest/}.

\appendix

\section{Consistency of \pmwdsuite Suite}
\label{appx:pmwd_checks}

\begin{figure}[ht]
    \centering
    \begin{subfigure}[b]{0.48\textwidth}
        \centering
        \includegraphics[width=\textwidth]{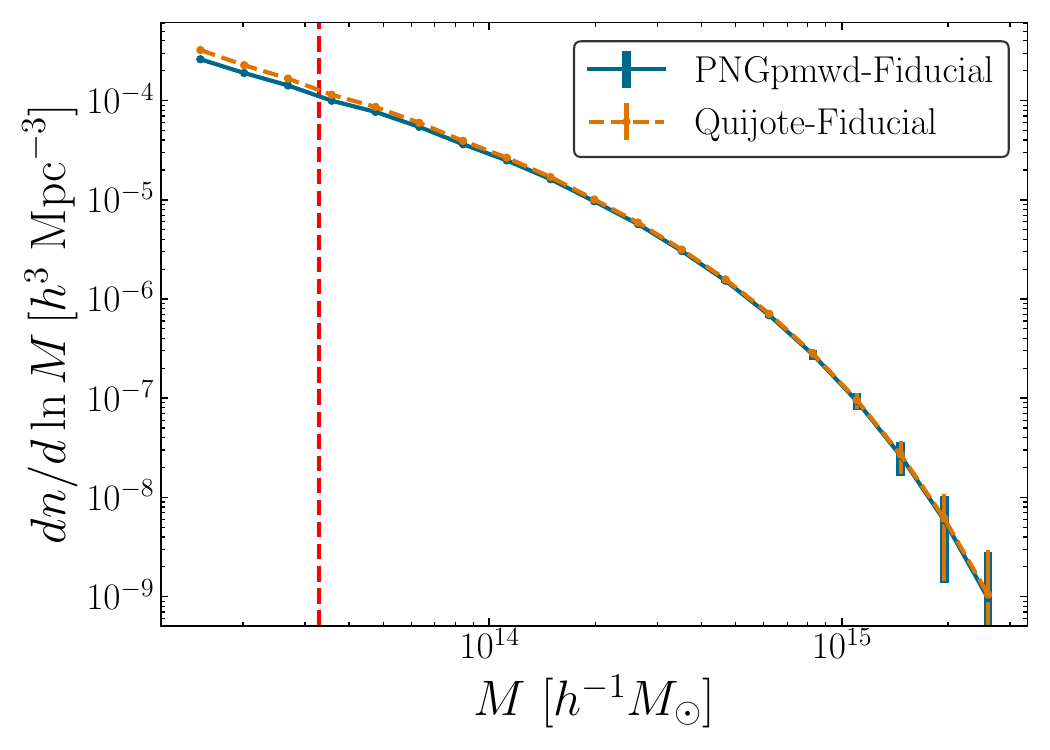}
        \label{fig:HMF_sub_a}
    \end{subfigure}
    \hfill
    \begin{subfigure}[b]{0.48\textwidth}
        \centering
        \includegraphics[width=\textwidth]{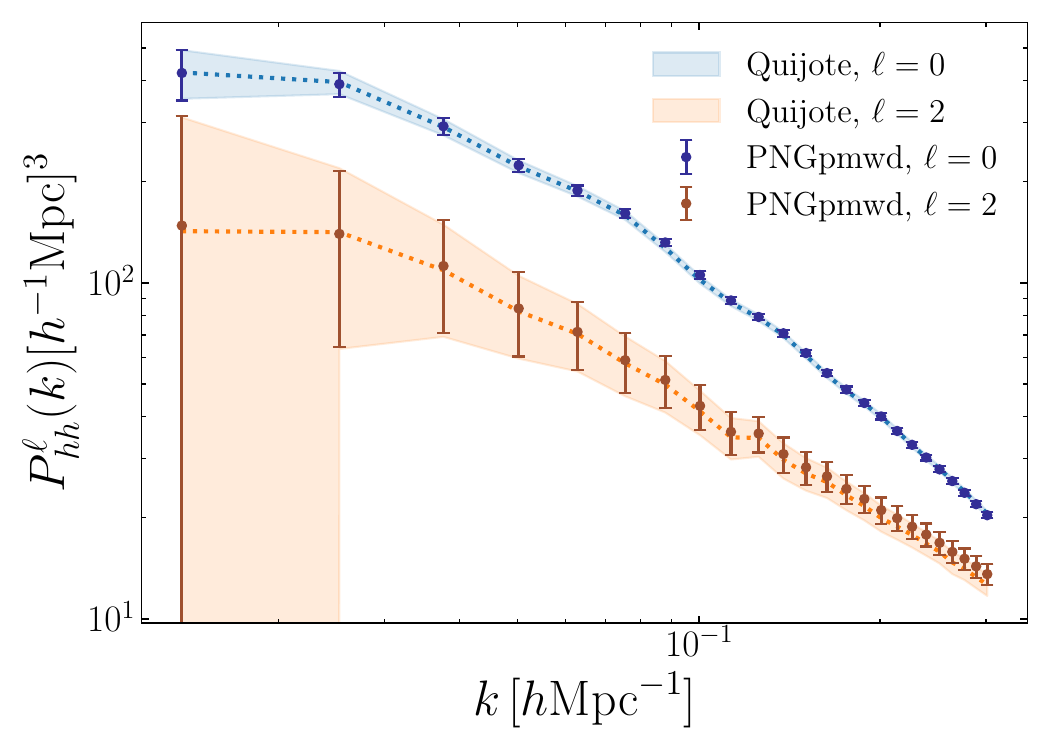}
        \label{fig:power_sub_b}
    \end{subfigure}
    \caption{Comparison between the fiducial cosmologies of \pmwdsuite and \quijote. Left: halo mass function, with the vertical red dotted line marking the lower mass threshold at $3.28 \times 10^{13}\,h^{-1}\rm{M}_\odot$. Right: redshift-space powerspectrum monopole and quadrupole for halos above the same mass threshold.}
    \label{fig:HMF_comparison}
\end{figure}

To check the similarity between the \pmwdsuite Suite and the \quijote Suite, we compared several standard halo and clustering statistics. Figure~\ref{fig:HMF_comparison} displays side-by-side comparisons of the halo mass function (left) and the redshift-space powerspectrum monopole and quadrupole (right), computed for the \texttt{HLow} mass bin, in the fiducial cosmology.

The halo mass function (left panel) shows that \pmwdsuite exhibits a mild suppression of halo abundance near the mass threshold (dashed red line), indicating an underdensity of low-mass halos relative to \quijote that gradually diminishes as one approaches the mass resolution. While the fiducial redshifts differ slightly ($z=0.503$ vs. $z=0.5$), the primary discrepancy comes from simulation methodology as fast simulations tend to diffuse particles within halos, leading to fragmentation of large halos and under-detection of smaller ones. Minor differences in the linear matter powerspectrum, computed via \texttt{CAMB} for \quijote and Eisenstein \& Hu fitting formulae for \pmwdsuite, may also subtly affect halo populations.

The right panel showcases the redshift-space powerspectrum. We find excellent agreement in both the monopole and quadrupole, with \pmwdsuite error bars closely tracking the variance envelope from \quijote, though small differences begin to emerge at higher wavenumbers (smaller scales).

Table~\ref{tab:bias_qjt} reports the linear bias, $b_1$, across mass bins. The $b_1$ values were estimated from the ratio of the powerspectrum measurements to the linear powerspectrum of the fiducial cosmology in real space, and averaged over $2000$ realizations. The slightly larger $b_1$ values in \pmwdsuite, may reflect halo fragmentation effects, which can enhance clustering and therefore bias estimates.

Overall, these comparisons provide confidence in the use of \pmwdsuite for cosmological inference, while also highlighting systematics that should be considered in downstream analyses.

\begin{table}[ht]
    \centering
    \begin{tabular}{l |c | c | l}
    \textbf{Label} & \quijote $(b_1)$ & \pmwdsuite $(b_1)$ &\textbf{Mass Range [$10^{13}\,M_\odot/h$]} \\
    \hline\hline
    \texttt{HLow} & $2.44$ & $2.46$ & $[3.28,\ \infty)$ \\
    \texttt{HLow-A} & $2.04$ & $2.03$ & $[3.28,\ 4.46)$ \\
    \texttt{HLow-B} & $2.28$ & $2.29$ & $[4.46,\ 7.09)$ \\
    \texttt{HMid} & $3.06$ & $3.09$ & $[7.09,\ \infty)$ \\
    \texttt{HMid-A} & $2.57$ & $2.59$ & $[7.09,\ 9.06)$ \\
    \texttt{HMid-B} & $2.87$ & $2.89$ & $[9.06,\ 13.26)$ \\
    \texttt{HHigh} & $3.78$ & $3.82$ & $[13.26,\ \infty)$ \\
    \end{tabular}
    \caption{Comparison of linear bias values from the fiducial cosmology for both \pmwdsuite and \quijotepng, estimated from the ratio of the powerspectrum to the linear powerspectrum in the fiducial cosmology in real-space.}
    \label{tab:bias_qjt}
\end{table}

\section{Summarizing the topology}
\label{appx:topological_summaries}

Although persistence diagrams provide an interpretable summary of the multiscale topology of the dataset, their high dimensionality (in number of points) and variable size make them difficult to use with traditional methods. To overcome this, a range of compression techniques have been developed to convert persistence diagrams into fixed-size vector representations that retain specific topological information while enabling integration into standard pipelines. These vectorization methods have been extensively discussed in the literature and implemented across various languages and software packages. For a comprehensive overview, we refer the reader to~\cite{2019arXiv190407768C,2022arXiv221209703A,2024arXiv240902901C,2024arXiv241117340L}. We briefly describe below several commonly used approaches that were explored in this work.

\subsection*{Persistence images (PI)}

This is a direct, structured, and robust method for converting persistence diagrams into two-dimensional image representation~\cite{JMLR:v18:16-337}. Persistence images map each point to a smooth kernel density estimate, $\phi$, centered at its birth and death coordinates, $(b,d)$, given some width $\sigma$. These kernels are typically weighted by a function of persistence $(d-b)$ to emphasize long- or short-lived topological features. The result is a continuous surface, $\mu=\sum_iw(d_i-b_i)\phi_i$, known as the persistence surface, which encodes the overall shape and relative importance of topological features given in the original persistence diagram. This surface is then discretized into a grayscale image by summing the kernel contributions over a two-dimensional grid of pixels. This summary has been used as input to convolutional neural networks for inferring cosmological parameters~\cite{Yip:2023vud,Calles:2024cxl}. We emphasize that, in our case, this summary results in 18 images for a single simulation realization (or halo catalog), one per persistence diagram, and image-size of $64\times64$ pixels. 

\subsection*{Histogram of count (PD-Histogram)}

This vector representation captures the marginal distributions of birth and death feature scales, by constructing six histograms per diagram, three for birth distributions $(B_0, B_1, B_2)$ and three for death distributions $(D_0, D_1, D_2)$, each corresponding to $0-$, $1-$, and $2-$cycles, with 35 bins each. The procedure is simple: discretize the birth and death values into fixed bins and count the number of features falling into each bin, separately for births and deaths. This summary was used in Fisher forecast on primordial non-Gaussianities amplitudes and cosmological parameters in~\cite{Biagetti:2022qjl,Yip:2024hlz}.

\subsection*{Betti curve (PD-Betti)}

For a given homological dimension $p$, the Betti curve $\beta_p(\nu)$ records the number of $p$-cycles present at scale $\nu$. Equivalently, it counts the intervals $[b_i, d_i)$ in the persistence diagram that contain $\nu$:
\begin{equation}
\label{eq:betticurve}
    \beta_p(\nu) = \sum_{(b_i, d_i) \in \mathrm{PD}} \bm{1}\{\,b_i \leq \nu < d_i\,\},
\end{equation}
where $\bm{1}\{\cdot\}$ denotes the indicator function. In practice, we discretize the filtration range into 35 uniformly spaced bins, producing a binned Betti curve for each homology dimension. Betti curves are relatively robust to noise, as short-lived features contribute minimally to their overall shape. Betti curves are perhaps the most commonly used topological summary, given their computational efficiency and interpretability. We highlight some application to large-scale structure analyses~\cite{10.1093/mnras/stw2862,Tymchyshyn:2023czh,Abedi:2024ajq}.

\subsection*{Persistent Betti (PD-PerBetti)}
A slight variation of the Betti curve is the \emph{Persistent Betti Curve}, with the same discretization, which focuses on the most persistent topological features on the diagram. Instead of counting all features alive at scale $\nu$, this curve restricts the sum to those with the highest persistence values. Thus, it emphasizes the most robust topological structures in the data and reducing the influence of small-scale fluctuations. It was recently apply to real data and compared to Betti curves in~\cite{DES:2025akz}.

\subsection*{Persistence landscape (PD-Landscape)}

The $k$-th order persistence landscape~\cite{2018arXiv181004963B,2020arXiv200202778K} is a functional summary of a persistence diagram, defined as
\begin{equation}
    \lambda_k(\nu) = k\text{-max}_{1 \leq i \leq N} \Lambda_i(\nu), \quad k \in \mathbb{N},
\end{equation}
where $k$-max denotes the $k$-th largest value among the functions $\Lambda_i(\nu)$, each corresponding to a topological feature indexed by $i$. The function $\Lambda_i(\nu)$ is a piecewise-linear function, ``tend'', with endpoints at $(b_i, 0)$ and $(d_i, 0)$ and peaks at the midpoint $\left(\frac{b_i + d_i}{2}, \frac{d_i - b_i}{2}\right)$. Intuitively, lower-order landscapes (e.g., $\lambda_1$, $\lambda_2$, $\lambda_3$) highlight the most persistent features in the diagram, whereas higher-order landscapes incorporate finer topological details. Increasing the landscape order enriches the representation but also raises computational complexity and sensitivity to noise. In this work, we retain the first three landscape orders.

\subsection*{Persistence silhouette (PD-Silhouette)}
Although persistence landscapes emphasize dominant long-lived features, they may underpin short-lived ones. The $p$-persistence silhouette~\cite{chazal2014stochastic} mitigates this by introducing a weighted average over the tend functions according to persistence:
\begin{equation}
    \Psi_p(\nu) = 
    \frac{\displaystyle\sum_{i=1}^{N} (d_i - b_i)^p \, \Lambda_i(\nu)}
         {\displaystyle\sum_{i=1}^{N} (d_i - b_i)^p},
\end{equation}
with $p \geq 0$ controlling the relative emphasis. Larger values of $p$ (e.g., $p > 1$) accentuate long-lived features, while smaller values (e.g., $p < 0.5$) increase sensitivity to short-lived ones.

\subsection*{Persistence statistics (PD-statistic)}

This is a parameter-free vectorization that aggregates basic statistical measures computed from birth $(b)$, death $(d)$, midpoint $([b+d]/2)$, and persistence $(d-b)$ values from each persistence diagram~\cite{2018arXiv181100252P}. It comprises the following features:

\emph{Descriptive Statistics}: Mean, standard deviation, median, inter-quartile range (IQR), full range, and percentiles (10\textsuperscript{th}, 25\textsuperscript{th}, 75\textsuperscript{th}, 90\textsuperscript{th}) for births, deaths, midpoints, and persistence, together with the total number of features.

\emph{Entropy}: A persistence-weighted entropy defined on the persistence diagram by,
\begin{equation}
    E_\mu = -\sum_{[b,d] \in B} \mu_{b,d} \cdot \left(\frac{d-b}{L_\mu}\right) \cdot \log\left(\frac{d-b}{L_\mu}\right),
\end{equation}
where $L_\mu = \sum \mu_{b,d} \cdot (d-b)$ is the total weighted persistence. For visualization purposes, Table~\ref{tab:pd_stats} displays the numerical values derived from the persistence diagram of the $0$-cycle, realization $0$, corresponding to a fiducial cosmology in the ``1P'' dataset. This summary statistic is robust to small perturbations in the persistence diagram and benefits from the intuitive nature of vectorized features, which align with familiar statistical descriptors. However, it also has limitations, it may overlook fine-grained topological information in the diagram and implicitly assumes that these coarse summary statistics retain sufficient discriminative information for downstream tasks.

\begin{table}
\begin{tabular}{l|rrrrrrrrr}
     & Mean & STD & Median & IQR & Range & P10 & P25 & P75 & P90 \\
     \hline\hline
    Births & 11.61 & 7.62 & 9.66 & 10.51 & 37.83 & 3.74 & 5.50 & 16.01 & 22.89 \\
    Deaths & 14.25 & 7.45 & 13.46 & 10.78 & 38.23 & 5.16 & 8.25 & 19.03 & 24.34 \\
    Midpoints & 12.93 & 7.21 & 11.42 & 9.21 & 37.90 & 4.25 & 6.58 & 17.82 & 23.70 \\
    Lifespans & 2.64 & 4.38 & 0.77 & 1.84 & 34.08 & 0.07 & 0.30 & 2.14 & 9.34 \\
    \hline
    Count & 27871 &  &  &  &  &  &  &  & \\
    Entropy & 9.31 &  &  &  &  &  &  &  & \\
\end{tabular}
\caption{PD-statistic for a single fiducial realization. Notices that birth, death, midpoint, and lifespans are in units of Mpc$/h$.}
\label{tab:pd_stats}
\end{table}

Finally, for downstream tasks, the feature vector is often constructed by concatenating the chosen representation across all $p$-dimensional homology groups. Alternatively, specialized architectures have been developed to operate directly on persistence diagrams. These methods optimize representation learning either by selecting the most effective vectorization from a broad design space, as in \perslay, or by treating persistence diagrams as point clouds, as in \deepsets (see Appendix~\ref{appx:architectures}).

\section{Practical considerations for topological summaries}
\label{app:transfer}

When building a binned summary derived from persistence diagrams, it is essential for comparison between realizations that each bin in one summary corresponds to the same physical scale in another dataset. This requires choosing bounds (minimum and maximum feature scales) and discretizing them uniformly. Here we show how this choice affects cross-simulation inference. We train a MLP on Betti curves from the \texttt{LH\_LC300} \pmwdsuite sample and evaluate its ability to infer $\fnlloc$ for its own \texttt{LH\_LC300} test-set (in blue) and on the \texttt{Q-LH}\_$\fnlloc$ subset of \quijotepng (in orange).

\begin{figure}[ht]
  \centering
    \textbf{Betti curves vectorization}\\[0.5em]

  \begin{subfigure}[t]{\textwidth}
    \centering
      \includegraphics[height=5cm]{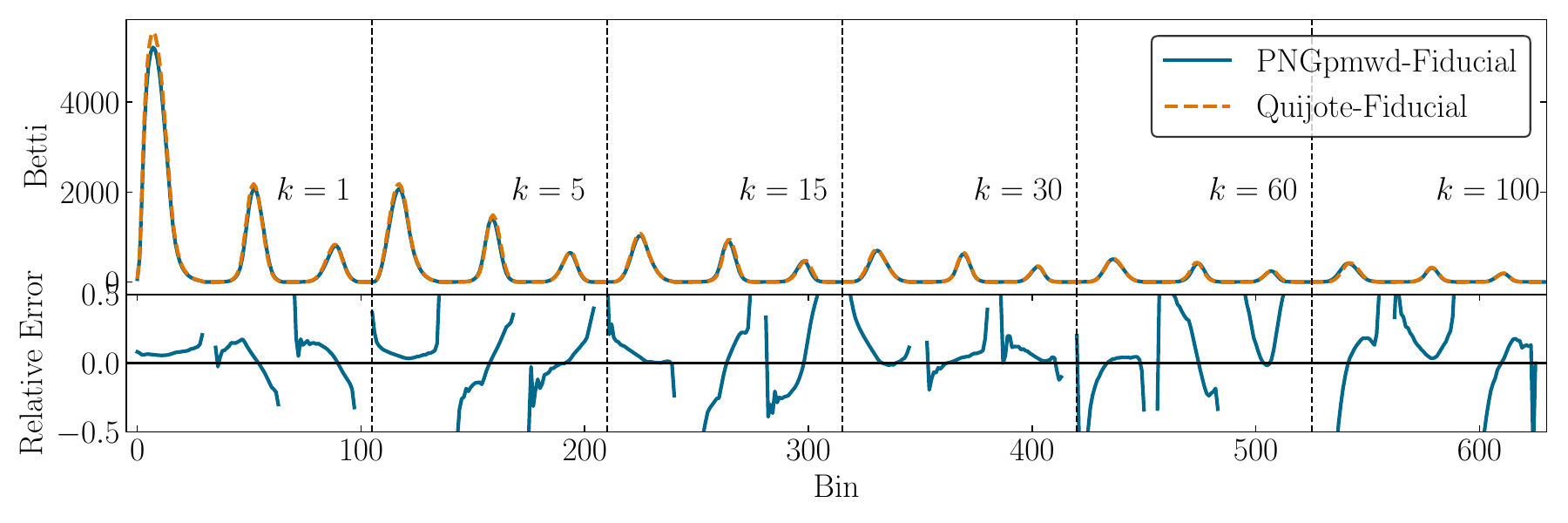}
    \caption{Normalize \pmwdsuite and \quijote independently}
    \label{fig:betti_selfbounds}
  \end{subfigure}
  
  \vspace{1em}

  \begin{subfigure}[t]{\textwidth}
    \centering
      \includegraphics[height=5cm]{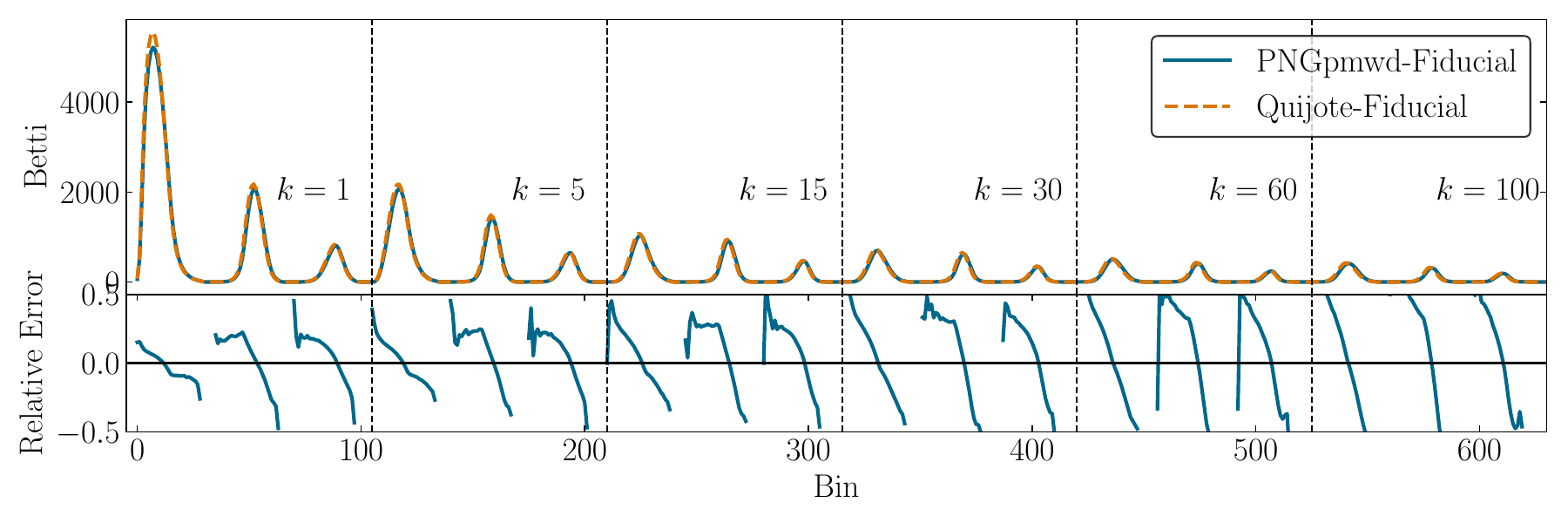}
    \caption{\pmwdsuite fiducial bounds on both \pmwdsuite and \quijote}
    \label{fig:betti_pmwdfidbounds}
  \end{subfigure}

  \vspace{1em}
  
  \begin{subfigure}[t]{\textwidth}
    \centering
      \includegraphics[height=5cm]{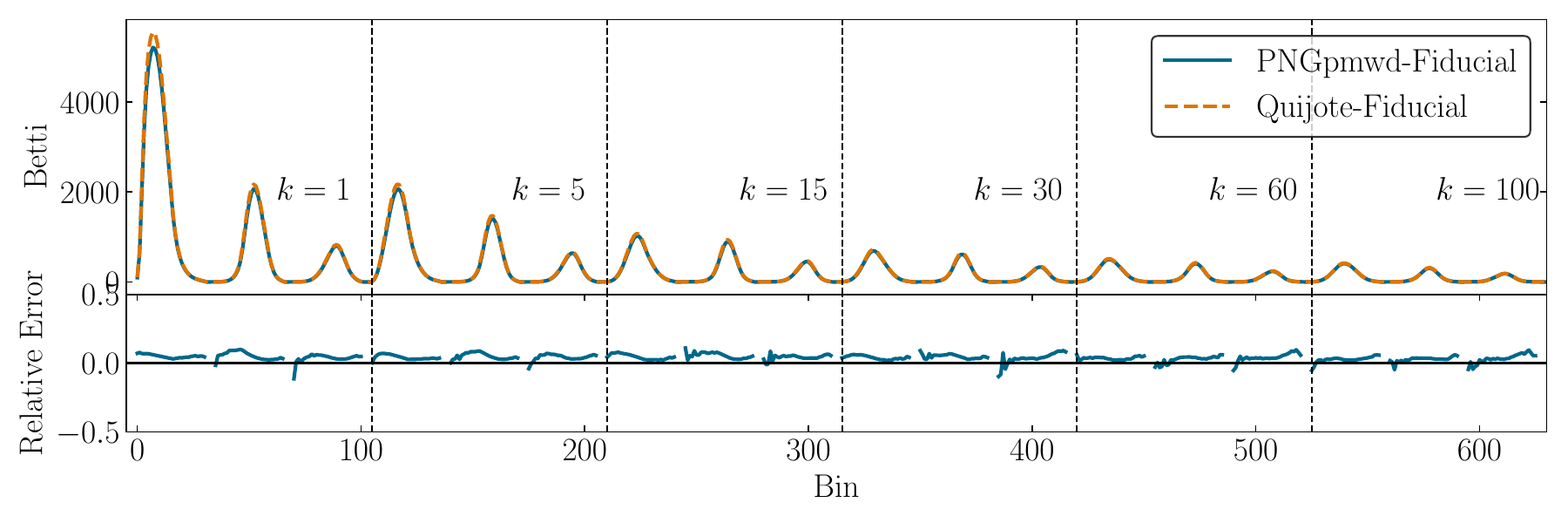}
    \caption{Standardize each PD run individually}
    \label{fig:betti_scaleperrun}
  \end{subfigure}

  \caption{Impact of bound choices on Betti curves. Each panel shows on top the mean Betti curves over $2000$ fiducial realizations and bottom the relative difference between \quijote and \pmwdsuite for the \texttt{HMid} mass bin. (a) Dataset-specific bounds: each suite is normalized using its own min-max filtration ranges, preserving feature distributions per-bin number. (b) \pmwdsuite bounds applied to both suites: \quijote features are misaligned because its mean inter-halo distances differ, leading to biased curves. (c) Per-run normalization: rescaling each persistence diagram restores bin alignment but removes absolute scale information.}
  \label{fig:betti_bounds}
\end{figure}

We explore three approaches:

\begin{itemize}
    \item Normalize \pmwdsuite and \quijotepng independently:
    As described in Section~\ref{sec:persistencepipeline}, we compute min-max birth and death scales for each homology dimension and each $k$-value across the full \texttt{LH\_LC300} (\pmwdsuite) training set, and apply these consistently to its training/validation/test splits. Separately, we compute analogous bounds for the \texttt{Q-LH}\_$\fnlloc$ (\quijotepng) sample. This procedure preserves the internal distribution of features within the bound range of each dataset, as shown in Figure~\ref{fig:betti_selfbounds}, although the bounds physical scales differ between suites, the relative variations in Betti curves induced by $\fnlloc$ are preserved.

    \item \pmwdsuite bounds on both \pmwdsuite and \quijotepng:
    Instead of computing separate bounds for \texttt{Q-LH}\_$\fnlloc$ (\quijotepng), we apply the bounds learned from \texttt{LH\_LC300} (\pmwdsuite) to both test-sets. Because \pmwdsuite and \quijotepng differ in mean inter-halo distances, the physical scales at which topological features are born are not matched between suites. As a result, features in \quijotepng are mapped into bins corresponding to earlier filtration scales, distorting their distribution and shifting the resulting Betti curves, Figure~\ref{fig:betti_pmwdfidbounds}. Notices that the relative variations due to $\fnlloc$ are still preserved between datasets but off-set.

    \item Standardize each PD run individually:
    In this approach, each persistence diagram, regardless of dataset, is independently rescaled to the interval $[0,1]$. This removes bin misalignment and produces Betti curves with matched support, Figure~\ref{fig:betti_scaleperrun}. However, because all absolute scales are normalized away, this procedure also washes some of the relative impact of $\fnlloc$, compressing or stretching its physical variations.

\end{itemize}

\begin{figure}[ht]
  \centering

  \begin{subfigure}[t]{\textwidth}
    \centering
    \textbf{Betti curves model transferring}\\[0.5em]
    \begin{minipage}[t]{0.45\textwidth}
      \centering
      \includegraphics[height=5cm]{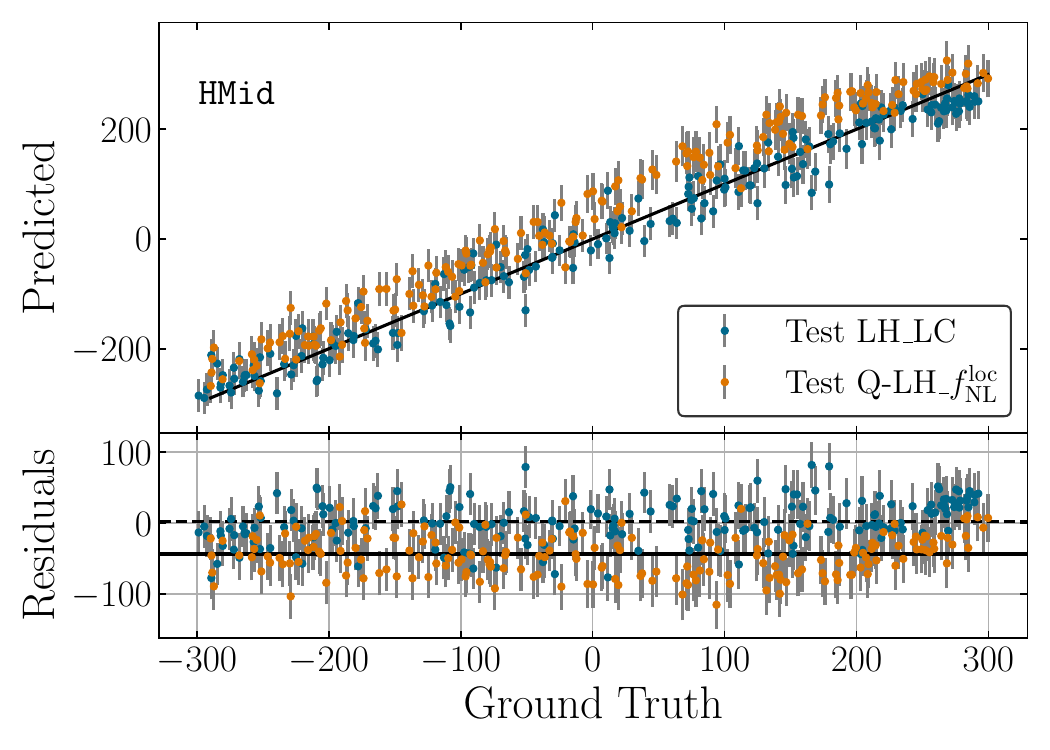}
    \end{minipage}
    \hfill
    \begin{minipage}[t]{0.45\textwidth}
      \centering
      \includegraphics[height=5cm]{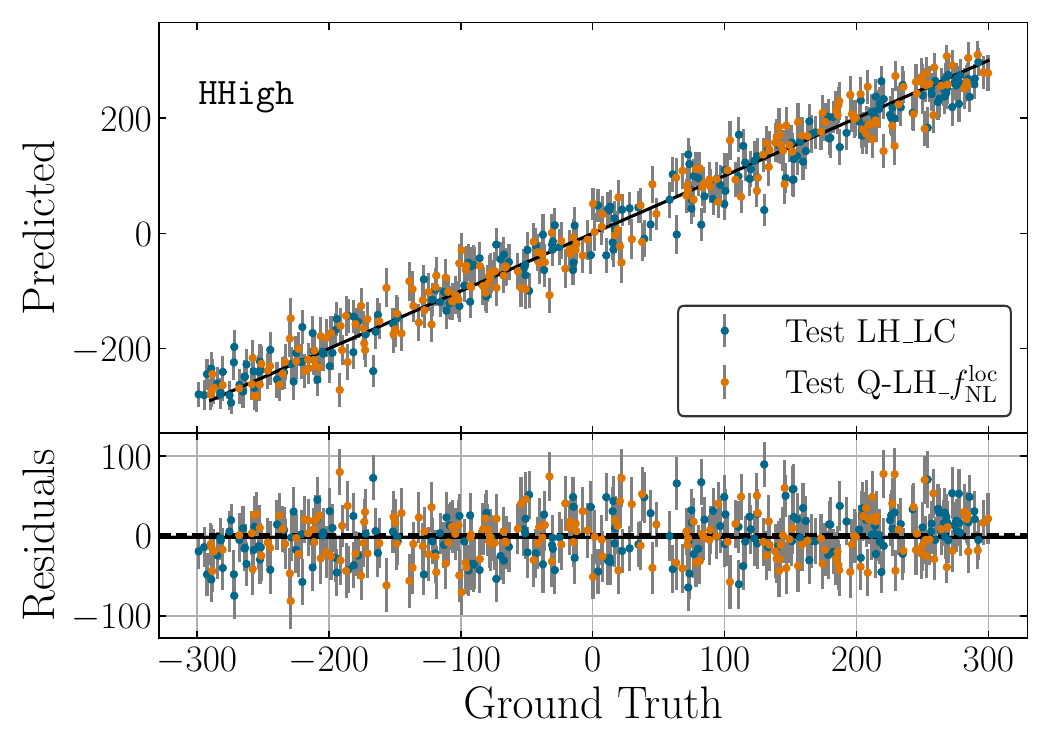}
    \end{minipage}
    \caption{Normalize \pmwdsuite and \quijote independently}
  \end{subfigure}
  
  \vspace{1em}

  \begin{subfigure}[t]{\textwidth}
    \centering
    \begin{minipage}[t]{0.45\textwidth}
      \centering
      \includegraphics[height=5cm]{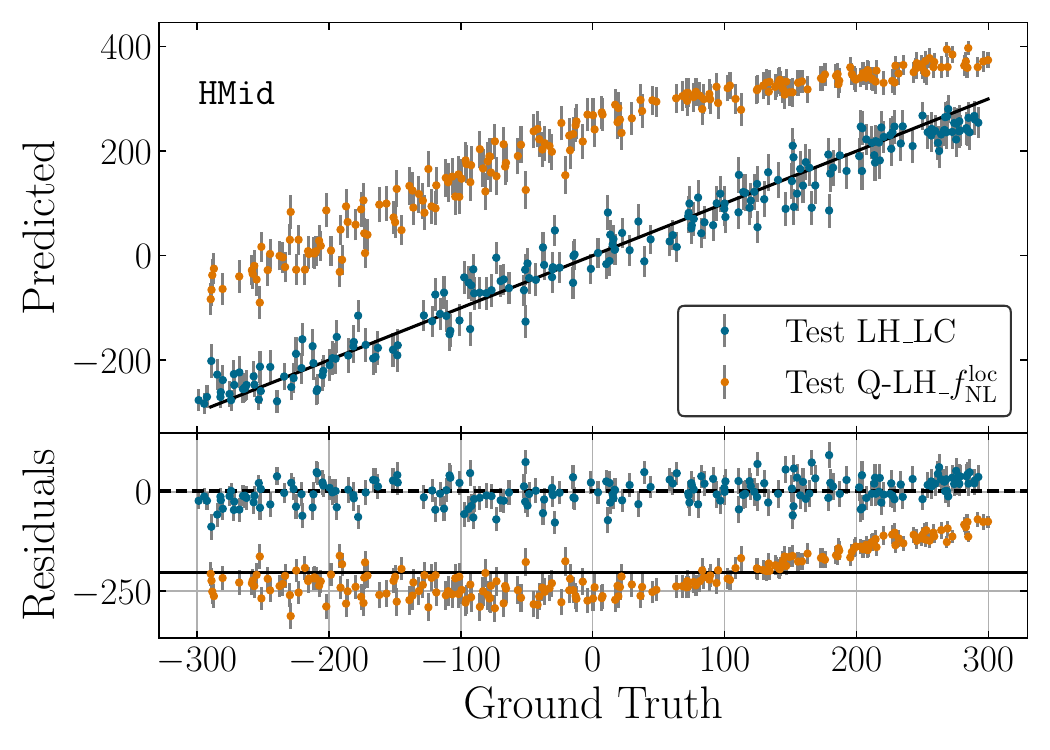}
    \end{minipage}
    \hfill
    \begin{minipage}[t]{0.45\textwidth}
      \centering
      \includegraphics[height=5cm]{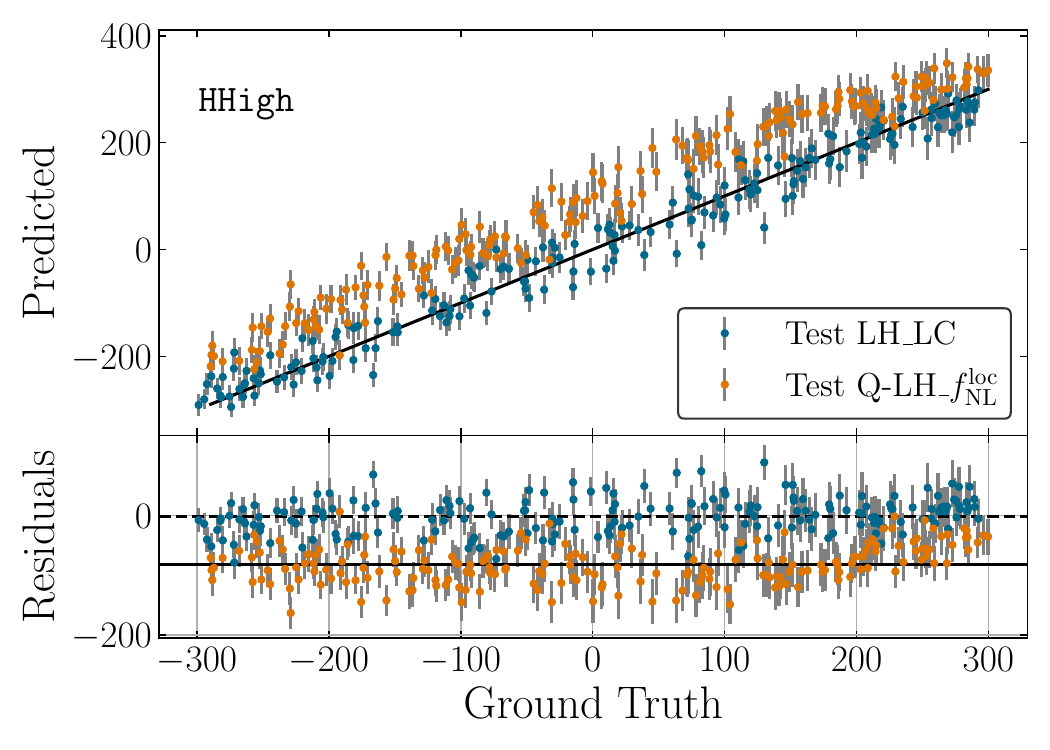}
    \end{minipage}
    \caption{\pmwdsuite fiducial bounds on both \pmwdsuite and \quijote}
  \end{subfigure}

  \vspace{1em}

  \begin{subfigure}[t]{\textwidth}
    \centering
    \begin{minipage}[t]{0.45\textwidth}
      \centering
      \includegraphics[height=5cm]{plots/MLP_betti_LC_mbin3_scaleperrun.pdf}
    \end{minipage}
    \hfill
    \begin{minipage}[t]{0.45\textwidth}
      \centering
      \includegraphics[height=5cm]{plots/MLP_betti_LC_mbin6_scaleperrun.pdf}
    \end{minipage}
    \caption{Standardize each PD run individually}
  \end{subfigure}

  \caption{Performance of a MLP trained on the \texttt{LH\_LC300} dataset under the three standardization schemes shown in Figure~\ref{fig:betti_bounds}, evaluated on both \pmwdsuite and \quijotepng test sets for \texttt{HMid} and \texttt{HHigh} mass bins. (a) Dataset-specific bounds: summarizing each suite with its own internally consistent min-max filtration ranges enables transferability. (b) Applying \pmwdsuite bounds to \quijote: shift  the distribution of features into different bins due to differences in the mean inter-halo distances, producing biased vectors and predictions. (c) Per-run scaling: transfer is possible but constraints degrade due to the loss of absolute scale information.}
    \label{fig:betti_inference_bounds}
\end{figure}

Figure~\ref{fig:betti_inference_bounds} shows how these choices propagate to inference.

\begin{itemize}

    \item Normalize \pmwdsuite and \quijotepng independently:
    This strategy enables transferability because each dataset is summarized using bounds tailored to its own distribution of topological features. As a result, the bin populations remain internally consistent, and variations in the persistence diagrams, such as those induced by changes in $\fnlloc$, appear as genuine physical shifts rather than artifacts of the discretization.

    \item \pmwdsuite bounds on both \pmwdsuite and \quijotepng:
    By contrast, fixing the same bounds for both test-sets introduces mismatches in the bin-population, e.g. features are displaced into bins corresponding to earlier physical scales, producing biased Betti curves and consequently biased predictions. The overall trend in predictions for $\fnlloc$ is still visible because genuine $\fnlloc$ physical shifts are not erased, but become systematically shifted.
    
    \item Standardize each PD run individually:
    Finally, per-run scaling avoids inter-dataset misalignment and enable transferability, but at the cost of discarding absolute scale information, which substantially degrades the strength of the constraints.

\end{itemize}

If one simulation (or, in a general sense, a set of points) is sparser than another following the same underlying distribution, the typical separation between halos will increase, delaying the inclusion of simplices into the filtration. Larger nearest-neighbour separations delay the first nonzero DTM values for vertices, push edge connections to larger radii, and overall shift birth scales in the persistence diagrams toward higher filtration values. Figures~\ref{fig:betti_loc_unbound},~\ref{fig:betti_eq_unbound} shows this effect across the \{\texttt{HLow}, \texttt{HMid}, \texttt{HHigh}\} mass bins. Although sparsity alters absolute scales, the relative $\fnl$ signal is preserved, with $\fnl>0$ causing topological features to emerge earlier in the filtration compared to the fiducial cosmology.

\begin{figure}[ht]
    \centering
    \textbf{PD-statistic with \xgb}\\[0.5em]

  \begin{subfigure}[t]{\textwidth}
    \centering
    \begin{minipage}[t]{0.45\textwidth}
      \centering
      \includegraphics[height=5cm]{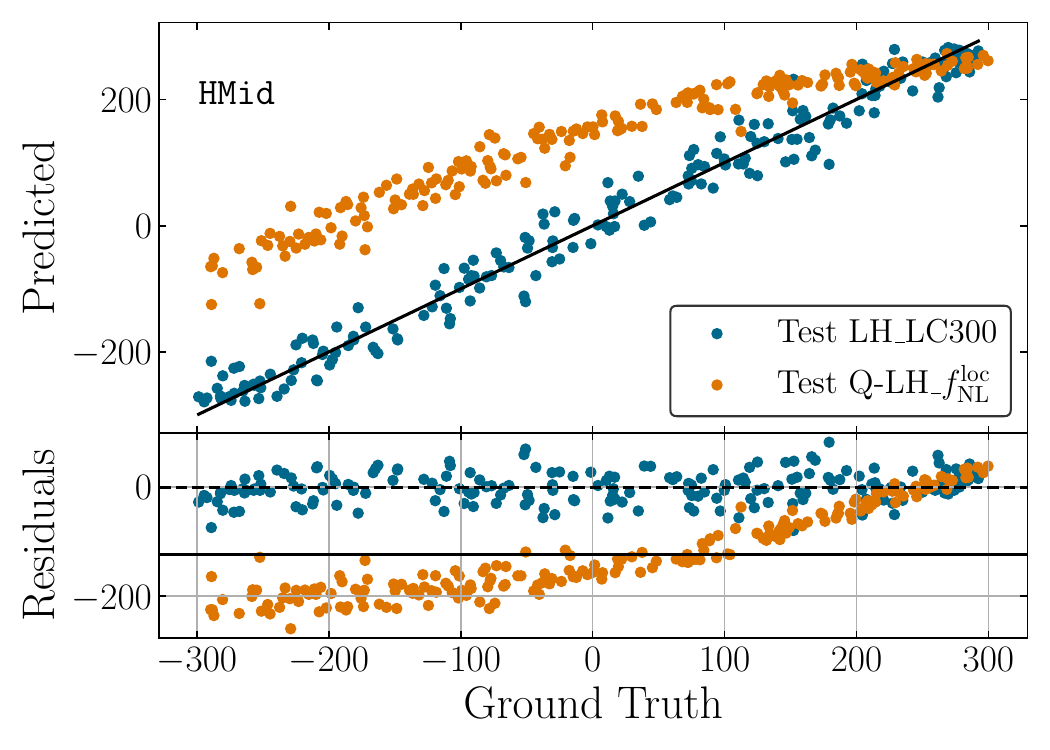}
    \end{minipage}
    \hfill
    \begin{minipage}[t]{0.45\textwidth}
      \centering
      \includegraphics[height=5cm]{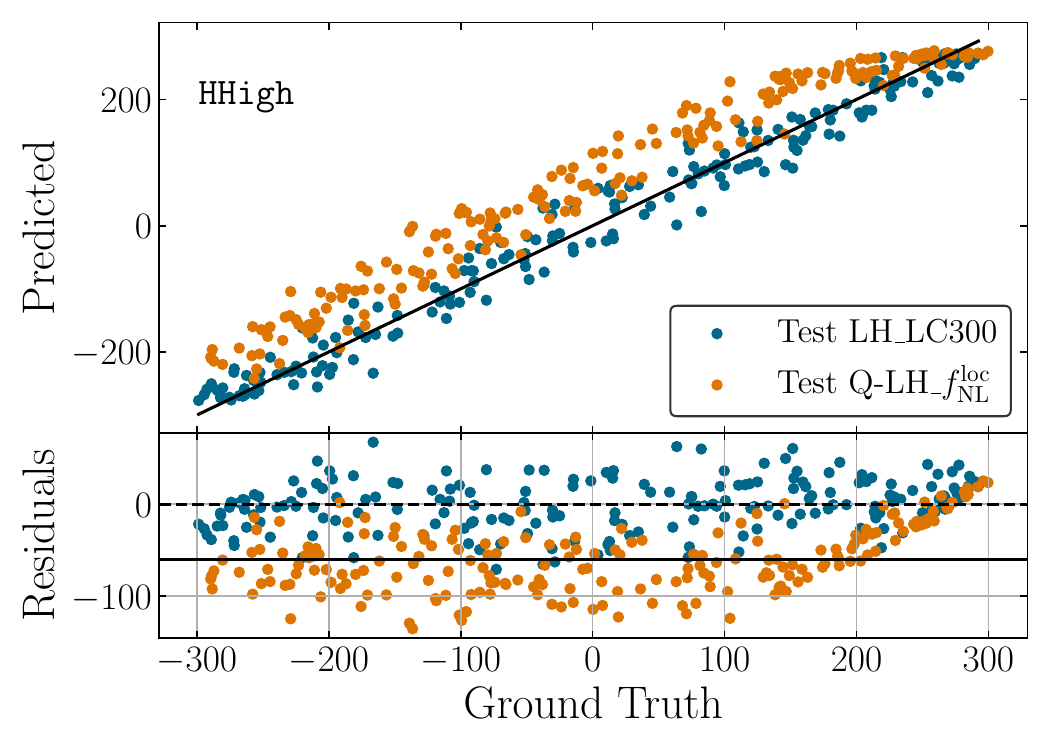}
    \end{minipage}
    \caption{Normalize \pmwdsuite and \quijote independently}
  \end{subfigure}
  
  \vspace{1em}

  \begin{subfigure}[t]{\textwidth}
    \centering
    \begin{minipage}[t]{0.45\textwidth}
      \centering
      \includegraphics[height=5cm]{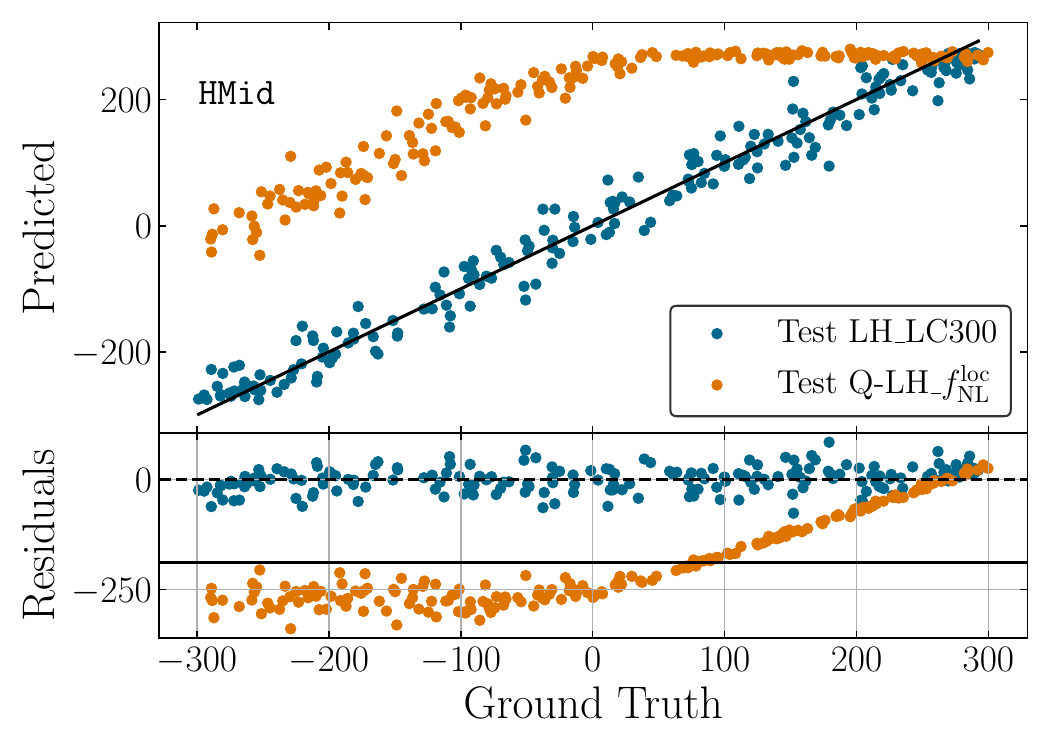}
    \end{minipage}
    \hfill
    \begin{minipage}[t]{0.45\textwidth}
      \centering
      \includegraphics[height=5cm]{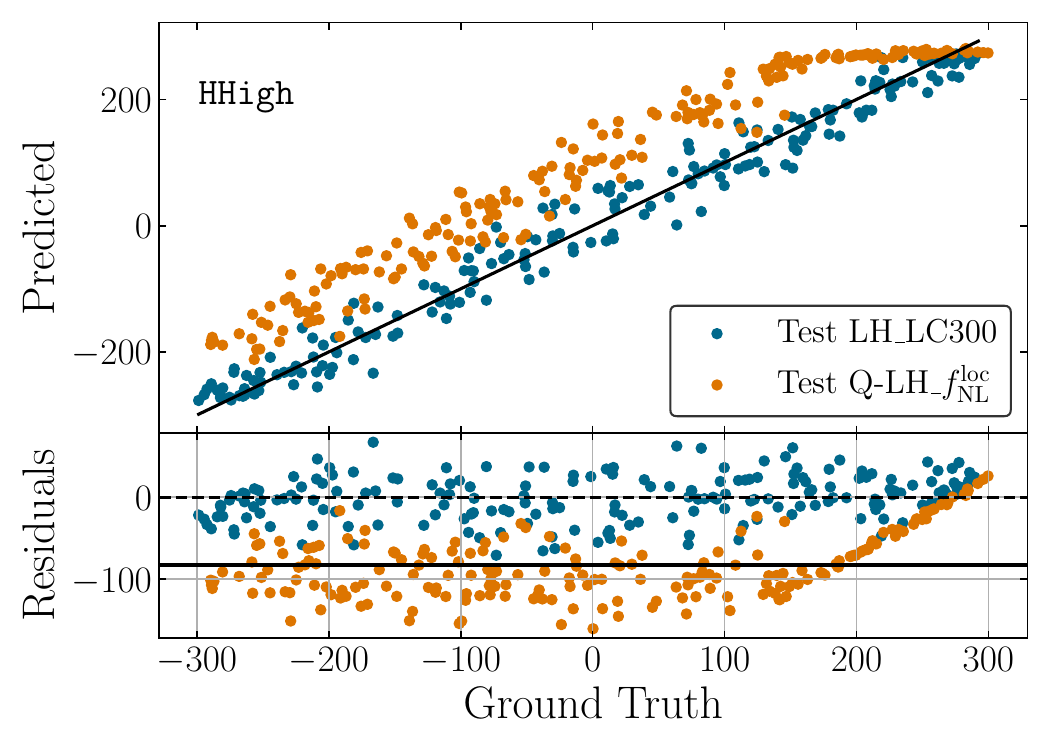}
    \end{minipage}
    \caption{\pmwdsuite fiducial bounds on both \pmwdsuite and \quijote}
  \end{subfigure}

  \vspace{1em}

  \begin{subfigure}[t]{\textwidth}
    \centering
    \begin{minipage}[t]{0.45\textwidth}
      \centering
      \includegraphics[height=5cm]{plots/xgb_pd_stat_scaleperrun_LC_mbin3.pdf}
    \end{minipage}
    \hfill
    \begin{minipage}[t]{0.45\textwidth}
      \centering
      \includegraphics[height=5cm]{plots/xgb_pd_stat_scaleperrun_LC_mbin6.pdf}
    \end{minipage}
    \caption{Standardize each PD run individually}
  \end{subfigure}
  
    \caption{Performance of models for the PD-statistic trained on the \texttt{LH\_LC300} dataset for the three standardization schemes~\ref{app:transfer} evaluated on \pmwdsuite and \quijotepng test sets for the \texttt{HMid} and \texttt{HHigh} mass bins. In this case, rescaling the persistence diagrams has a smaller impact on transferability than differences in the mean halo density between the two suites.}
    \label{fig:pdstat_bounds}
\end{figure}

\section{Impact of $\fnl$ on the summaries}

\begin{figure}[ht]
    \centering
    \includegraphics[width=0.95\linewidth]{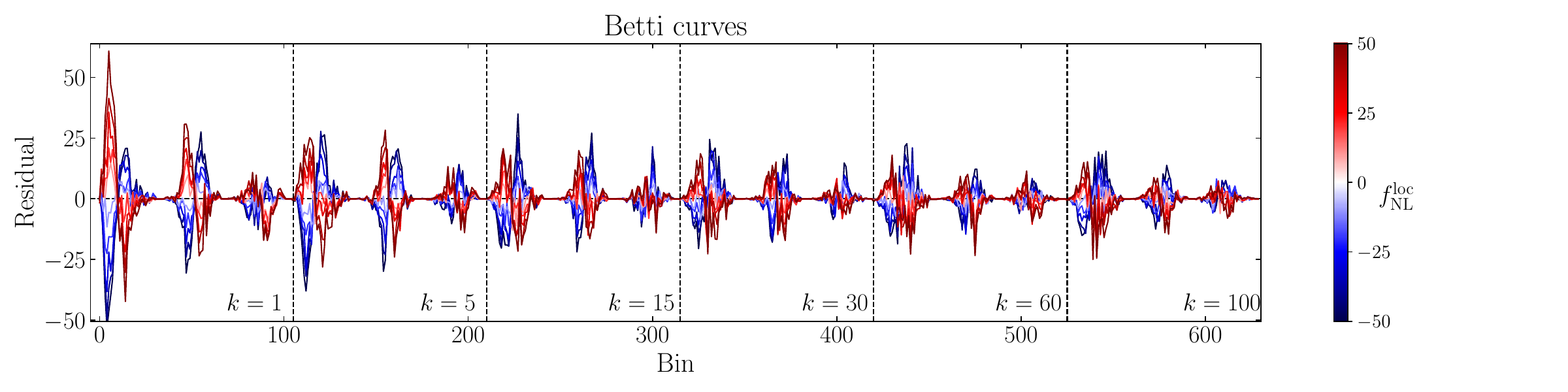}
    \includegraphics[width=0.95\linewidth]{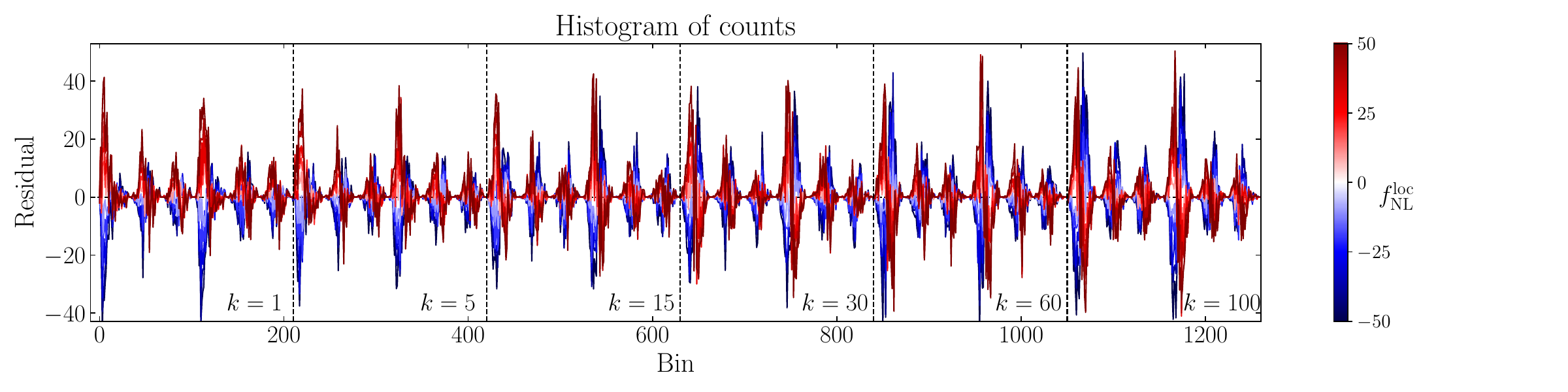}
    \includegraphics[width=0.95\linewidth]{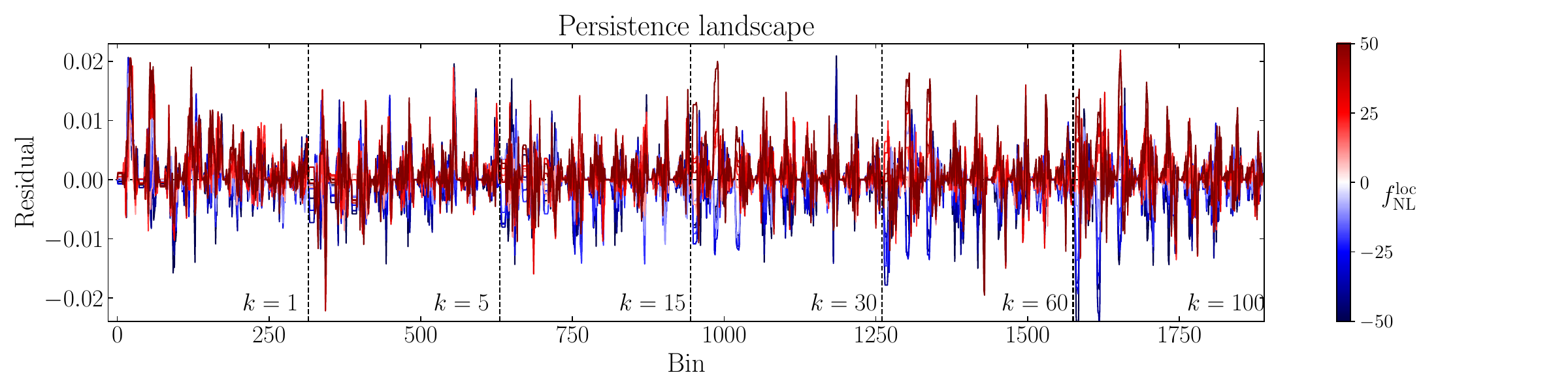}
    \includegraphics[width=0.95\linewidth]{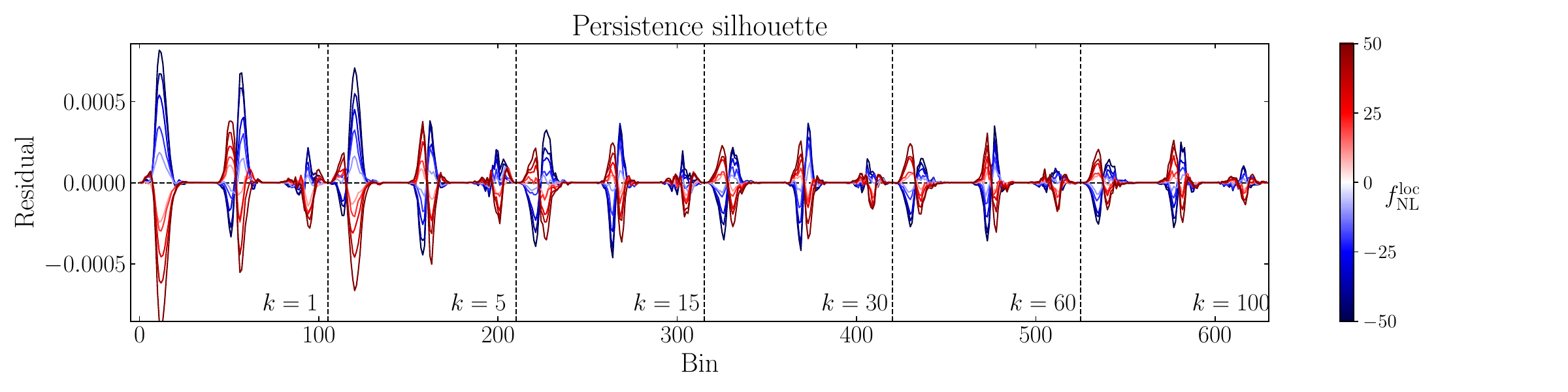}
    \includegraphics[width=0.95\linewidth]{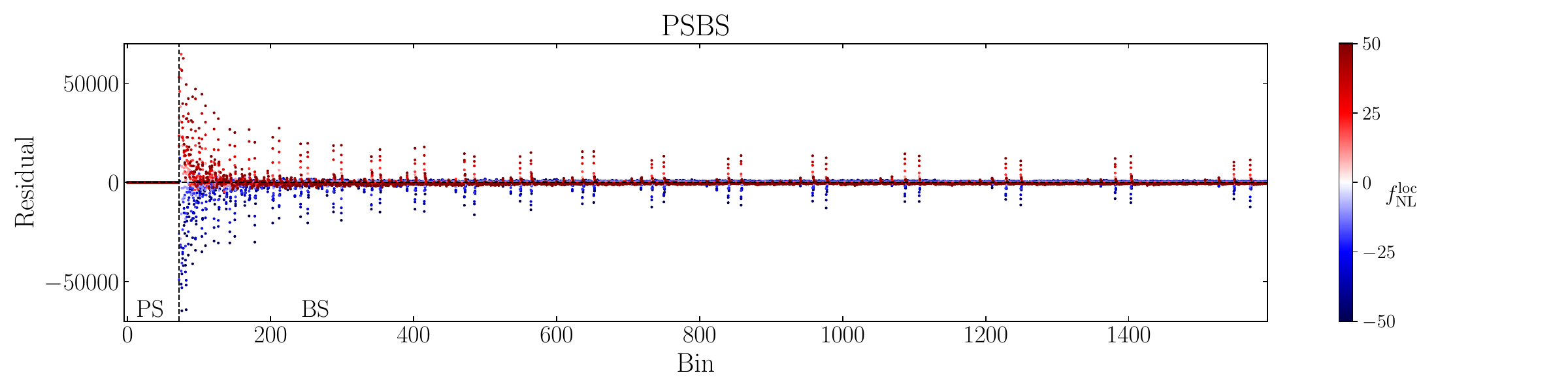}
    \caption{Comparison of data vectors for different $\Delta f_{\rm NL}^{loc}$ in the \texttt{LH\_1P} dataset in the \texttt{HMid} mass bin. Each panel shows the mean residual obtained by subtracting the fiducial from the corresponding $f_{\rm NL}$-step and averaging over 10 realizations with matched random seeds to suppress cosmic variance.}
\label{fig:vectors_all}
\end{figure}

In Figure \ref{fig:vectors_all} we show how the different topological summaries respond to changes in $\fnlloc$. We vary $\fnlloc$ across ten evenly spaced steps from $-50$ to $+50$ using the LH\_\texttt{1P} simulation setup in a fixed mass bin, \texttt{HMid}. To isolate the effect of local primordial non-Gaussianity, each $\fnlloc$ value is paired with a fiducial realization generated with the same random seed to suppress most of the cosmic variance, leaving a clean residual signal that reflects only the impact of $\fnlloc$ on each summary. For comparison, we also include the powerspectrum and bispectrum, whose strongest response consistently involves large wave-length modes, aligning with the expected signature of local-type primordial non-Gaussianity.

\subsection{Dependence across mass bins}
\label{app:massdependence}
\begin{figure}[ht]
    \centering
    \begin{subfigure}{\textwidth}
        \centering
        \includegraphics[width=\textwidth]{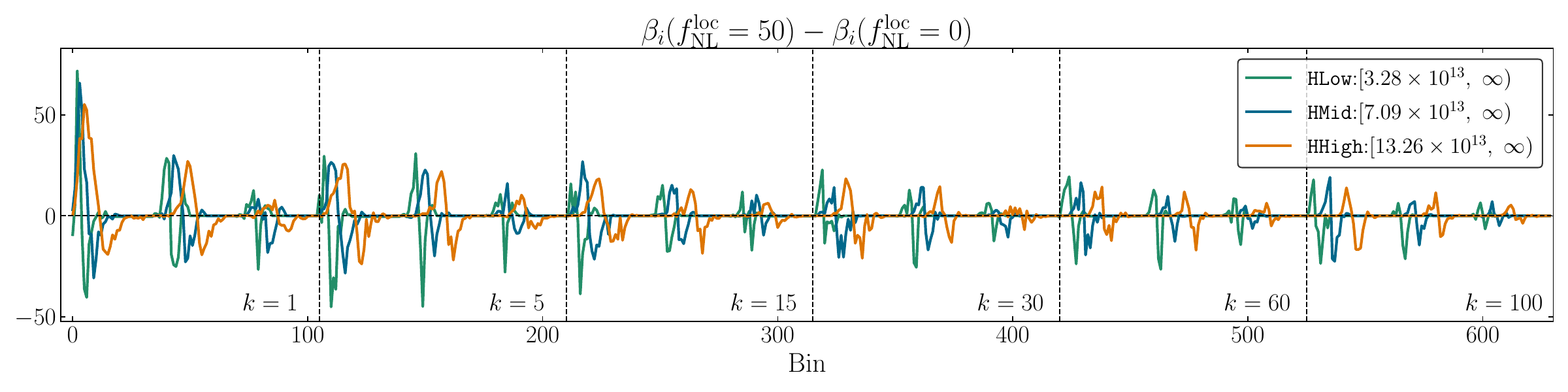}
        \caption{Residual Betti curves for $\fnlloc$ and different mass cuts.}
        \label{fig:betti_loc_unbound}
    \end{subfigure}

    \begin{subfigure}{\textwidth}
        \centering
        \includegraphics[width=\textwidth]{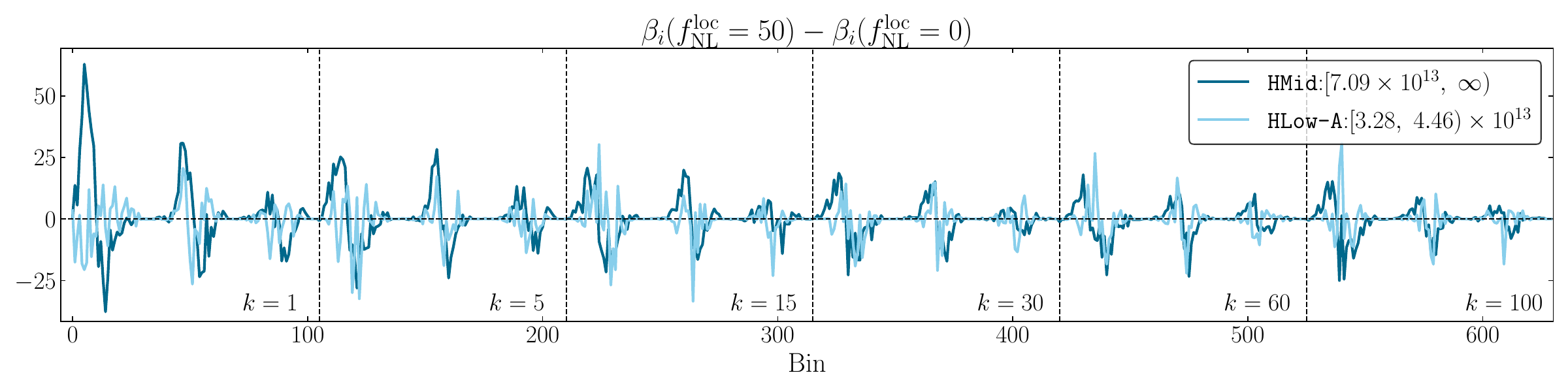}
        \caption{Residual Betti curves for $\fnlloc$, equal-density bins, but different mass-selected tracers.}
        \label{fig:betti_loc_equal}
    \end{subfigure}

    \begin{subfigure}{\textwidth}
        \centering
        \includegraphics[width=\textwidth]{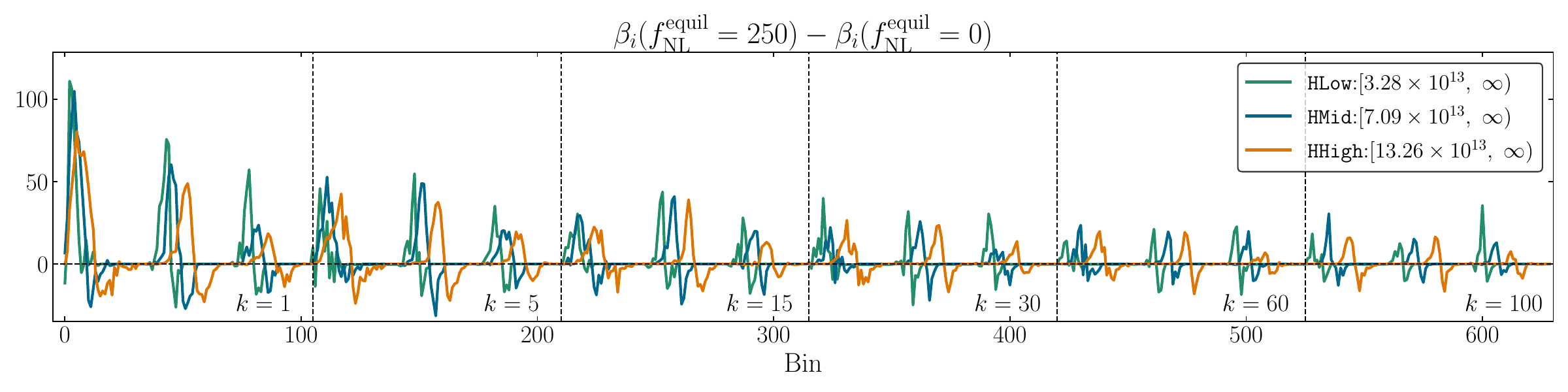}
        \caption{Residual Betti curves for $\fnleq$ and different mass cuts.}
        \label{fig:betti_eq_unbound}
    \end{subfigure}

    \begin{subfigure}{\textwidth}
        \centering
        \includegraphics[width=\textwidth]{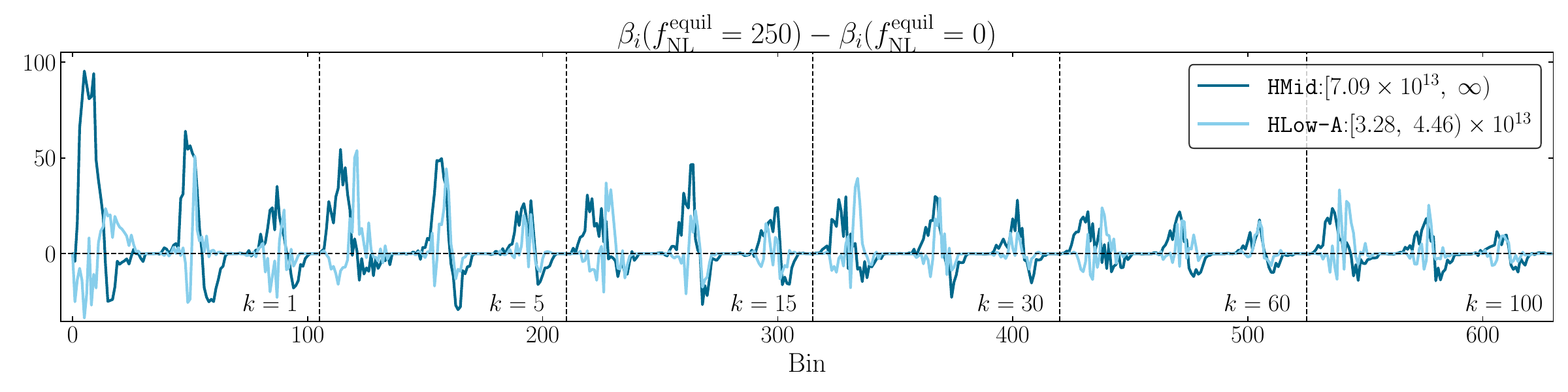}
        \caption{Residual Betti curves for $\fnleq$, equal-density bins, but different mass-selected tracers.}
        \label{fig:betti_eq_equal}
    \end{subfigure}

    \caption{Each panel shows residual Betti curves, $\beta_i$, computed relative to the $\fnl = 0$ fiducial baseline with reduced cosmic variance. The top half shows results for $\fnlloc$, first for the unbound mass bins and then for different mass tracers. The bottom half shows the corresponding residuals for $\fnleq$, following the same panel layout.}
    \label{fig:betti_curves_mbin_fnl}
\end{figure}

In Figure~\ref{fig:betti_curves_mbin_fnl} we examine how the topological features respond to $\fnl$ across halo mass bins. For clarity and interpretability, we focus on Betti-curves, and plot the residuals between $\fnl=0$ and $\Delta\fnl$, using matched seeds to suppress cosmic variance and averaging over $10$ realizations from the LH\_\texttt{1P} dataset. Each $k$-division displays $0$-, $1$-, and $2$-cycles (first, second, and third peaks, respectively), computed at their $k$-nearest neighbor in the DTM  function. In the top panel, positive $\fnl$ shifts features toward earlier scales, along with an overall displacement caused by the different halo density (see Figure~\ref{fig:filtration_massbins}). The next panel, which compares bins with similar $\bar{n}_h$, highlights the mass-dependent sign flip: low-mass halos shift in response to the opposite direction of high-mass halos, with the contrast decreasing as $k$ increases. This flip seems more pronounced for the equilateral shape (Figure~\ref{fig:betti_eq_equal}), whereas for the local shape it is present but less apparent at this particular redshift (Figure~\ref{fig:betti_loc_equal}).

This flip behavior mirrors trends reported in other summary statistics. The distribution of $k$-nearest neighbor distances shows the same mass-dependent inversion~\cite{Coulton:2023ouk}, and because the birth scale of $0$-cycles is set by the DTM, constructed directly from these distances, it suggests a closely related origin~\cite{Ouellette:2025nll,Gangopadhyay:2025kct}. This trend is also consistent with the shift in the halo mass function under PNG~\cite{Jung:2023kjh}. Persistence images similarly move toward smaller scales for positive $\fnlloc$~\cite{Biagetti:2020skr,Biagetti:2022qjl}. Finally, the response of $2$-cycles aligns with the expected reduction in void sizes under local PNG~\cite{Kamionkowski:2008sr}, as also observed in~\cite{Biagetti:2020skr}.

\section{Model architectures}
\label{appx:architectures}
\subsection*{Multilayer Perceptrons (MLP)}  

For one-dimensional vector representations (Betti curves, histogram of counts, persitence landscape, and persistence silhouettes), we use fully connected multilayer perceptrons. The architecture follows a symmetric encoder-decoder design inspired by autoencoders, which consistently outperformed alternatives such as increasing-decreasing or constant-width configurations for most topological vectorizations.

The network consists of sequential dense layers split into two stages, an encoder that progressively compresses the input, and a decoder that symmetrically reconstructs it toward the output. Neuron counts follow a halving pattern in the encoder and a doubling pattern in the decoder, with a minimum bottleneck width of 4 neurons to preserve expressiveness. Each hidden layer is followed by a non-linear activation function, selected from a predefined set during hyperparameter tuning, along with the total number of layers. A dropout layer with fixed rate $0.4$ is applied before the final output to mitigate overfitting. The output layer is a dense layer with linear activation that predicts the cosmological parameters.

For the PSBS vector, we adopt a constant-width architecture for the hidden layers, each followed by an activation function and dropout. Batch normalization is applied to the input features, and the output is produced by a final dense layer with linear activation.

\subsection*{Convolutional Neural Networks (CNN)}

For persistence image inputs, we treat each combination of filtration values and homology
dimension as a separate input channel, resulting in images with fixed width and height and
$18$ channels. The architecture is built from two main convolutional blocks, an initial max-pooling block for early feature aggregation, followed by a deeper convolutional block for refined feature extraction.

The first block includes multiple convolutional layers, each followed by a $2$D max-pooling operation. All convolutions use $3 \times 3$ kernels with stride $1$ and ``same'' padding, activated by a non-linear function. The second block consists of additional convolutional layers with identical kernel and activation settings, but without pooling. The number of layers, channels, and activation function are treated as hyperparameters and optimized per dataset via dedicated search.

Following the convolutional stages, the output is flattened and passed through a dropout layer with fixed rate $0.4$. A final dense layer with linear activation produces the predicted cosmological parameters.

\subsection*{DeepSets}
\label{sec:deepsets}
\deepsets are neural architectures designed for set-structured inputs, ensuring permutation invariance, that is, the output remains unchanged under any reordering of input elements \cite{zaheer2017deep}. For persistence diagrams $\mathcal{D}$, viewed as multisets of points $x \in \mathbb{R}^2$, this property is achieved via the canonical decomposition:
\begin{equation}
\label{eq:deepsets}
    f(\mathcal{D}) = \rho\left(\sum_{x \in \mathcal{D}} \phi(x)\right),
\end{equation}
where $\phi$ maps each topological feature to a latent vector in a shared embedding space, and the summation acts as a symmetric aggregation operator. The pooled embedding is then transformed by $\rho$ to produce the final output. This models has found its place in some cosmological applications~\cite{Wang:2022zpv,Wang:2025ofq}.

Each persistence diagram is processed by a dedicated $\phi$ network, a constant-width stack of dense layers, shared across all features within the diagram but instantiated separately for each of the $18$ input channels to enable dimension-specific representations. The depth, width, and activation function of each $\phi$ network are selected via hyperparameter search.

After feature transformation, three pooling operations, sum, mean, and max, are applied along the feature axis of each diagram. The resulting vectors are concatenated across all $p$-cycle channels and passed to the $\rho$ network, implemented as a constant-width MLP. Its architecture is likewise tuned via hyperparameter optimization. A dropout layer with fixed rate $0.4$ precedes the final dense layer with linear activation, which outputs the predicted cosmological parameters.

\subsection*{PersLay}
\label{sec:perslay}

\perslay~\cite{2019arXiv190409378C} extends the \deepsets framework with a specialized neural layer for persistence diagrams, combining permutation invariance with learnable vectorization. It constructs embeddings via a weighted transformation:
\begin{equation}
	f(\mathcal{D}) = \rho\left( \sum_{x \in \mathcal{D}} w_{\theta_1}(x) \cdot \phi_{\theta_2}(x) \right),
\end{equation}
where both $w_{\theta_1}$ and $\phi_{\theta_2}$ are parameterized functions with analytically defined gradients, enabling end-to-end training via backpropagation. The embedding function $\phi$ can be instantiated using common choices such as Gaussian kernels, continuous basis expansions, or piecewise linear functions. This formulation generalizes many classical vectorizations, such as persistence images, landscapes, and Betti curves, by learning task-specific representations rather than relying on fixed encodings.

Unlike raw \deepsets applied directly to persistence diagrams, which can produce discontinuous representations, \perslay ensures continuity and stability through its structured weighting and embedding design.

\section{Hyperparameter space}
\label{appx:hyperparameter}

All neural-based models were implemented in TensorFlow~\cite{tensorflow2015-whitepaper}, with hyperparameter optimization performed via KerasTuner~\cite{omalley2019kerastuner} on an NVIDIA QUADRO RTX 6000 GPU. Due to the computational cost of exhaustive searches across all datasets, mass bins, vectorization methods, and architectures, we adopted a two-stage reduction strategy.

In the first stage, we explored a broad hyperparameter space using the \texttt{Q-LH}\_$\fnlloc$ (from \quijotepng) and \texttt{LH\_LC300} from the \pmwdsuite suites, focusing on the local-type non-Gaussianity dataset and the full mass bin (\texttt{HLow}). This search space mirrored that of~\cite{Calles:2024cxl}, with separate searches conducted for each vectorization method and its associated architecture.

Based on convergence trends in Stage~1, in the second stage we fixed the learning rate to $1\times10^{-4}$, the value around which most models had converged, and restricted the search to variations in layer depth, width, and activation functions. These were drawn from the most frequently optimal configurations identified earlier. This refined search was applied to both \texttt{HLow} and \texttt{HMid}, supported by the observation that top-performing configurations were consistent across mass bins within each dataset and input type.

For the final results shown in Table~\ref{tab:fnl_loc_metrics}--\ref{tab:fnl_equil_metrics}, we selected the most frequently optimal configuration for each architecture and vectorization method across mass bins and datasets, and applied it consistently throughout. In what follows, we summarize the restricted hyperparameter search conducted in Stage~2.

\paragraph{Multilayer Perceptrons (MLP)}  
A grid search strategy was used to evaluate both flat-MLPs for the PSBS vector and the ``encoder-decoder'' type for the topological vectors. For the flat MLPs the number of fully connected layers ranged from two to four; activation functions were chosen from ReLU, leaky ReLU, ELU, or inverse hyperbolic sine (asinh); and layer widths were selected from 64, 256, or 512 units. While for the encoder-decoder, the width is fixed to 512 and the number of layers is chosen between 4 and 7 with the same non-linear activation-space. The objective function minimized the validation loss across the defined configuration space.

\paragraph{Convolutional Neural Networks (CNN)}  
This search performs Bayesian optimization. It explores from two to five max-pool layers, with the secondary block depth varying from one to four layers. The channel count is selected from 16, 32, 64, or 128. Activation functions are chosen from ReLU, leaky ReLU, ELU, and asinh. The search evaluates up to 50 configurations, aiming to minimize validation loss.

\paragraph{DeepSets}  
For the permutation-invariant block $\phi$, the number of layers varied discretely from one to three, with internal widths of either 32 or 64 units. The output block $\rho$ was tuned across depths of three to five layers, with layer widths chosen from 8, 16, 64, or 256 units. Nonlinearities were selected from ReLU, leaky ReLU, and asinh. These parameters were embedded in a Bayesian optimization framework initialized with 20 random trials and guided exploration for 10 more configurations, targeting the validation loss.

\paragraph{Boosted Trees}  

We tuned the XGBoost regressor via sequential grid searches over learning rate, boosting rounds, maximum depth, subsample ratio, and column sub-sampling ratio. Starting with a base learning rate of 0.1 and 1000 boosting rounds, the number of estimators was fine-tuned using a custom early-stopping criterion with 50 rounds. The maximum tree depth was tested from 1 to 6, after which subsample and column subsample ratios were explored in the range [0.6, 0.9]. The learning rate was then re-optimized over seven values $(0.5,0.2,0.15,0.1,0.05,0.01,0.001)$. After each update, the optimal number of boosting rounds was re-evaluated.

\section{Complete tables}
\label{appx:full_results}

In this appendix, we present the inference results for both $\fnlloc$ and $\fnleq$ across the bounded halo mass bins, \{\texttt{HLow-A}, \texttt{HLow-B}, \texttt{HMid-A}, \texttt{HMid-B}\}, completing the set of results for all mass bins. For each model and mass selection, we report the RMSE and $R^2$ scores. Overall, \texttt{HLow-A} yields the weakest constraints for both $\fnlloc$ and $\fnleq$, this bin correspond to halos with $50-67$ particles in the fiducial cosmology. The \texttt{HLow-B} and \texttt{HMid-A} bins, containing halos with $68-107$ and $108-137$ particles, respectively, exhibit comparable performance. Finally, \texttt{HMid-B}, based on halos with $138-201$ particles, provides the best constraints within these subsets, representing the best-performing tracer before including the most massive halos.

\begin{table}[ht!]
    \small
    \centering
    \begin{tabular}{|c|l|c c|c c|c c|c c|}
    \hline
    Model & Mass bin 
    & \multicolumn{2}{c|}{\texttt{Q-LH}\_$\fnlloc$} 
    & \multicolumn{2}{c|}{\texttt{LH\_LC300}} 
    & \multicolumn{2}{c|}{\texttt{LH\_LC50}} 
    & \multicolumn{2}{c|}{\texttt{LH\_LC\_om\_s8}} \\
    & & RMSE & {R2} 
      & RMSE & {R2} 
      & RMSE & {R2} 
      & RMSE & {R2} \\[0.5ex] 
    \hline\hline
    \multirow{3}{*}{\makecell[t]{PSBS \\
    (\textbf{MLP})}}
    &\texttt{HLow-A}       & $85.0$ &$\cca{75.6}$  & $106.1$ &$\cca{66.8}$ & $30.2$ &$\cca{3.2}$ & $29.4$ &$\cca{1.9}$ \\ 
    &\texttt{HLow-B}       & $76.5$ &$\cca{80.2}$  & $87.7$ &$\cca{77.3}$ & $29.7$ &$\cca{6.5}$ & $29.1$ &$\cca{3.9}$ \\ 
    &\texttt{HMid-A}       & $79.4$ &$\cca{78.7}$  & $92.8$ &$\cca{74.6}$ & $30.4$ &$\cca{2.0}$ & $29.2$ &$\cca{3.0}$ \\ 
    &\texttt{HMid-B}       & $56.8$ &$\cca{89.1}$  & $65.2$ &$\cca{87.5}$ & $29.1$ &$\cca{9.9}$ & $29.2$ &$\cca{3.2}$ \\ 
    \hline
    \multirow{3}{*}{\makecell[t]{PD \\
    (\textbf{\deepsets})}}
    &\texttt{HLow-A}       & $122.9$ &$\cca{49.0}$  & $161.9$ &$\cca{22.4}$ & $30.8$ &$-0.6$ & $28.2$ &$\cca{0.2}$ \\ 
    &\texttt{HLow-B}       & $106.9$ &$\cca{61.4}$  & $122.4$ &$\cca{55.8}$ & $30.7$ &$\cca{0.1}$ & $28.2$ &$\cca{0.3}$ \\ 
    &\texttt{HMid-A}       & $102.8$ &$\cca{64.3}$  & $111.9$ &$\cca{63.0}$ & $30.8$ &$-0.6$ & $28.2$ &$\cca{0.3}$ \\ 
    &\texttt{HMid-B}       & $79.1$ &$\cca{78.9}$  & $85.9$ &$\cca{78.2}$ & $30.6$ &$\cca{0.5}$ & $28.1$ &$\cca{0.6}$ \\ 
    \hline
    \multirow{3}{*}{\makecell[t]{PI \\
    (\textbf{\cnn})}}
    &\texttt{HLow-A}       & $94.5$ &$\cca{69.9}$  & $106.8$ &$\cca{66.3}$ & $30.6$ &$\cca{0.8}$ & $28.2$ &$\cca{0.4}$ \\ 
    &\texttt{HLow-B}       & $83.3$ &$\cca{76.6}$  & $96.8$ &$\cca{72.3}$ & $30.4$ &$\cca{1.8}$ & $28.0$ &$\cca{1.6}$ \\ 
    &\texttt{HMid-A}       & $88.4$ &$\cca{73.6}$  & $95.9$ &$\cca{72.9}$ & $30.6$ &$\cca{0.8}$ & $28.0$ &$\cca{1.5}$ \\ 
    &\texttt{HMid-B}       & $66.2$ &$\cca{85.2}$  & $76.7$ &$\cca{82.6}$ & $30.1$ &$\cca{3.9}$ & $27.8$ &$\cca{2.7}$ \\ 
    \hline
    \multirow{3}{*}{\makecell[t]{PD-Statistics \\
    (\textbf{XGB})}}
    &\texttt{HLow-A}       & $86.4$ &$\cca{74.8}$  & $100.7$ &$\cca{70.0}$ & $29.9$ &$\cca{5.3}$ & $29.2$ &$\cca{3.0}$ \\ 
    &\texttt{HLow-B}       & $75.0$ &$\cca{81.0}$  & $89.4$ &$\cca{76.4}$ & $29.7$ &$\cca{6.0}$ & $29.1$ &$\cca{3.5}$ \\ 
    &\texttt{HMid-A}       & $79.1$ &$\cca{78.9}$  & $84.7$ &$\cca{78.8}$ & $30.1$ &$\cca{3.5}$ & $29.5$ &$\cca{1.4}$ \\ 
    &\texttt{HMid-B}       & $56.9$ &$\cca{89.1}$  & $66.4$ &$\cca{87.0}$ & $29.3$ &$\cca{8.7}$ & $29.2$ &$\cca{3.0}$ \\ 
    \hline
    \multirow{3}{*}{\makecell[t]{PD-Statistics/PSBS \\
    (\textbf{XGB})}}
    &\texttt{HLow-A}       & $80.6$ &$\cca{78.1}$  & $99.4$ &$\cca{70.9}$ & $30.1$ &$\cca{4.0}$ & $29.2$ &$\cca{3.1}$ \\ 
    &\texttt{HLow-B}       & $70.8$ &$\cca{83.1}$  & $84.5$ &$\cca{78.9}$ & $29.6$ &$\cca{7.0}$ & $29.1$ &$\cca{3.9}$ \\ 
    &\texttt{HMid-A}       & $78.5$ &$\cca{79.2}$  & $83.3$ &$\cca{79.5}$ & $29.7$ &$\cca{6.5}$ & $29.3$ &$\cca{2.5}$ \\ 
    &\texttt{HMid-B}       & $53.1$ &$\cca{90.5}$  & $65.2$ &$\cca{87.4}$ & $29.1$ &$\cca{10.2}$ & $29.1$ &$\cca{3.9}$ \\ 
    \hline
    \multirow{3}{*}{\makecell[t]{PD-Histogram \\
    (\textbf{MLP})}}
    &\texttt{HLow-A}       & $83.3$ &$\cca{76.6}$  & $98.1$ &$\cca{71.6}$ & $29.8$ &$\cca{5.4}$ & $29.2$ &$\cca{3.1}$ \\ 
    &\texttt{HLow-B}       & $75.4$ &$\cca{80.8}$  & $84.2$ &$\cca{79.1}$ & $29.8$ &$\cca{5.5}$ & $29.0$ &$\cca{4.3}$ \\ 
    &\texttt{HMid-A}       & $76.7$ &$\cca{80.1}$  & $81.9$ &$\cca{80.2}$ & $30.1$ &$\cca{3.5}$ & $29.3$ &$\cca{2.8}$ \\ 
    &\texttt{HMid-B}       & $58.1$ &$\cca{88.6}$  & $66.1$ &$\cca{87.1}$ & $29.2$ &$\cca{9.3}$ & $29.3$ &$\cca{2.5}$ \\ 
    \hline
    \multirow{3}{*}{\makecell[t]{PD-Betti \\
    (\textbf{MLP})}}
    &\texttt{HLow-A}       & $84.5$ &$\cca{75.9}$  & $99.1$ &$\cca{71.0}$ & $30.8$ &$-0.7$ & $29.7$ &$-0.2$ \\ 
    &\texttt{HLow-B}       & $75.2$ &$\cca{80.9}$  & $85.9$ &$\cca{78.2}$ & $30.8$ &$-0.7$ & $29.7$ &$-0.3$ \\ 
    &\texttt{HMid-A}       & $75.1$ &$\cca{81.0}$  & $81.4$ &$\cca{80.4}$ & $30.8$ &$-0.5$ & $29.7$ &$-0.3$ \\ 
    &\texttt{HMid-B}       & $59.3$ &$\cca{88.1}$  & $67.2$ &$\cca{86.7}$ & $30.6$ &$\cca{0.5}$ & $29.7$ &$-0.2$ \\ 
    \hline
    \multirow{3}{*}{\makecell[t]{PD-PerBetti \\
    (\textbf{MLP})}}
    &\texttt{HLow-A}       & $85.1$ &$\cca{75.6}$  & $99.7$ &$\cca{70.7}$ & $29.8$ &$\cca{5.7}$ & $29.1$ &$\cca{4.0}$ \\ 
    &\texttt{HLow-B}       & $74.6$ &$\cca{81.2}$  & $85.5$ &$\cca{78.4}$ & $29.5$ &$\cca{7.7}$ & $29.0$ &$\cca{4.7}$ \\ 
    &\texttt{HMid-A}       & $75.3$ &$\cca{80.9}$  & $81.1$ &$\cca{80.6}$ & $30.0$ &$\cca{4.7}$ & $29.2$ &$\cca{2.8}$ \\ 
    &\texttt{HMid-B}       & $58.4$ &$\cca{88.5}$  & $66.5$ &$\cca{87.0}$ & $29.0$ &$\cca{10.3}$ & $29.2$ &$\cca{3.0}$ \\ 
    \hline
    \multirow{3}{*}{\makecell[t]{PD-Silhouette \\
    (\textbf{XGB})}}
    &\texttt{HLow-A}       & $91.5$ &$\cca{71.8}$  & $109.1$ &$\cca{64.8}$ & $30.4$ &$\cca{1.6}$ & $29.4$ &$\cca{1.7}$ \\ 
    &\texttt{HLow-B}       & $86.1$ &$\cca{75.0}$  & $100.9$ &$\cca{70.0}$ & $30.2$ &$\cca{3.4}$ & $29.5$ &$\cca{1.4}$ \\ 
    &\texttt{HMid-A}       & $81.2$ &$\cca{77.7}$  & $95.3$ &$\cca{73.2}$ & $30.5$ &$\cca{1.4}$ & $29.5$ &$\cca{1.1}$ \\ 
    &\texttt{HMid-B}       & $64.1$ &$\cca{86.1}$  & $73.9$ &$\cca{83.9}$ & $30.0$ &$\cca{4.1}$ & $29.6$ &$\cca{0.7}$ \\ 
    \hline
    \multirow{3}{*}{\makecell[t]{PD-Landscape \\
    (\textbf{XGB})}}
    &\texttt{HLow-A}       & $104.9$ &$\cca{62.9}$  & $114.7$ &$\cca{61.1}$ & $30.5$ &$\cca{1.4}$ & $29.7$ &$-0.2$ \\ 
    &\texttt{HLow-B}       & $97.4$ &$\cca{68.0}$  & $111.0$ &$\cca{63.6}$ & $30.6$ &$\cca{0.5}$ & $29.5$ &$\cca{1.0}$ \\ 
    &\texttt{HMid-A}       & $91.8$ &$\cca{71.6}$  & $102.2$ &$\cca{69.2}$ & $30.5$ &$\cca{1.2}$ & $29.6$ &$\cca{0.8}$ \\ 
    &\texttt{HMid-B}       & $71.0$ &$\cca{83.0}$  & $81.6$ &$\cca{80.3}$ & $30.4$ &$\cca{1.8}$ & $29.6$ &$\cca{0.7}$ \\ 
    \hline
    \end{tabular}
    \caption{Evaluation metrics for $\fnlloc$ inference across the bounded mass bins. RMSE scores are rounded to one decimal place, while R2 values are presented as percentiles and visually highlighted using colored cells. The defined mass bins (in units of $10^{13}\ \rm{M}_\odot/h$) are as follows: \texttt{HLow-A} ranges $[3.28,\ 4.46)$, \texttt{HLow-B} in $[4.46,\ 7.09)$, \texttt{HMid-A} between $[7.09,\ 9.06)$, and \texttt{HMid-B} includes $[9.06,\ 13.26)$.}
    \label{tab:fnl_loc_metrics_full_1}
\end{table}

\begin{table}[ht!]
    \small
    \centering
    \begin{tabular}{|c|l|c c|c c|c c|}
    \hline
    Model & Mass bin 
    & \multicolumn{2}{c|}{\texttt{LH\_EQ600}} 
    & \multicolumn{2}{c|}{\texttt{LH\_EQ250}} 
    & \multicolumn{2}{c|}{\texttt{LH\_EQ\_om\_s8}} \\
    & & RMSE & {R2} 
      & RMSE & {R2} 
      & RMSE & {R2} \\[0.5ex] 
    \hline\hline
    \multirow{3}{*}{\makecell[t]{PSBS \\
    (\textbf{MLP})}}
    &\texttt{HLow-A}       & $270.8$ &$\cca{45.9}$  & $149.0$ &$\cca{5.6}$ & $144.0$ &$\cca{5.9}$ \\ 
    &\texttt{HLow-B}       & $219.5$ &$\cca{64.5}$  & $136.9$ &$\cca{20.3}$ & $147.0$ &$\cca{1.8}$ \\ 
    &\texttt{HMid-A}       & $207.6$ &$\cca{68.2}$  & $135.2$ &$\cca{22.4}$ & $148.1$ &$\cca{0.4}$ \\ 
    &\texttt{HMid-B}       & $160.7$ &$\cca{81.0}$  & $118.8$ &$\cca{40.0}$ & $147.7$ &$\cca{0.8}$ \\ 
    \hline
    \multirow{3}{*}{\makecell[t]{PD \\
    (\textbf{\deepsets})}}
    &\texttt{HLow-A}       & $368.8$ &$-0.4$  & $154.1$ &$-0.9$ & $141.2$ &$-0.1$ \\ 
    &\texttt{HLow-B}       & $369.2$ &$-0.6$  & $153.8$ &$-0.5$ & $141.2$ &$\cca{-0.0}$ \\ 
    &\texttt{HMid-A}       & $320.2$ &$\cca{24.0}$  & $153.7$ &$-0.3$ & $141.1$ &$\cca{-0.0}$ \\ 
    &\texttt{HMid-B}       & $237.6$ &$\cca{58.0}$  & $153.5$ &$-0.1$ & $140.7$ &$\cca{0.6}$ \\ 
    \hline
    \multirow{3}{*}{\makecell[t]{PI \\
    (\textbf{\cnn})}}
    &\texttt{HLow-A}       & $333.7$ &$\cca{17.8}$  & $153.9$ &$-0.6$ & $141.4$ &$-0.4$ \\ 
    &\texttt{HLow-B}       & $286.9$ &$\cca{39.2}$  & $151.4$ &$\cca{2.6}$ & $141.6$ &$-0.6$ \\ 
    &\texttt{HMid-A}       & $267.7$ &$\cca{47.1}$  & $149.4$ &$\cca{5.1}$ & $141.2$ &$-0.1$ \\ 
    &\texttt{HMid-B}       & $211.2$ &$\cca{67.1}$  & $141.2$ &$\cca{15.3}$ & $140.1$ &$\cca{1.5}$ \\ 
    \hline
    \multirow{3}{*}{\makecell[t]{PD-Statistics \\
    (\textbf{XGB})}}
    &\texttt{HLow-A}       & $245.5$ &$\cca{55.5}$  & $145.2$ &$\cca{10.4}$ & $143.3$ &$\cca{6.8}$ \\ 
    &\texttt{HLow-B}       & $200.8$ &$\cca{70.2}$  & $136.5$ &$\cca{20.8}$ & $147.5$ &$\cca{1.2}$ \\ 
    &\texttt{HMid-A}       & $205.2$ &$\cca{68.9}$  & $135.5$ &$\cca{22.0}$ & $147.7$ &$\cca{0.9}$ \\ 
    &\texttt{HMid-B}       & $159.4$ &$\cca{81.2}$  & $121.3$ &$\cca{37.5}$ & $146.9$ &$\cca{2.0}$ \\ 
    \hline
    \multirow{3}{*}{\makecell[t]{PD-Statistics/PSBS \\
    (\textbf{XGB})}}
    &\texttt{HLow-A}       & $241.6$ &$\cca{56.9}$  & $143.8$ &$\cca{12.1}$ & $143.8$ &$\cca{6.1}$ \\ 
    &\texttt{HLow-B}       & $199.2$ &$\cca{70.7}$  & $133.8$ &$\cca{23.9}$ & $147.1$ &$\cca{1.7}$ \\ 
    &\texttt{HMid-A}       & $199.2$ &$\cca{70.7}$  & $133.4$ &$\cca{24.3}$ & $147.5$ &$\cca{1.2}$ \\ 
    &\texttt{HMid-B}       & $161.8$ &$\cca{80.7}$  & $120.0$ &$\cca{38.8}$ & $147.5$ &$\cca{1.2}$ \\ 
    \hline
    \multirow{3}{*}{\makecell[t]{PD-Histogram \\
    (\textbf{MLP})}}
    &\texttt{HLow-A}       & $270.6$ &$\cca{46.0}$  & $150.2$ &$\cca{4.0}$ & $146.1$ &$\cca{3.1}$ \\ 
    &\texttt{HLow-B}       & $227.6$ &$\cca{61.8}$  & $147.0$ &$\cca{8.1}$ & $147.9$ &$\cca{0.6}$ \\ 
    &\texttt{HMid-A}       & $242.2$ &$\cca{56.7}$  & $143.5$ &$\cca{12.5}$ & $147.7$ &$\cca{0.9}$ \\ 
    &\texttt{HMid-B}       & $189.8$ &$\cca{73.4}$  & $130.6$ &$\cca{27.5}$ & $147.3$ &$\cca{1.5}$ \\ 
    \hline
    \multirow{3}{*}{\makecell[t]{PD-Betti \\
    (\textbf{MLP})}}
    &\texttt{HLow-A}       & $276.1$ &$\cca{43.7}$  & $151.4$ &$\cca{2.5}$ & $145.9$ &$\cca{3.2}$ \\ 
    &\texttt{HLow-B}       & $226.6$ &$\cca{62.1}$  & $145.0$ &$\cca{10.6}$ & $148.3$ &$\cca{0.1}$ \\ 
    &\texttt{HMid-A}       & $244.6$ &$\cca{55.8}$  & $143.2$ &$\cca{12.9}$ & $146.8$ &$\cca{2.2}$ \\ 
    &\texttt{HMid-B}       & $190.6$ &$\cca{73.2}$  & $128.4$ &$\cca{29.9}$ & $146.8$ &$\cca{2.1}$ \\ 
    \hline
    \multirow{3}{*}{\makecell[t]{PD-PerBetti \\
    (\textbf{MLP})}}
    &\texttt{HLow-A}       & $276.3$ &$\cca{43.7}$  & $150.1$ &$\cca{4.2}$ & $143.7$ &$\cca{6.2}$ \\ 
    &\texttt{HLow-B}       & $227.4$ &$\cca{61.9}$  & $142.3$ &$\cca{13.9}$ & $147.1$ &$\cca{1.7}$ \\ 
    &\texttt{HMid-A}       & $246.2$ &$\cca{55.3}$  & $143.0$ &$\cca{13.1}$ & $146.3$ &$\cca{2.8}$ \\ 
    &\texttt{HMid-B}       & $191.4$ &$\cca{73.0}$  & $129.7$ &$\cca{28.5}$ & $146.8$ &$\cca{2.1}$ \\ 
    \hline
    \multirow{3}{*}{\makecell[t]{PD-Silhouette \\
    (\textbf{XGB})}}
    &\texttt{HLow-A}       & $298.7$ &$\cca{34.2}$  & $152.8$ &$\cca{0.7}$ & $145.6$ &$\cca{3.7}$ \\ 
    &\texttt{HLow-B}       & $252.7$ &$\cca{52.9}$  & $145.8$ &$\cca{9.6}$ & $147.7$ &$\cca{0.9}$ \\ 
    &\texttt{HMid-A}       & $258.9$ &$\cca{50.5}$  & $145.9$ &$\cca{9.5}$ & $148.8$ &$-0.6$ \\ 
    &\texttt{HMid-B}       & $214.1$ &$\cca{66.2}$  & $136.6$ &$\cca{20.6}$ & $148.1$ &$\cca{0.3}$ \\ 
    \hline
    \multirow{3}{*}{\makecell[t]{PD-Landscape \\
    (\textbf{XGB})}}
    &\texttt{HLow-A}       & $327.8$ &$\cca{20.7}$  & $154.1$ &$-0.9$ & $148.6$ &$-0.4$ \\ 
    &\texttt{HLow-B}       & $293.7$ &$\cca{36.3}$  & $151.0$ &$\cca{3.1}$ & $148.7$ &$-0.5$ \\ 
    &\texttt{HMid-A}       & $280.9$ &$\cca{41.8}$  & $148.2$ &$\cca{6.7}$ & $148.6$ &$-0.3$ \\ 
    &\texttt{HMid-B}       & $228.0$ &$\cca{61.6}$  & $138.8$ &$\cca{18.2}$ & $148.8$ &$-0.6$ \\ 
    \hline
    \end{tabular}
    \caption{Inference accuracy for $\fnleq$ evaluated across bunded mass bins. RMSE and R2 metrics are rounded to a single decimal; R2 values are percentile-based and highlighted with colored formatting. The mass bins (in units of $10^{13}\ \rm{M}_\odot/h$) are defined as follows: \texttt{HLow-A} spans $[3.28,\ 4.46)$, \texttt{HLow-B} covers $[4.46,\ 7.09)$, \texttt{HMid-A} includes $[7.09,\ 9.06)$, and \texttt{HMid-B} corresponds to $[9.06,\ 13.26)$.}
    \label{tab:fnl_equil_metrics_full_1}
\end{table}

\bibliographystyle{JHEP}
\bibliography{biblio.bib}

\end{document}